\pdfoutput=1
\documentclass[12pt,a4paper]{article}
\usepackage{ifthen} % for conditional statements
\usepackage{overpic} % 
\usepackage[utf8]{inputenc}

\newboolean{pdflatex}
\setboolean{pdflatex}{true} % False for eps figures 

\newboolean{articletitles}
\setboolean{articletitles}{true} % False removes titles in references

\newboolean{uprightparticles}
\setboolean{uprightparticles}{false} %True for upright particle symbols

\def\paperauthors{LHCb collaboration} % Leave as is for PAPER and CONF
\def\paperasciititle{Search for lepton-universality violation in B->Kll decays} % Set ASCII title here
\def\papertitle{Search for lepton-universality violation in \BuKll decays} % Latex formatted title
\def\paperkeywords{{High Energy Physics}, {LHCb}, {Lepton Universality}, {rare decays}} % Comma separated list
\def\papercopyright{\the\year\ CERN for the benefit of the LHCb collaboration} % new since 9/Apr/2018
\def\paperlicence{CC-BY-4.0 licence}
\def\paperlicenceurl{https://creativecommons.org/licenses/by/4.0/}

\usepackage[top=1in, bottom=1.25in, left=1in, right=1in]{geometry}

\usepackage{microtype}
\usepackage{lineno}  % for line numbering during review
\usepackage{xspace} % To avoid problems with missing or double spaces after
\usepackage{caption} %these three command get the figure and table captions automatically small

\usepackage{graphicx}  % to include figures (can also use other packages)
\usepackage{color}
\usepackage{colortbl}
\graphicspath{{./figs/}} % Make Latex search fig subdir for figures
\DeclareGraphicsExtensions{.pdf,.PDF,png,.PNG}

\usepackage{amsmath} % Adds a large collection of math symbols
\usepackage{amssymb}
\usepackage{amsfonts}
\usepackage{upgreek} % Adds in support for greek letters in roman typeset

\newcommand*\patchAmsMathEnvironmentForLineno[1]{%
\expandafter\let\csname old#1\expandafter\endcsname\csname #1\endcsname
\expandafter\let\csname oldend#1\expandafter\endcsname\csname
end#1\endcsname
 \renewenvironment{#1}%
   {\linenomath\csname old#1\endcsname}%
   {\csname oldend#1\endcsname\endlinenomath}%
}
\newcommand*\patchBothAmsMathEnvironmentsForLineno[1]{%
  \patchAmsMathEnvironmentForLineno{#1}%
  \patchAmsMathEnvironmentForLineno{#1*}%
}
\AtBeginDocument{%
\patchBothAmsMathEnvironmentsForLineno{equation}%
\patchBothAmsMathEnvironmentsForLineno{align}%
\patchBothAmsMathEnvironmentsForLineno{flalign}%
\patchBothAmsMathEnvironmentsForLineno{alignat}%
\patchBothAmsMathEnvironmentsForLineno{gather}%
\patchBothAmsMathEnvironmentsForLineno{multline}%
\patchBothAmsMathEnvironmentsForLineno{eqnarray}%
}

\usepackage{hyperxmp}

\usepackage[pdftex,
            pdfauthor={\paperauthors},
            pdftitle={\paperasciititle},
            pdfkeywords={\paperkeywords},
            pdfcopyright={Copyright (C) \papercopyright},
            pdflicenseurl={\paperlicenceurl}]{hyperref}

\usepackage[colorinlistoftodos,textsize=scriptsize]{todonotes}

\usepackage[all]{hypcap} % Internal hyperlinks to floats.

\usepackage{xspace} 
\usepackage{upgreek}

\newcommand{\offsetoverline}[2][0.1em]{\kern #1\overline{\kern -#1 #2}}%

\def\lhcb   {\mbox{LHCb}\xspace}

\def\MagUp {\mbox{\em Mag\kern -0.05em Up}\xspace}

\ifthenelse{\boolean{uprightparticles}}%
{

 \def\Pmu         {\ensuremath{\upmu}\xspace}                 
 \def\Pnu         {\ensuremath{\upnu}\xspace}                 
                  
 \def\Ppi         {\ensuremath{\uppi}\xspace}

 \def\Ppsi        {\ensuremath{\uppsi}\xspace}

 \def\PDelta      {\ensuremath{\Delta}\xspace}                 
 \def\PXi         {\ensuremath{\Xi}\xspace}                 
 \def\PLambda     {\ensuremath{\Lambda}\xspace}                 
 \def\PSigma      {\ensuremath{\Sigma}\xspace}                 
 \def\POmega      {\ensuremath{\Omega}\xspace}                 
 \def\PUpsilon    {\ensuremath{\Upsilon}\xspace}

 \def\PB      {\ensuremath{\mathrm{B}}\xspace}                 
                  
 \def\PD      {\ensuremath{\mathrm{D}}\xspace}

 \def\PJ      {\ensuremath{\mathrm{J}}\xspace}                 
 \def\PK      {\ensuremath{\mathrm{K}}\xspace}

 \def\Pb      {\ensuremath{\mathrm{b}}\xspace}                 
 \def\Pc      {\ensuremath{\mathrm{c}}\xspace}                 
                  
 \def\Pe      {\ensuremath{\mathrm{e}}\xspace}

 \def\Pi      {\ensuremath{\mathrm{i}}\xspace}

 \def\Ps      {\ensuremath{\mathrm{s}}\xspace}

}
{

 \def\Pmu         {\ensuremath{\mu}\xspace}                 
 \def\Pnu         {\ensuremath{\nu}\xspace}                 
                  
 \def\Ppi         {\ensuremath{\pi}\xspace}

 \def\Ppsi        {\ensuremath{\psi}\xspace}                 
                  
 \mathchardef\PDelta="7101
 \mathchardef\PXi="7104
 \mathchardef\PLambda="7103
 \mathchardef\PSigma="7106
 \mathchardef\POmega="710A
 \mathchardef\PUpsilon="7107
                  
 \def\PB      {\ensuremath{B}\xspace}                 
                  
 \def\PD      {\ensuremath{D}\xspace}

 \def\PJ      {\ensuremath{J}\xspace}                 
 \def\PK      {\ensuremath{K}\xspace}

 \def\Pb      {\ensuremath{b}\xspace}                 
 \def\Pc      {\ensuremath{c}\xspace}                 
                  
 \def\Pe      {\ensuremath{e}\xspace}

 \def\Pi      {\ensuremath{i}\xspace}

 \def\Ps      {\ensuremath{s}\xspace}

}

\makeatletter
\ifcase \@ptsize \relax% 10pt
  \newcommand{\miniscule}{\@setfontsize\miniscule{4}{5}}% \tiny: 5/6
\or% 11pt
  \newcommand{\miniscule}{\@setfontsize\miniscule{5}{6}}% \tiny: 6/7
\or% 12pt
  \newcommand{\miniscule}{\@setfontsize\miniscule{5}{6}}% \tiny: 6/7
\fi
\makeatother

\DeclareRobustCommand{\optbar}[1]{\shortstack{{\miniscule (\rule[.5ex]{1.25em}{.18mm})}
  \\ [-.7ex] $#1$}}

\def\en         {{\ensuremath{\Pe^-}}\xspace}   % electron negative (\em is taken)
\def\ep         {{\ensuremath{\Pe^+}}\xspace}
\def\epm        {{\ensuremath{\Pe^\pm}}\xspace} 
 
\def\epem       {{\ensuremath{\Pe^+\Pe^-}}\xspace}
\def\mumu       {{\ensuremath{\Pmu^+\Pmu^-}}\xspace}

\def\ellm       {{\ensuremath{\ell^-}}\xspace}
\def\ellp       {{\ensuremath{\ell^+}}\xspace}
\def\ellell     {\ensuremath{\ell^+ \ell^-}\xspace}

\def\neu        {{\ensuremath{\Pnu}}\xspace}
\def\neub       {{\ensuremath{\overline{\Pnu}}}\xspace}
\def\neue       {{\ensuremath{\neu_e}}\xspace}
\def\neueb      {{\ensuremath{\neub_e}}\xspace}
\def\squark    {{\ensuremath{\Ps}}\xspace}

\def\cquark    {{\ensuremath{\Pc}}\xspace}

\def\bquark    {{\ensuremath{\Pb}}\xspace}

\def\pion   {{\ensuremath{\Ppi}}\xspace}

\def\pip    {{\ensuremath{\pion^+}}\xspace}
\def\pim    {{\ensuremath{\pion^-}}\xspace}

\def\kaon    {{\ensuremath{\PK}}\xspace}
  \def\Kbar    {{\kern 0.2em\overline{\kern -0.2em \PK}{}}\xspace}

\def\KorKbar {\kern 0.18em\optbar{\kern -0.18em K}{}\xspace}

\def\Kp      {{\ensuremath{\kaon^+}}\xspace}
\def\Km      {{\ensuremath{\kaon^-}}\xspace}

\def\Kstarz  {{\ensuremath{\kaon^{*0}}}\xspace}

\def\Kstar   {{\ensuremath{\kaon^*}}\xspace}

  \def\Dbar    {{\kern 0.2em\overline{\kern -0.2em \PD}{}}\xspace}
\def\D       {{\ensuremath{\PD}}\xspace}

\def\DorDbar {\kern 0.18em\optbar{\kern -0.18em D}{}\xspace}
\def\Dz      {{\ensuremath{\D^0}}\xspace}
\def\Dzb     {{\ensuremath{\Dbar{}^0}}\xspace}
\def\Dp      {{\ensuremath{\D^+}}\xspace}

\def\Dstarp  {{\ensuremath{\D^{*+}}}\xspace}

\def\Ds      {{\ensuremath{\D^+_\squark}}\xspace}

\def\B       {{\ensuremath{\PB}}\xspace}
\def\Bbar    {{\ensuremath{\kern 0.18em\overline{\kern -0.18em \PB}{}}}\xspace}

\def\BorBbar    {\kern 0.18em\optbar{\kern -0.18em B}{}\xspace}
\def\Bz      {{\ensuremath{\B^0}}\xspace}

\def\Bu      {{\ensuremath{\B^+}}\xspace}

\def\Bp      {{\ensuremath{\Bu}}\xspace}

\def\Bd      {{\ensuremath{\B^0}}\xspace}
\def\Bs      {{\ensuremath{\B^0_\squark}}\xspace}

\def\jpsi     {{\ensuremath{{\PJ\mskip -3mu/\mskip -2mu\Ppsi\mskip 2mu}}}\xspace}
\def\psitwos  {{\ensuremath{\Ppsi{(2S)}}}\xspace}

\def\Y#1S{\ensuremath{\PUpsilon{(#1S)}}\xspace}

\def\Lz          {{\ensuremath{\PLambda}}\xspace}

\def\LorLbar     {\kern 0.18em\optbar{\kern -0.18em \PLambda}{}\xspace}

\def\Lc          {{\ensuremath{\Lz^+_\cquark}}\xspace}

\def\Lb           {{\ensuremath{\Lz^0_\bquark}}\xspace}

\def\BF         {{\ensuremath{\mathcal{B}}}\xspace}
\def\BR         {\BF}

\newcommand{\decay}[2]{\mbox{\ensuremath{#1\!\to #2}}\xspace}         % {\Pa}{\Pb \Pc}

\def\to                 {\ensuremath{\rightarrow}\xspace}

\def\qsq       {{\ensuremath{q^2}}\xspace}

\def\BdKstmm  {\decay{\Bd}{\Kstarz\mumu}}

\def\bsll     {\decay{\bquark}{\squark \ell^+ \ell^-}}

\def\AT#1     {\ensuremath{A_{\mathrm{T}}^{#1}}\xspace}           % 2

\def\C#1      {\ensuremath{\mathcal{C}_{#1}}\xspace}                       % 9
\def\Cp#1     {\ensuremath{\mathcal{C}_{#1}^{'}}\xspace}                    % 7
\def\Ceff#1   {\ensuremath{\mathcal{C}_{#1}^{\mathrm{(eff)}}}\xspace}        % 9  
\def\Cpeff#1  {\ensuremath{\mathcal{C}_{#1}^{'\mathrm{(eff)}}}\xspace}       % 7
\def\Ope#1    {\ensuremath{\mathcal{O}_{#1}}\xspace}                       % 2
\def\Opep#1   {\ensuremath{\mathcal{O}_{#1}^{'}}\xspace}                    % 7

\newcommand{\aunit}[1]{\ensuremath{\text{\,#1}}}       
\newcommand{\tev}{\aunit{Te\kern -0.1em V}\xspace}
\newcommand{\gev}{\aunit{Ge\kern -0.1em V}\xspace}
\newcommand{\mev}{\aunit{Me\kern -0.1em V}\xspace}
\newcommand{\kev}{\aunit{ke\kern -0.1em V}\xspace}
\newcommand{\ev}{\aunit{e\kern -0.1em V}\xspace}
\newcommand{\mevc}{\ensuremath{\aunit{Me\kern -0.1em V\!/}c}\xspace}
\newcommand{\gevc}{\ensuremath{\aunit{Ge\kern -0.1em V\!/}c}\xspace}
\newcommand{\mevcc}{\ensuremath{\aunit{Me\kern -0.1em V\!/}c^2}\xspace}
\newcommand{\gevcc}{\ensuremath{\aunit{Ge\kern -0.1em V\!/}c^2}\xspace}
\newcommand{\gevgevcccc}{\ensuremath{\gev^2\!/c^4}\xspace} % for q^2

\def\fb   {\ensuremath{\aunit{fb}}\xspace}
\def\invfb   {\ensuremath{\fb^{-1}}\xspace}

\newcommand{\chisq}{\ensuremath{\chi^2}\xspace}

\newcommand{\chisqip}{\ensuremath{\chi^2_{\text{IP}}}\xspace}

\def\deriv {\ensuremath{\mathrm{d}}}

\def\gsim{{~\raise.15em\hbox{$>$}\kern-.85em
          \lower.35em\hbox{$\sim$}~}\xspace}
\def\lsim{{~\raise.15em\hbox{$<$}\kern-.85em
          \lower.35em\hbox{$\sim$}~}\xspace}

\def\pt         {\ensuremath{p_{\mathrm{T}}}\xspace}

\def\photosplusplus     {\mbox{\textsc{Photos}++}\xspace}

\xspace

\def\tell1  {TELL1\xspace}
\def\ukl1   {UKL1\xspace}

\newcommand{\eg}{\mbox{\itshape e.g.}\xspace}
\newcommand{\ie}{\mbox{\itshape i.e.}\xspace}

\newcommand{\etc}{\mbox{\itshape etc.}\xspace}

\def\RK         {\ensuremath{R_{\kaon}}\xspace}
\def\RPsitwos   {\ensuremath{R_{\kaon}^{\psitwos}}\xspace}
\def\rjpsi      {\ensuremath{r_{\jpsi}}\xspace}
\def\RH         {\ensuremath{R_{H}}\xspace}
\def\RKstar     {\ensuremath{R_{\Kstarz}}\xspace}
\def\Kee        {\ensuremath{\Kp\epem}\xspace}

\def\Kll        {\ensuremath{\Kp\ellell}\xspace}

\def\mKee       {\ensuremath{m({\Kp\epem})}\xspace}
\def\mKmm       {\ensuremath{m({\Kp\mumu})}\xspace}
\def\mKll       {\ensuremath{m({\Kp\ellell})}\xspace}

\def\mKllconst  {\ensuremath{m_{\jpsi}{(\Kp\ellell)}}\xspace}

\def\mKllgeneric  {\ensuremath{m_{(\jpsi)}{(\Kp\ellell)}}\xspace}

\def\BuJpsiK  {\decay{\Bu}{\jpsi\Kp}}

\def\BuJpsiKll  {\decay{\Bu}{\jpsi(\to\ellell)\Kp}}

\def\BuPsiKll  {\decay{\Bu}{\psitwos(\to\ellell)\Kp}}
\def\BuKll  {\decay{\Bu}{\Kp\ellell}}

\def\BuJpsiKmm  {\decay{\Bu}{\jpsi(\to\mumu)\Kp}}
\def\BuPsiKmm  {\decay{\Bu}{\psitwos(\to\mumu)\Kp}}
\def\BuKmm  {\decay{\Bu}{\Kp\mumu}}

\def\BuzHmm  {\decay{\B}{H\mumu}}
\def\BuJpsiKee  {\decay{\Bu}{\jpsi(\to\epem)\Kp}}
\def\BuPsiKee  {\decay{\Bu}{\psitwos(\to\epem)\Kp}}
\def\BuKee  {\decay{\Bu}{\Kp\epem}}

\def\BuzHee  {\decay{\B}{H\epem}}
\def\BuKPhill {\decay{\Bu}{\phi(\to\ellell)\Kp}}
\def\BuJpsipi  {\decay{\Bu}{\jpsi\pip}}
\def\BdorBu  {{\ensuremath{\B^{0,+}}}\xspace}
\def\KstarzorKstarp  {{\ensuremath{\kaon^{*}(892)^{(0,+)}}}\xspace}
\def\BuBdKstarzorKstplusee {\decay{\BdorBu}{\KstarzorKstarp(\to \Kp\pi^{(-,0)})\epem}}
\def\BuBdKstarzorKstplusjpsi {\decay{\BdorBu}{\jpsi(\to\epem)\KstarzorKstarp(\to \Kp\pi^{(-,0)})}}

\def\BuDzenu {\decay{\Bu}{\Dzb(\to \Kp\en\neueb)\ep\neue}}
\def\Hb {{\ensuremath{H_b}}\xspace}
\def\Hc {{\ensuremath{H_c}}\xspace}
\def\HbtoHc {\decay{\Hb}{\Hc(\to\Kp\ellm\neub X)\ellp\nu Y}}
\def\BuKpipi  {\decay{\Bu}{\Kp\pip\pim}}

\def\RKvalue {\ensuremath{0.846\,^{+\,0.060}_{-\,0.054}\,^{+\,0.016}_{-\,0.014}}}

\def\significance {{2.5}\xspace}

\def\RKrunonevalue {\ensuremath{0.717\,^{+\,0.083}_{-\,0.071}\,^{+\,0.017}_{-\,0.016}}}

\def\RKruntwovalue {\ensuremath{0.928\,^{+\,0.089}_{-\,0.076}\,^{+\,0.020}_{-\,0.017}}}

\def\RKrunone         {\ensuremath{R_{\kaon}^{{7 \mathrm{\,and\,} 8\tev }}}\xspace}
\def\RKruntwo   {\ensuremath{R_{\kaon}^{{\mathrm{13\tev }}}}\xspace}
\usepackage{cite} % Allows for ranges in citations
\usepackage{mciteplus}
\usepackage{multirow}
\usepackage{booktabs}

\begin{document}

%%%%%%%%%%%%%%%%%%%%%%%%%
%%%%% Title     %%%%%%%%%
%%%%%%%%%%%%%%%%%%%%%%%%%
\renewcommand{\thefootnote}{\fnsymbol{footnote}}
\setcounter{footnote}{1}

% %%%%%%% CHOOSE TITLE PAGE--------
%\onecolumn
%\input{title-LHCb-INT}
%\input{title-LHCb-ANA}
%\input{title-LHCb-CONF}
% $Id: title-LHCb-PAPER.tex 122889 2018-08-17 17:59:55Z pkoppenb $
% ===============================================================================
% Purpose: LHCb-PAPER journal paper title page template
% Author: 
% Created on: 2010-09-25
% ===============================================================================

%%%%%%%%%%%%%%%%%%%%%%%%%
%%%%%  TITLE PAGE  %%%%%%
%%%%%%%%%%%%%%%%%%%%%%%%%
\begin{titlepage}
\pagenumbering{roman}

% Header ---------------------------------------------------
\vspace*{-1.5cm}
\centerline{\large EUROPEAN ORGANIZATION FOR NUCLEAR RESEARCH (CERN)}
\vspace*{1.5cm}
\noindent
\begin{tabular*}{\linewidth}{lc@{\extracolsep{\fill}}r@{\extracolsep{0pt}}}
\ifthenelse{\boolean{pdflatex}}% Logo format choice
{\vspace*{-1.5cm}\mbox{\!\!\!\includegraphics[width=.14\textwidth]{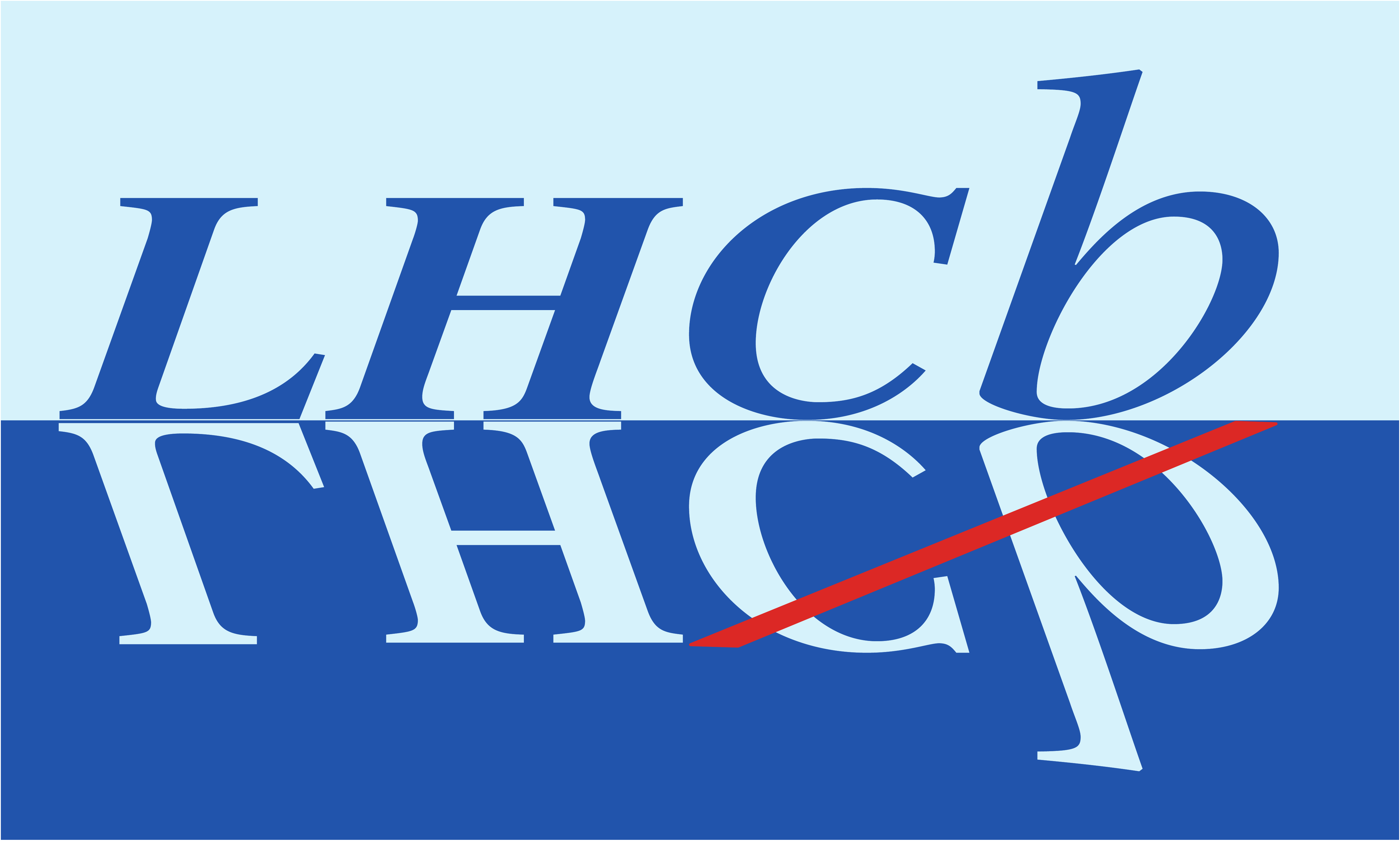}} & &}%
{\vspace*{-1.2cm}\mbox{\!\!\!\includegraphics[width=.12\textwidth]{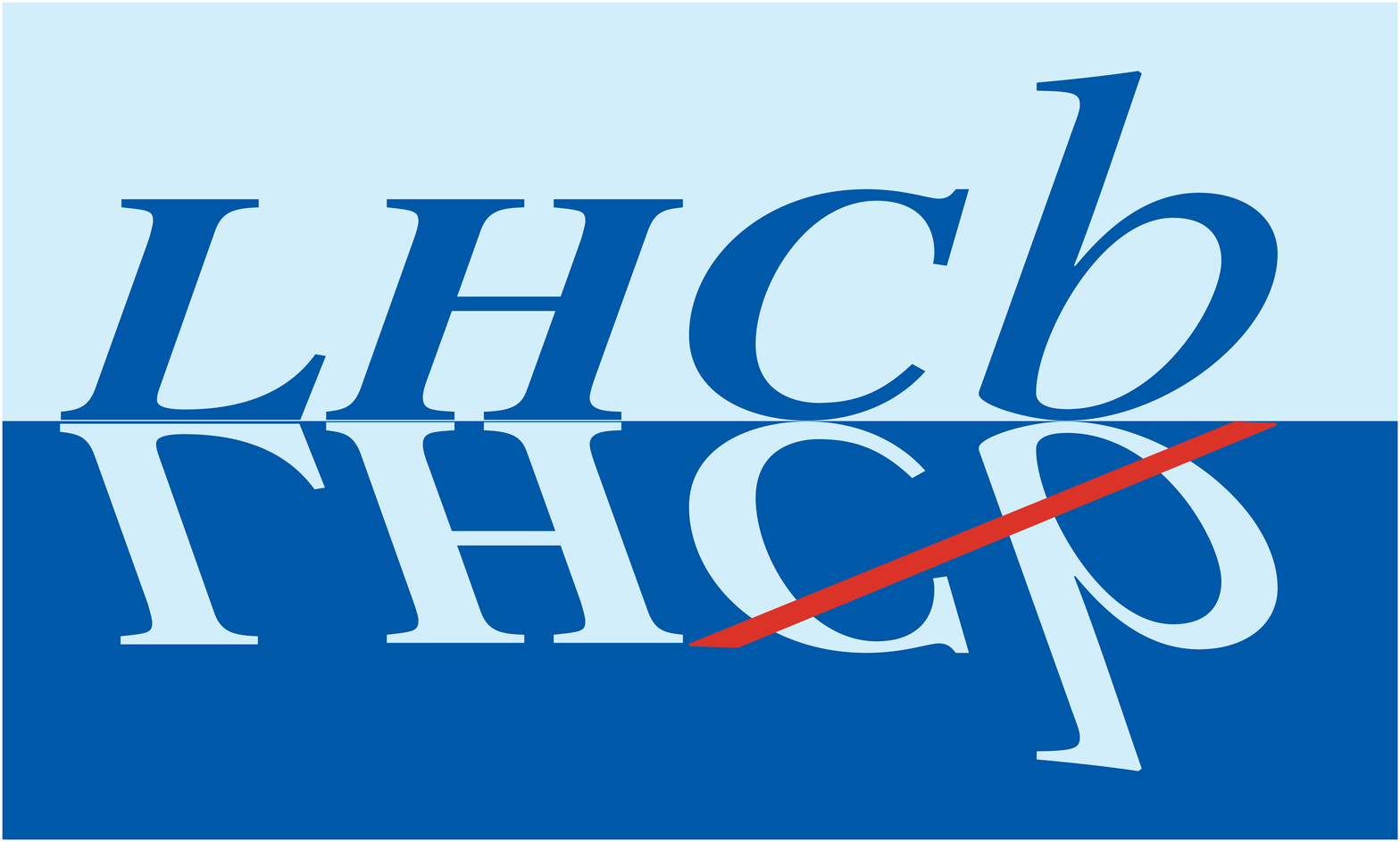}} & &}%
\\
 & & CERN-EP-2019-043 \\  % ID 
 & & LHCb-PAPER-2019-009 \\  % ID 
 & & 22 March 2019 \\ % Date - Can also hardwire e.g.: 23 March 2010
 & & \\
% not in paper \hline
\end{tabular*}

\vspace*{4.0cm}

% Title --------------------------------------------------
{\normalfont\bfseries\boldmath\huge
\begin{center}
% DO NOT EDIT HERE. Instead edit macro in main.tex to keep metadata correct
  \papertitle 
\end{center}
}

\vspace*{2.0cm}

% Authors -------------------------------------------------
\begin{center}
%In the footnote, replace 'paper' by 'Letter' in case of submission to PRL or PLB 
% Edit macro in main.tex to keep metadata correct
\paperauthors\footnote{Full author list given at the end of the Letter.}
\end{center}

\vspace{\fill}

% Abstract -----------------------------------------------
\begin{abstract}
  \noindent
A measurement  of the ratio of branching fractions of the decays \BuKmm and \BuKee is presented. The proton-proton collision data used correspond to
 an integrated luminosity of $5.0\invfb$ recorded with the LHCb experiment at centre-of-mass energies of $7$, $8$ and $13\tev$.
For the dilepton mass-squared range $1.1 < \qsq < 6.0 \gevgevcccc$ the ratio of branching fractions is measured to be \mbox{$R_K = \RKvalue$},
where the first uncertainty is statistical and the second systematic.
This is the most precise measurement of \RK to date and is compatible with the Standard Model at the level of \significance standard deviations.

\end{abstract}

\vspace*{2.0cm}

\begin{center}
  Published in Phys.~Rev.~Lett. 122 (2019) 191801.
\end{center}

\vspace{\fill}

{\footnotesize 
% Edit macro in main.tex to keep metadata correct
\centerline{\copyright~\papercopyright. \href{\paperlicenceurl}{\paperlicence}.}}
\vspace*{2mm}

\end{titlepage}

%%%%%%%%%%%%%%%%%%%%%%%%%%%%%%%%
%%%%%  EOD OF TITLE PAGE  %%%%%%
%%%%%%%%%%%%%%%%%%%%%%%%%%%%%%%%

%  empty page follows the title page ----
\newpage
\setcounter{page}{2}
\mbox{~}

\cleardoublepage

%\twocolumn
% %%%%%%%%%%%%% ---------

\renewcommand{\thefootnote}{\arabic{footnote}}
\setcounter{footnote}{0}

%%%%%%%%%%%%%%%%%%%%%%%%%%%%%%%%
%%%%%  Table of Content   %%%%%%
%%%%%%%%%%%%%%%%%%%%%%%%%%%%%%%%
%%%% Uncomment next 2 lines if desired
%\tableofcontents
%\cleardoublepage

%%%%%%%%%%%%%%%%%%%%%%%%%
%%%%% Main text %%%%%%%%%
%%%%%%%%%%%%%%%%%%%%%%%%%

\pagestyle{plain} % restore page numbers for the main text
\setcounter{page}{1}
\pagenumbering{arabic}

%% Uncomment during review phase. 
%% Comment before a final submission.
% \linenumbers

% You can include short sections directly in the main tex file.
% However, for larger papers it is desirable to split the text into
% several semiautonomous files, which can be revised independently.
% This is especially useful when developing a document in
% collaboration with several people, since then different parts can be
% edited independently.  This type of file organization is shown here.
% 

Decays involving \bsll transitions, where $\ell$ represents a lepton, are mediated by flavour-changing neutral currents. Such decays are suppressed in the Standard Model~(SM), as they proceed only through amplitudes that involve electroweak loop diagrams. 
 These processes are sensitive to virtual contributions from new particles, which could have masses that are inaccessible to direct searches for resonances, even at Large Hadron Collider experiments.

 Theoretical predictions for exclusive \bsll decays rely on the calculation of hadronic effects, and recent measurements have therefore focused on quantities where the uncertainties from such effects are reduced to some extent, such as angular observables and ratios of branching fractions. 
 The results of the angular analysis of the decay \BdKstmm~\cite{LHCb-PAPER-2015-051,Aaboud:2018krd,Aubert:2006vb,Lees:2015ymt,Wei:2009zv,Wehle:2016yoi,Aaltonen:2011ja,Khachatryan:2015isa,Sirunyan:2017dhj} and measurements of the branching fractions of several $\bsll$ decays~\cite{LHCb-PAPER-2016-012, LHCb-PAPER-2015-023, LHCb-PAPER-2014-006, LHCb-PAPER-2015-009} are in some tension with SM predictions~\cite{Altmannshofer:2017yso,Capdevila:2017bsm,Hurth:2017hxg,DAmico:2017mtc,Geng:2017svp,Ciuchini:2017mik}. However, the treatment of the hadronic effects in the theoretical predictions is still the subject of considerable debate~\cite{Jager:2014rwa, Descotes-Genon:2015uva,Lyon:2014hpa,Khodjamirian:2012rm,Khodjamirian:2010vf,Descotes-Genon:2014uoa,Horgan:2013pva, Beaujean:2013soa, Hambrock:2013zya, Altmannshofer:2013foa,Bobeth:2017vxj}.

The electroweak couplings of all three charged leptons are identical in the SM and, consequently, the decay properties (and the hadronic effects) are expected to be the same up to corrections related to the lepton mass, regardless of the lepton flavour (referred to as {\em lepton universality}). 
The ratio of branching fractions for \BuzHmm and \BuzHee decays, where $H$ is a hadron,  can be predicted precisely in an appropriately chosen range of the dilepton mass squared $q^{2}_{\rm min} < \qsq < q^{2}_{\rm max}$~\cite{Hiller:2003js, Wang:2003je}. This ratio is defined by
     \begin{equation}
\label{eq:rh}
\RH  =  \dfrac{\displaystyle\int_{q^2_\mathrm{min}}^{q^2_{\rm max}} \dfrac{\deriv\Gamma[\BuzHmm]}{\deriv\qsq} \deriv\qsq}{\displaystyle\int_{q^2_{\rm min}}^{q^2_{\rm max}} \dfrac{\deriv\Gamma[\BuzHee]}{\deriv\qsq} \deriv\qsq} ~,
\end{equation} 
     \noindent where $\Gamma$ is the \qsq-dependent partial width of the decay.  
     In the range $1.1 < \qsq < 6.0 \gevgevcccc$, such ratios are predicted to  be unity with $\mathcal{O}(1\%)$ precision~\cite{Bordone:2016gaq}. 
     The inclusion of charge-conjugate processes is implied throughout this Letter. 
     
      The most precise measurements of \RK in the region $1.0 < \qsq < 6.0 \gevgevcccc$ and \RKstar in the regions $0.045 < \qsq < 1.1 \gevgevcccc$ and $1.1 < \qsq < 6.0 \gevgevcccc$ have been made by the \lhcb collaboration and, depending on the theoretical prediction used,  are 2.6~\cite{LHCb-PAPER-2014-024}, $2.1$--$2.3$ and $2.4$--$2.5$ standard deviations~\cite{LHCb-PAPER-2017-013} below their respective SM expectations~\cite{Descotes-Genon:2015uva,Bobeth:2007,Bordone:2016gaq,Capdevila:2016ivx,Capdevila:2017ert,Serra:2016ivr,EOS,Straub:2015ica,Straub:2018kue,Altmannshofer:2017fio,Jager:2014rwa}. 
      These tensions and those observed in the angular and branching-fraction measurements can all be accommodated simultaneously in models with an additional heavy neutral gauge boson~\cite{Altmannshofer:2014cfa, Crivellin:2015mga, Celis:2015ara, Falkowski:2015zwa} or with leptoquarks~\cite{Hiller:2014yaa,Gripaios:2014tna,Varzielas:2015iva,Barbieri:2016las,Bordone2018}.
     
      This Letter presents the most precise measurement of the ratio \RK in the range $1.1 < \qsq < 6.0 \gevgevcccc$. The analysis is performed using 5.0\invfb of proton-proton collision data collected with the \lhcb detector during three data-taking periods in which the centre-of-mass energy of the collisions was $7$, $8$ and $13\tev$. The data were taken in the years 2011, 2012 and 2015--2016, respectively.
       Compared to the previous LHCb \RK measurement~\cite{LHCb-PAPER-2014-024}, the analysis benefits from a larger data sample (an additional $2.0\invfb$ collected in 2015--2016) and an improved reconstruction; moreover the lower limit of the \qsq range is increased, in order to be compatible with other \lhcb \bsll analyses and to suppress further the contribution from \BuKPhill decays. The results supersede those of Ref.~\cite{LHCb-PAPER-2014-024}.
       
Throughout this Letter \BuKll refers only to decays with \mbox{$1.1<\qsq<6.0\gevgevcccc$}, which are denoted nonresonant, whereas \BuJpsiKll decays are referred to as resonant. 
     The nonresonant \qsq range excludes the resonant \BuJpsiKll region and the high-\qsq region that contains contributions from excited charmonium resonances.

The analysis strategy is designed to  reduce systematic uncertainties induced by the markedly different reconstruction  of decays with muons in the final state compared to decays with electrons. These differences arise due to the significant bremsstrahlung emission of the electrons and the different signatures exploited in the online {\em trigger} selection.  
    Systematic uncertainties that would otherwise affect the calculation of the efficiencies of the \BuKmm and \BuKee decay modes are suppressed by measuring \RK as a double ratio of branching fractions, 
    \begin{equation}
    \label{eq:doubleratio}
       \RK = {\frac{\BR(\BuKmm)}{\BR(\BuJpsiKmm)}} \bigg{/} {\frac{\BR(\BuKee)}{\BR(\BuJpsiKee)}} \, .
    \end{equation}
    \noindent 
    The measurement requires knowledge of the observed yield and the efficiency to trigger, reconstruct and select each decay mode. 
    The use of this double ratio exploits the fact that $\jpsi \to \ell^+\ell^-$ decays are observed to have lepton-universal branching fractions within 0.4\%~\cite{Ablikim:2013pqa,PDG2018}. 
    Using Eq.~\eqref{eq:doubleratio} then requires the nonresonant \BuKee detection efficiency to be known only relative to that of the resonant \BuJpsiKee decay, rather than the \BuKmm decay. As the detector signatures of each resonant decay are similar to those of the corresponding nonresonant decay, systematic effects are  reduced and the precision on \RK is dominated by the statistical uncertainty.

After the application of selection criteria, which are discussed below, the four decay modes \BuJpsiKmm, \BuJpsiKee, \BuKmm and \BuKee are separated from the background  on a statistical basis, using fits to the \mKll distributions. 
For the resonant decays, the mass  \mKllconst 
is computed by constraining the dilepton system to the known \jpsi mass~\cite{PDG2018}. This improves the electron-mode mass resolution (full width at half maximum) from 140 to 24.5\mevcc and the muon-mode mass resolution from 30 to 17.5\mevcc. 
The \mKll fit ranges and the \qsq selection used for the different decay modes are shown in Table~\ref{tab:q2ranges}.
The selection requirements applied to the resonant and nonresonant decays are otherwise identical. 
The two ratios of efficiencies required to form  Eq.~\eqref{eq:doubleratio} are taken from simulation.
 The simulation is calibrated using data-derived control channels, including \BuJpsiKmm and  \BuJpsiKee. Correlations arising from the use of these decay modes both for this calibration and in the determination of the double ratio of Eq.~\eqref{eq:doubleratio} are taken into account. 
 A further feature of the analysis strategy is that the results were not inspected until all analysis procedures were finalised.

\begin{table}[t]
\centering
\caption{Resonant and nonresonant mode $\qsq$ and \mKll ranges. The variables \mKll and \mKllconst are used for nonresonant and resonant decays, respectively.}\label{tab:q2ranges}
\begin{tabular}{ccc}
\toprule
Decay mode & \qsq & \mKllgeneric \\
           & $[\gevgevcccc]$ & $[\gevcc]$ \\
\midrule
{\begin{tabular}{r@{\;}l}nonresonant&$e^+e^-$\\resonant&$e^+e^-$\\nonresonant&$\mu^+\mu^-$\\resonant&$\mu^+\mu^-$\end{tabular}}     & {\begin{tabular}{r@{.}l@{\;--\;}r@{.}l}1&1 & 6&0\\6&00 & 12&96\\1&1 & 6&0\\8&68&10&09\end{tabular}}    & {\begin{tabular}{r@{\;--\;}l}4.88&6.20\\5.08&5.70\\5.18&5.60\\5.18&5.60\end{tabular}}\\
\bottomrule
\end{tabular}
\end{table}

The LHCb detector is a single-arm forward spectrometer covering the pseudorapidity range $2 < \eta < 5$, described in detail in Refs.~\cite{Alves:2008zz,LHCb-DP-2014-002}. The detector includes a silicon-strip vertex detector surrounding the proton-proton interaction region, tracking stations on either side of a dipole magnet, ring-imaging Cherenkov (RICH) detectors, calorimeters and muon chambers. The simulation used in this analysis is produced using the software described in Refs.~\cite{Sjostrand:2006za,*Sjostrand:2007gs,LHCb-PROC-2010-056,Lange:2001uf,Golonka:2005pn,Allison:2006ve, *Agostinelli:2002hh,LHCb-PROC-2011-006}. Final-state radiation is simulated using \photosplusplus 3.61 in the default configuration~\cite{Golonka:2005pn, Davidson:2010ew}, which is observed to agree with a full quantum electrodynamics calculation at the level of $1\%$~\cite{Bordone:2016gaq}. 

Candidate events are first required to pass a hardware trigger that selects either a high transverse momentum (\pt) muon; or an electron, hadron or photon with high transverse energy deposited in the calorimeters. In this analysis, it is required that \BuKmm and \BuJpsiKmm candidates are triggered by one of the muons, whereas \BuKee and \BuJpsiKee candidates are required to be triggered in one of three ways: by either one of the electrons; by the kaon from the \Bp decay; or by  particles in the event that are not part of the signal candidate. 
In the software trigger, the tracks of the final-state particles are required to form a vertex that is significantly displaced from any of the primary proton-proton interaction vertices (PVs) in the event. A multivariate algorithm is used for the identification of secondary vertices consistent with the decay of a \bquark hadron~\cite{BBDT, LHCb-PROC-2015-018}.

Candidates are formed from a particle identified as a charged kaon, together with a pair of well-reconstructed oppositely charged particles identified as either electrons or muons.
Each particle is required to have sizeable \pt and to be inconsistent with coming from a PV.
The particles must originate from a common vertex with good vertex-fit quality, which is displaced significantly from all of the PVs in the event. The \Bp momentum vector is required to be aligned with the vector connecting one of the PVs in the event (subsequently referred to as the associated PV) and the \Bp decay vertex.

Kaons and muons are identified using the output of multivariate classifiers that exploit information from the tracking system, the RICH detectors, the calorimeters and the muon chambers~\cite{LHCb-DP-2014-002, LHCb-DP-2014-001, LHCb-DP-2013-003,LHCb-DP-2013-001,LHCb-DP-2012-003,LHCb-DP-2012-002}.
Electrons are identified by matching tracks to electromagnetic calorimeter~(ECAL) showers and adding information from the RICH detectors. 
The ratio of the energy detected in the ECAL to the momentum measured by the tracking system is central to this identification. 
If an electron radiates a photon downstream of the dipole magnet, the photon and electron deposit their energy in the same ECAL cells and the original energy of the electron is measured. However, if an electron radiates a photon upstream of the magnet, the energy of the photon will not be deposited in the same ECAL cells as the electron. For each electron track, a search is therefore made for ECAL showers around the extrapolated track direction (before the magnet) that are not associated with any other charged tracks. 
The energy of any such shower is added to the electron energy that is derived from the measurements made in the tracker.

Backgrounds from exclusive decays of \bquark hadrons and the so-called combinatorial background, formed from the reconstructed fragments of multiple heavy-flavor hadron decays, are reduced using selection criteria that are discussed below.
The muon modes benefit from superior mass resolution so that a reduced mass range can be used (see Table~\ref{tab:q2ranges}). Consequently, the only remaining backgrounds after the application of the selection criteria are combinatorial and, for the resonant mode, from the Cabibbo-suppressed decay \BuJpsipi, where the pion is misidentified as a kaon.
For the electron modes, where a wider mass range is used, significant residual exclusive backgrounds also contribute.
Since higher-mass \Kstar resonances are suppressed in the mass range selected, the dominant exclusive backgrounds for the resonant and nonresonant modes are from partially reconstructed \BuBdKstarzorKstplusjpsi and \BuBdKstarzorKstplusee decays, respectively, where the pion is not included in the candidate.
At the level of $\mathcal{O}(1\%)$ of the \Kee signal, there are also exclusive background contributions from \BuDzenu decays and, at low \mKee, from the radiative tail of \BuJpsiKee decays.
This tail is visible in the distribution of \mKee versus {\qsq}, which is given in the Supplemental Material to this Letter~\cite{suppl}.

Cascade backgrounds of the form \HbtoHc, where \Hb is a beauty hadron (\Bu, \Bz, \Bs or \Lb), \Hc a charm  hadron (\Dz, \Dp, \Ds, \Lc), and $X$, $Y$ are particles that are not reconstructed, are suppressed by requiring that the kaon-lepton invariant mass satisfies the constraint $m(\Kp\ellm)>m_{\Dz}$, where $m_{\Dz}$ is the known $\Dz$ mass~\cite{PDG2018}. Cascade backgrounds with a misidentified particle are suppressed by applying a similar veto, but with the lepton-mass hypothesis changed to that of a pion (denoted $\ell[\to\pi]$). In the muon case, it is sufficient to reject $K\mu[\to \pi]$ combinations with a mass smaller than $m_{\Dz}$. In the electron case this veto is applied without the bremsstrahlung recovery, \ie  based on only the measured track momenta, and a window around the \Dz mass is used to reject candidates. The vetoes retain 97\% of \BuKmm and 95\% of \BuKee decays passing the full selection. The relevant mass distributions are given in the Supplemental Material~\cite{suppl}.

Other exclusive \bquark-hadron decays require at least two particles to be misidentified in order to form backgrounds. 
These include the decays \BuKpipi and misreconstructed \BuJpsiKll and \BuPsiKll decays, where the kaon is misidentified as a lepton and the lepton (of the same electric charge) as a kaon.
The particle-identification  criteria used in the selection render such backgrounds negligible. Backgrounds from decays with a photon converted into an \epem pair are also negligible.

Combinatorial background is reduced using Boosted Decision Tree (BDT) algorithms~\cite{Breiman}, which employ the gradient boosting technique~\cite{GradBoost}.
For the nonresonant muon mode and for each of the three different trigger categories of the 
nonresonant electron mode, a single BDT is trained for the 7 and $8\tev$ data, and an additional BDT is trained for the $13\tev$ data. The same BDTs are used to select the resonant decays. The BDT training uses nonresonant \Kll candidates selected from the data with $\mKll > 5.4\gevcc$ as a proxy for the background, and simulated nonresonant \Kll candidates as a proxy for the signal decays. The training and testing is performed using the $k$-folding technique with $k=10$~\cite{Blum:1999:BHB:307400.307439}. 
The variables used as input to these BDTs are:  the  \pt of the \Bp, \Kp and dilepton candidates, and the minimum and maximum \pt of the leptons;  the \Bp, dilepton and \Kp \chisqip with respect to the associated PV, where \chisqip is defined as the difference in the vertex-fit \chisq of the PV reconstructed with and without the particle being considered; the minimum and maximum \chisqip of the leptons; the \Bp vertex-fit quality; the significance of the \Bp flight distance; and the angle between the \Bp candidate momentum vector and the direction between the associated PV and the \Bp decay vertex. 
The selection applied to the BDT output variables is chosen to maximise the predicted significance of the nonresonant signal yield. The BDT selection reduces the combinatorial background by approximately $99\%$, while retaining $85\%$ of the signal modes. 
The efficiency of each BDT response is independent of \mKll in the regions used to determine the event yields.
After the full selection is applied, the fraction of signal candidates in each trigger category is consistent with the expectation from simulation.

An unbinned extended maximum-likelihood fit to the \mKee and \mKmm distributions of nonresonant candidates is used to determine \RK.
In order to take into account the correlation between the selection efficiencies, the different trigger categories and data-taking periods are fitted simultaneously.
The resonant decay mode yields are incorporated as constraints in this fit, such that the \BuKmm yield and \RK are fit parameters. The resonant yields are determined from separate unbinned extended maximum-likelihood fits to the \mKllconst distributions.
For all the mass-shape models described below, the parameters are derived from simulated decays that are calibrated using data control channels.

All four signal modes are modelled by functions with multi-Gaussian cores and power-law tails on both sides of the peak~\cite{Skwarnicki:1986xj,Santos:2013gra}. 
The electron-mode signal mass shapes are described with the sum of three distributions which model whether a bremsstrahlung photon cluster was added to neither, either or both of the \epm candidates. 
The fraction of signal decays in each of the bremsstrahlung categories is constrained to the value obtained from the simulation. 

The shape of the \BuJpsipi background is taken from simulation, while its size is constrained with respect to the \BuJpsiK mode using the known ratio of the relevant branching fractions~\cite{LHCb-PAPER-2016-051, PDG2018} and efficiencies. 
In each trigger category,
the shape and relative fraction of the background from partially reconstructed \BuBdKstarzorKstplusee or \BuBdKstarzorKstplusjpsi decays are also taken from simulation. 
The overall yield of these partially reconstructed decays is left free to vary in the fit, in order to accommodate possible lepton-universality violation in such decays.
In the fits to nonresonant \Kee candidates, the shape of the radiative tail of \BuJpsiKee decays is taken from simulation and its yield is constrained to the expected value within its uncertainty.   
In all fits, the combinatorial background is modelled with an exponential function with a freely varying yield and shape.

In order to evaluate the efficiencies accurately, weights are applied to simulated candidates to correct for the imperfect modelling of the \Bu production kinematics, the particle-identification performance, and the trigger response. The weights are computed sequentially, making use of control samples of $\decay{\jpsi}{\mumu}$,  $\decay{\Dstarp}{\Dz(\to\Km\pip)\pip}$ and \BuJpsiKll decays, and are applied  to both resonant and nonresonant simulated candidates. Only subsets of the  \BuJpsiKll samples are used to derive these corrections, which minimises the number of common candidates being used for both the determination of the corrections and the measurement. The correlations between samples are taken into account in the results and cross-checks presented below. The overall effect of the corrections on the \RK measurement is at the $0.02$ level, demonstrating the robustness of the double-ratio method in suppressing systematic biases that affect the resonant and nonresonant decay modes similarly.

Two classes of systematic uncertainty are considered: those that only affect the nonresonant decay yields, and those that affect the ratio of efficiencies for different trigger categories and data-taking periods in the fit for \RK. 
The uncertainty from the choice of mass-shape models falls into the former category and is estimated by fitting pseudoexperiments with alternative models that still describe the data well. The effect on \RK is at the $\pm0.01$ level. % 1.66%  -> 0.014
Systematic uncertainties in the latter category affect the ratios of efficiencies and hence the value of \RK that maximises the likelihood. These uncertainties are accounted for through constraints on the efficiency values used in the fit to determine \RK, taking into account the correlations between different trigger categories and data-taking periods.
The combined statistical and systematic uncertainty is then determined from a profile-likelihood scan. In order to isolate the statistical contribution to the uncertainty, the profile-likelihood scan is repeated with the efficiencies fixed to their fitted values.  
For the subsamples of the electron-mode data where the trigger is based on the kaon or on other particles in the event that are not part of the signal candidate, the dominant systematic uncertainties come from the (data-derived) calibration of the trigger efficiencies. For the electron trigger, there are comparable contributions from the statistical uncertainties associated with various calibration samples and the calibration of data-simulation differences. 

The migration of events in \qsq is studied in the simulation. The effect of the differing \qsq resolution between data and simulation, which alters the estimate of the migration, gives a negligible uncertainty in the determination of the ratio of efficiencies. 
The uncertainties on parameters used in the simulation decay model (Wilson coefficients, form factors, other hadronic uncertainties \etc) affect the \qsq distribution and hence the selection efficiencies determined from simulation. The variation caused by the  uncertainties on these parameters is propagated to an uncertainty on \RK using predictions from the {\sc{flavio}} software package~\cite{Straub:2018kue}. The resulting systematic effect on \RK is negligible, even when non-SM values of the Wilson coefficients are considered.

Several cross-checks are used to verify the analysis procedure. The single ratio \mbox{$\rjpsi=\BR(\BuJpsiKmm)/\BR(\BuJpsiKee)$} is known to be compatible with unity at the 0.4\% level~\cite{Ablikim:2013pqa,PDG2018}. This ratio does not benefit from the cancellation of systematic effects that the double ratio used to measure \RK exploits, and is therefore a stringent test of the control of the efficiencies. The corrections applied to the simulation do not force \rjpsi to be unity and some of the corrections shift \rjpsi in opposing directions. 
The value of \rjpsi is found to be $1.014\pm0.035$, where the uncertainty includes the statistical uncertainty and those systematic effects relevant to the \RK measurement.
It does not include additional subleading systematic effects that should be accounted for in a complete measurement of \rjpsi.  
As a further cross-check, the double ratio of branching fractions, \RPsitwos, defined by
\begin{equation*}
\label{eq:RPsitwos}
\RPsitwos  = 
{\frac{\BR(\BuPsiKmm)}{\BR(\BuJpsiKmm)}} \bigg{/} {\frac{\BR(\BuPsiKee)}{\BR(\BuJpsiKee)}}  \,,
\end{equation*}
\noindent is determined to be $0.986\pm 0.013$, where again the uncertainty includes the statistical uncertainty but only those systematic effects that are relevant to the \RK measurement. 
This ratio provides an independent validation of the analysis procedure. 

\begin{figure}[t] 
\begin{center} 
\includegraphics[width=0.45\linewidth,trim={0 0.55cm 0 0}, clip]{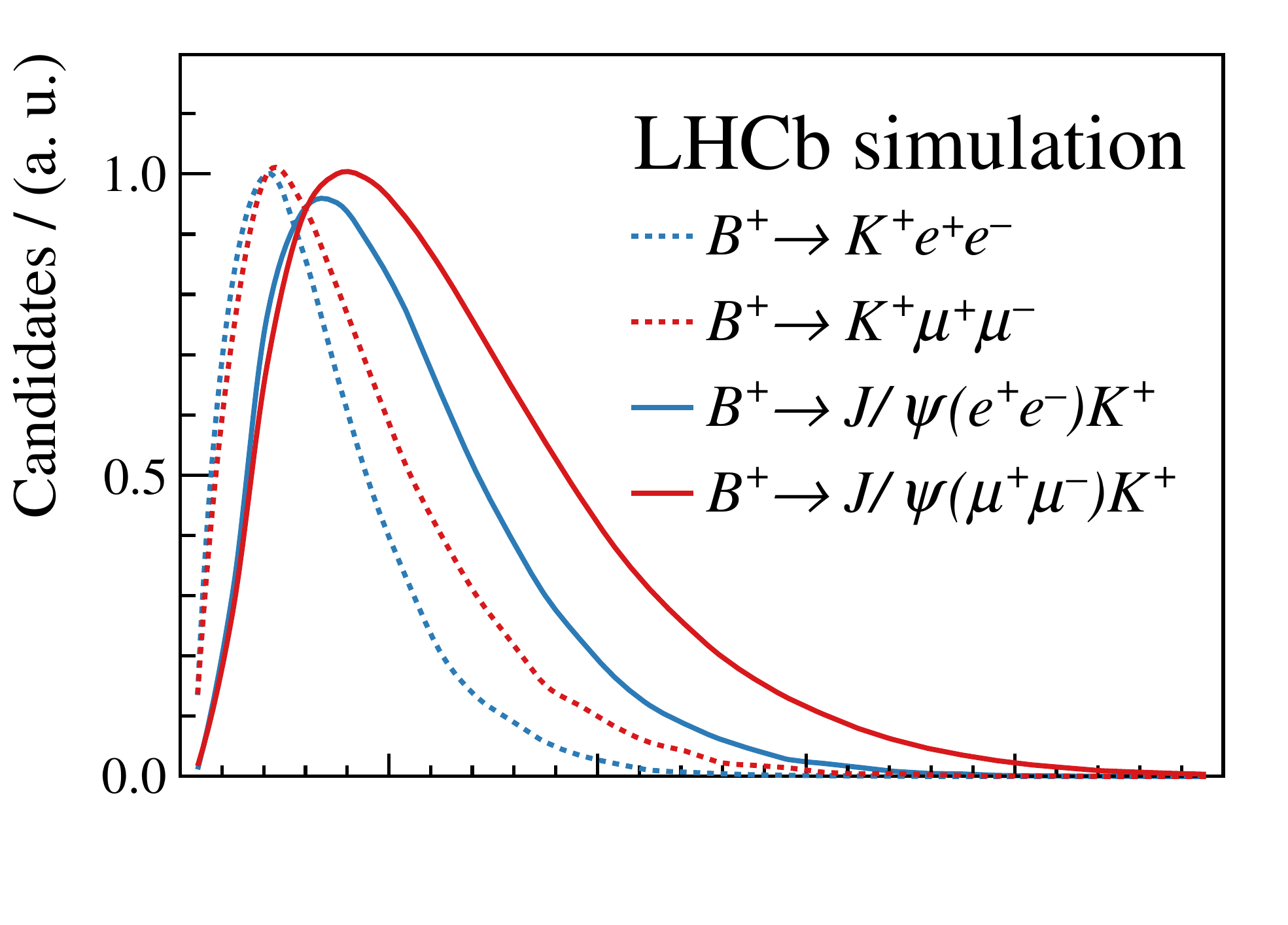}
 
\includegraphics[width=0.45\linewidth,trim={0 0 0 0.45cm}, clip]{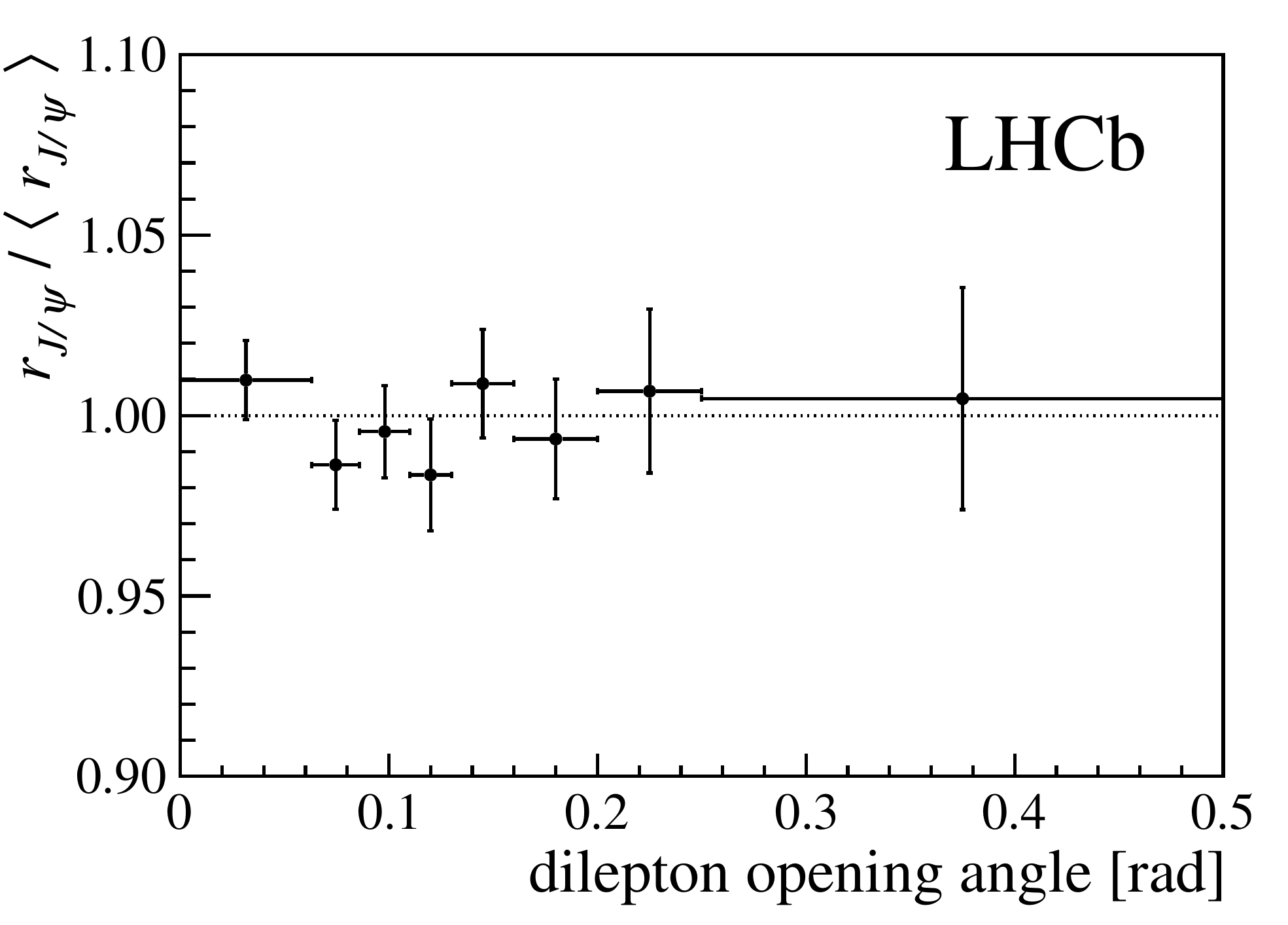} 
\end{center} 
\caption{(Top) expected distributions of the opening angle between the two leptons, in the laboratory frame, for the four modes in the double ratio used to determine \RK. (Bottom) the single ratio \rjpsi relative to its average value $\left< \rjpsi \right>$ as a function of the opening angle. } \label{fig:rjpsi_differential_main} \end{figure}

Leptons from \BuJpsiK decays have a different \qsq value than those from the nonresonant decay modes. However, the detector efficiency depends on laboratory-frame variables rather than on \qsq, \eg the momenta of the final-state particles, opening angles,~\etc In these laboratory variables there is significant overlap between the nonresonant and resonant modes, even if the decays do not overlap in \qsq (see the Supplemental Material~\cite{suppl}). 
The \rjpsi ratio is examined as a function of a number of reconstructed variables. Any trend would indicate an uncontrolled systematic effect that would only partially cancel in the double ratio.  For each of the variables examined, no significant trend is observed.  
Figure~\ref{fig:rjpsi_differential_main} shows the ratio as a function of the dilepton opening angle and other examples are provided in the Supplemental Material~\cite{suppl}. 
Assuming the deviations that are observed indicate genuine mismodelling of the efficiencies, rather than fluctuations, and taking into account the spectrum of the relevant variables in the nonresonant decay modes of interest, a total shift on \RK is computed for each of the variables examined. In each case, the resulting variation is within the estimated systematic uncertainty on \RK. 
The \rjpsi ratio is also computed in two- and three-dimensional bins of the considered variables. Again, no trend is seen and the deviations observed are consistent with the systematic uncertainties on \RK.
An example is shown in Fig.~\ref{fig:rjpsi_bin} in the Supplemental Material~\cite{suppl}. 
Independent studies of the electron reconstruction efficiency using control channels selected from the data also give consistent results.

\begin{figure}
    \centering
    \includegraphics[width=0.45\textwidth]{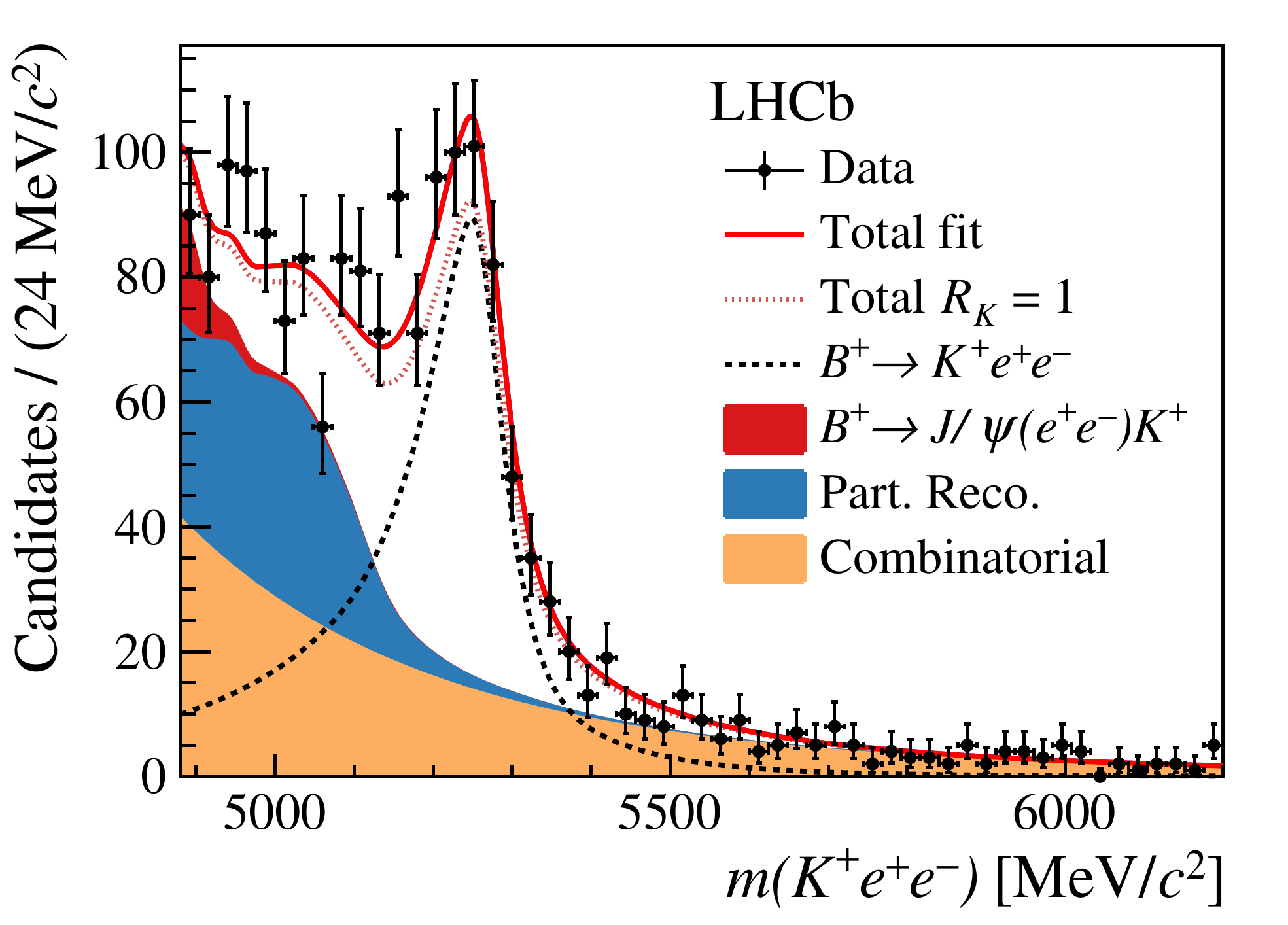}
    \includegraphics[width=0.45\textwidth]{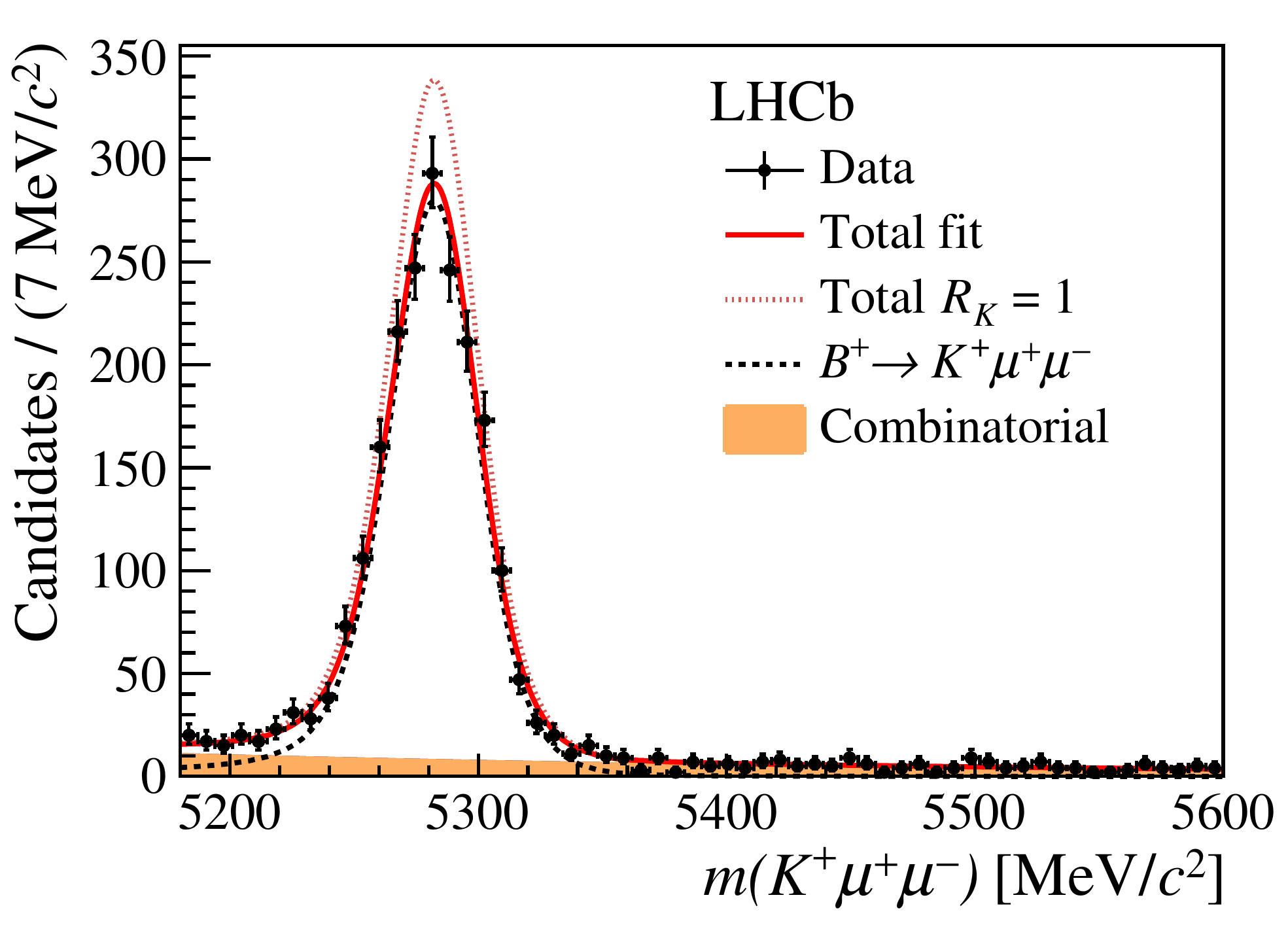}
    \includegraphics[width=0.45\textwidth]{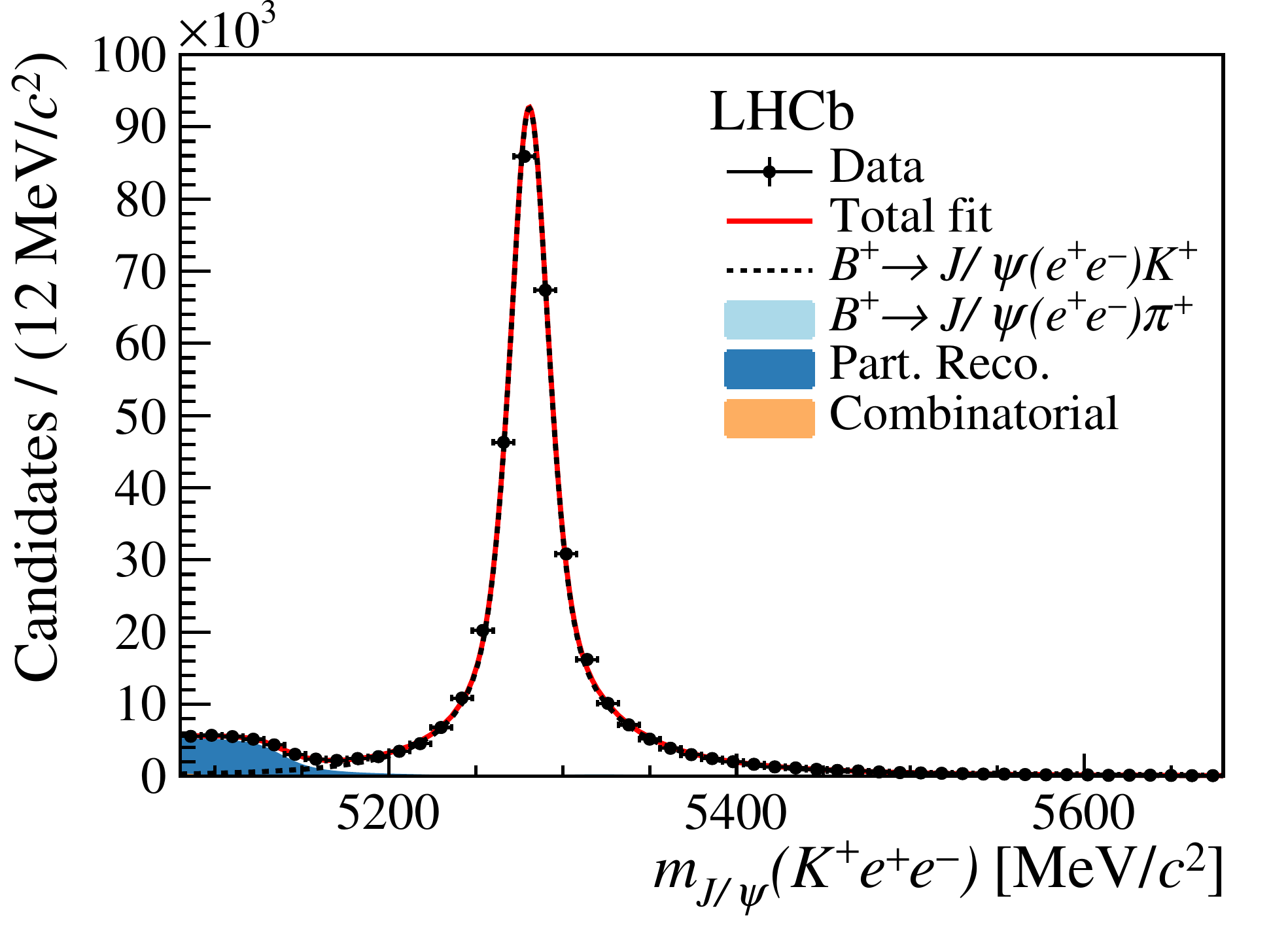}
    \includegraphics[width=0.45\textwidth]{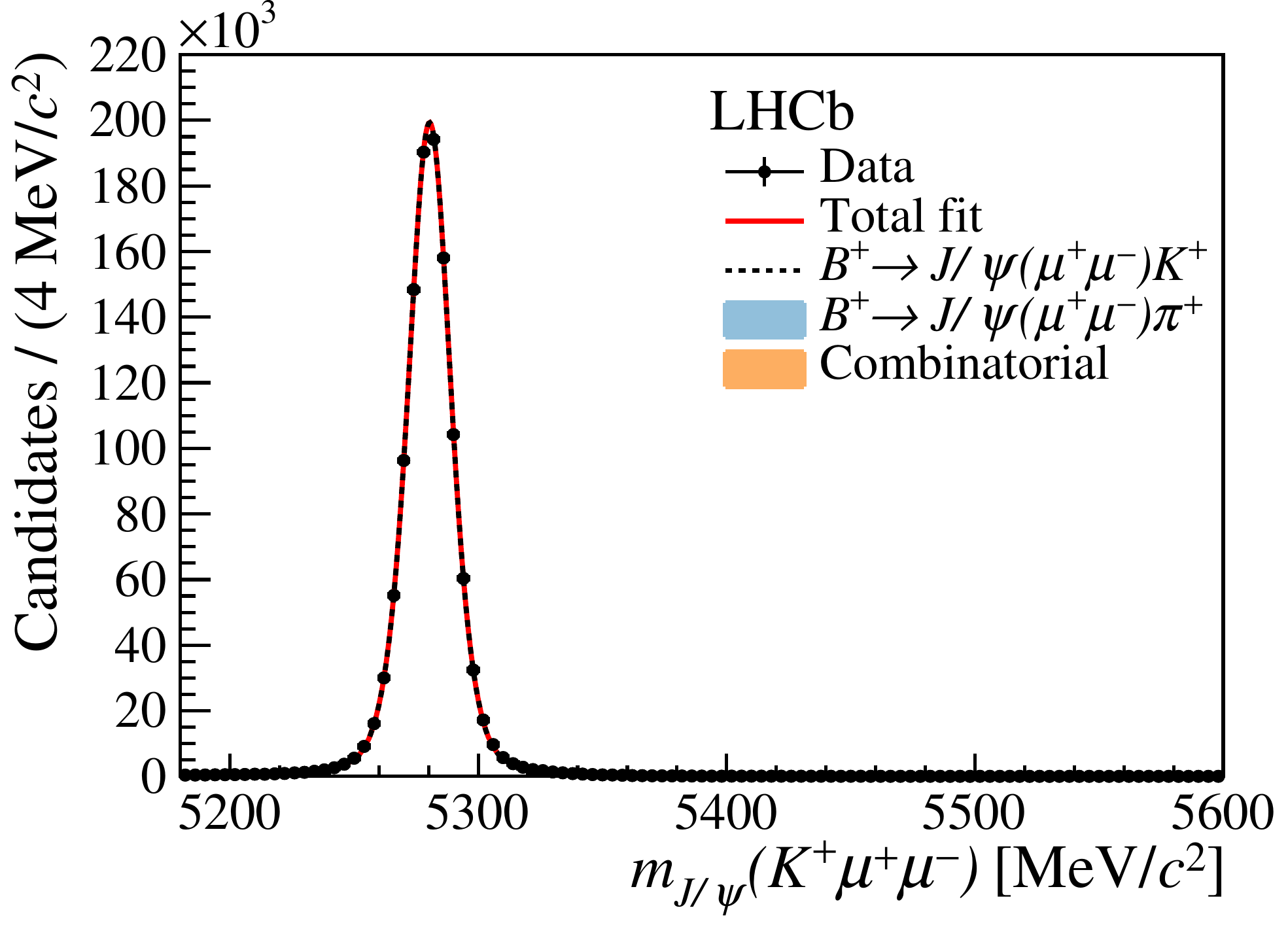}

    \caption{Fits to the \mKllgeneric invariant mass distribution for (left) electron and (right)~muon candidates for (top) nonresonant and (bottom) resonant decays. 
    For the electron (muon) nonresonant plots, the red-dotted line shows the distribution that would be expected from the observed number of \BuKmm (\BuKee) decays  and $\RK=1$.}
    \label{fig:fits}
\end{figure}

The results of the fits to the \mKll and \mKllconst distributions are shown in Fig.~\ref{fig:fits}. A total of 
$1943\pm 49$ \BuKmm decays are observed. 
A study of the \BuKmm differential branching fraction gives results that are consistent with previous \lhcb measurements~\cite{LHCb-PAPER-2014-006} but, owing to the selection criteria optimised for the precision on \RK, are less precise. The \BuKmm differential branching fraction observed is consistent between the 7~and~8\tev data and the 13\tev data. 

The value of \RK is measured to be 
\begin{displaymath}
\RK  = \RKvalue\,,
\end{displaymath}
\noindent where the first uncertainty is statistical and the second systematic.
This is the most precise measurement to date and is consistent with the SM expectation at the level of \significance~standard deviations~\cite{Descotes-Genon:2015uva,Bobeth:2007,Bordone:2016gaq,EOS,Straub:2018kue}. 
The likelihood profile as a function of \RK is given in the Supplemental Material~\cite{suppl}.
The value for \RK obtained is consistent across the different data-taking periods and trigger categories.  A fit to just the 7~and~8\tev data gives a value for \RK compatible with the previous LHCb measurement~\cite{LHCb-PAPER-2014-024} within one standard deviation. This level of consistency is evaluated using pseudoexperiments that take into account the overlap between the two data samples, which are not identical due to different reconstruction and selection procedures.
The result from just the 7 and 8\tev data is also compatible with that from only the 13\tev data at the 1.9 standard deviation level (see the Supplemental Material~\cite{suppl}).

The branching fraction of the \BuKee decay is determined in the nonresonant signal region \mbox{$1.1 < \qsq < 6.0 \gevgevcccc$} by combining the value of \RK with the value of $\BR(\BuKmm)$ from Ref.~\cite{LHCb-PAPER-2014-006}, taking into account correlated systematic uncertainties. This gives
\begin{displaymath}
\frac{d\BR(\BuKee)}{d\qsq}(1.1 < \qsq < 6.0 \gevgevcccc)  = (28.6\,^{+2.0}_{-1.7}\,\pm 1.4)\times 10^{-9}\,c^4\kern -0.1em/\kern -0.15em\gev^2\,.
\end{displaymath}
\noindent The dominant systematic uncertainty is from the limited knowledge of the \BuJpsiK branching fraction~\cite{PDG2018}. This is the most precise measurement to date and is consistent with predictions based on the SM~\cite{Khodjamirian:2017fxg,Straub:2018kue}. 

In summary, in the dilepton mass-squared region $1.1 < \qsq < 6.0 \gevgevcccc$, the ratio of the branching fractions for \BuKmm and \BuKee decays is measured to be $\RK=\RKvalue$. This is the most precise measurement of this ratio to date and is consistent with the SM prediction at the level of \significance standard deviations. 
Further reduction in the uncertainty on \RK can be anticipated when the data collected by LHCb in 2017 and 2018, which have a statistical power approximately equal to that of the full data set used here, are included in a future analysis. In the longer term, there are good prospects for high-precision measurements as much larger samples are collected with an upgraded LHCb detector~\cite{LHCb-PII-Physics}.

% \input{reference}

% Comment this in for paper drafts; do not include this in analysis note and conference reports
\section*{Acknowledgements}
%
% These Acknowledgements valid from 14-Aug-2018
%
\noindent We express our gratitude to our colleagues in the CERN
accelerator departments for the excellent performance of the LHC. We
thank the technical and administrative staff at the LHCb
institutes.
We acknowledge support from CERN and from the national agencies:
CAPES, CNPq, FAPERJ and FINEP (Brazil); 
MOST and NSFC (China); 
CNRS/IN2P3 (France); 
BMBF, DFG and MPG (Germany); 
INFN (Italy); 
NWO (Netherlands); 
MNiSW and NCN (Poland); 
MEN/IFA (Romania); 
MSHE (Russia); 
MinECo (Spain); 
SNSF and SER (Switzerland); 
NASU (Ukraine); 
STFC (United Kingdom); 
NSF (USA).
We acknowledge the computing resources that are provided by CERN, IN2P3
(France), KIT and DESY (Germany), INFN (Italy), SURF (Netherlands),
PIC (Spain), GridPP (United Kingdom), RRCKI and Yandex
LLC (Russia), CSCS (Switzerland), IFIN-HH (Romania), CBPF (Brazil),
PL-GRID (Poland) and OSC (USA).
We are indebted to the communities behind the multiple open-source
software packages on which we depend.
Individual groups or members have received support from
AvH Foundation (Germany);
EPLANET, Marie Sk\l{}odowska-Curie Actions and ERC (European Union);
ANR, Labex P2IO and OCEVU, and R\'{e}gion Auvergne-Rh\^{o}ne-Alpes (France);
Key Research Program of Frontier Sciences of CAS, CAS PIFI, and the Thousand Talents Program (China);
RFBR, RSF and Yandex LLC (Russia);
GVA, XuntaGal and GENCAT (Spain);
the Royal Society
and the Leverhulme Trust (United Kingdom);
Laboratory Directed Research and Development program of LANL (USA).

\addcontentsline{toc}{section}{References}
%\setboolean{inbibliography}{true}
\bibliographystyle{LHCb}
\bibliography{main,standard,LHCb-PAPER,LHCb-CONF,LHCb-DP,LHCb-TDR}

% \input{prl_justification}
 
% $Id: appendix.tex 124799 2018-11-23 12:14:45Z pkoppenb $
% ===============================================================================
% Purpose: appendix to the standard template: standard symbol alises from Ulrik
% Author: Tomasz Skwarnicki
% Created on: 2009-09-24
% ===============================================================================

%{\noindent\normalfont\bfseries\Large Appendices}
% \section*{Appendices}

\appendix

\clearpage
\renewcommand{\thefigure}{S\arabic{figure}}
\renewcommand{\thetable}{S\arabic{table}}
\renewcommand{\theequation}{S\arabic{equation}}

\setcounter{equation}{0}
\setcounter{figure}{0}
\setcounter{table}{0}
\setcounter{page}{1}

{\noindent\normalfont\bfseries\Large Supplemental Material}\\
\label{sec:Supplementary-App}

\begin{figure}[b!]
   \begin{center}
      \includegraphics[width=0.48\linewidth]{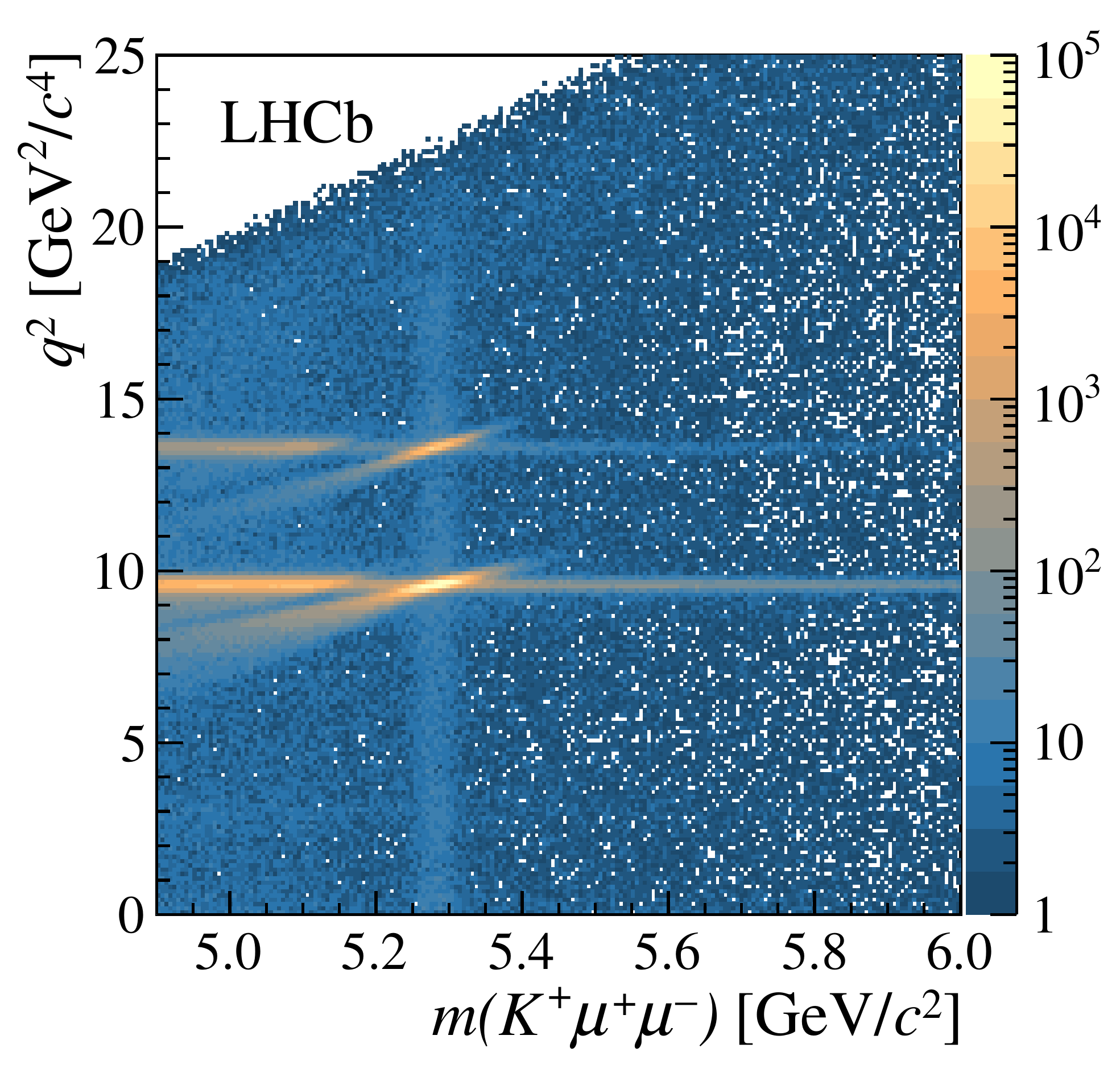}
      \hfill
      \includegraphics[width=0.48\linewidth]{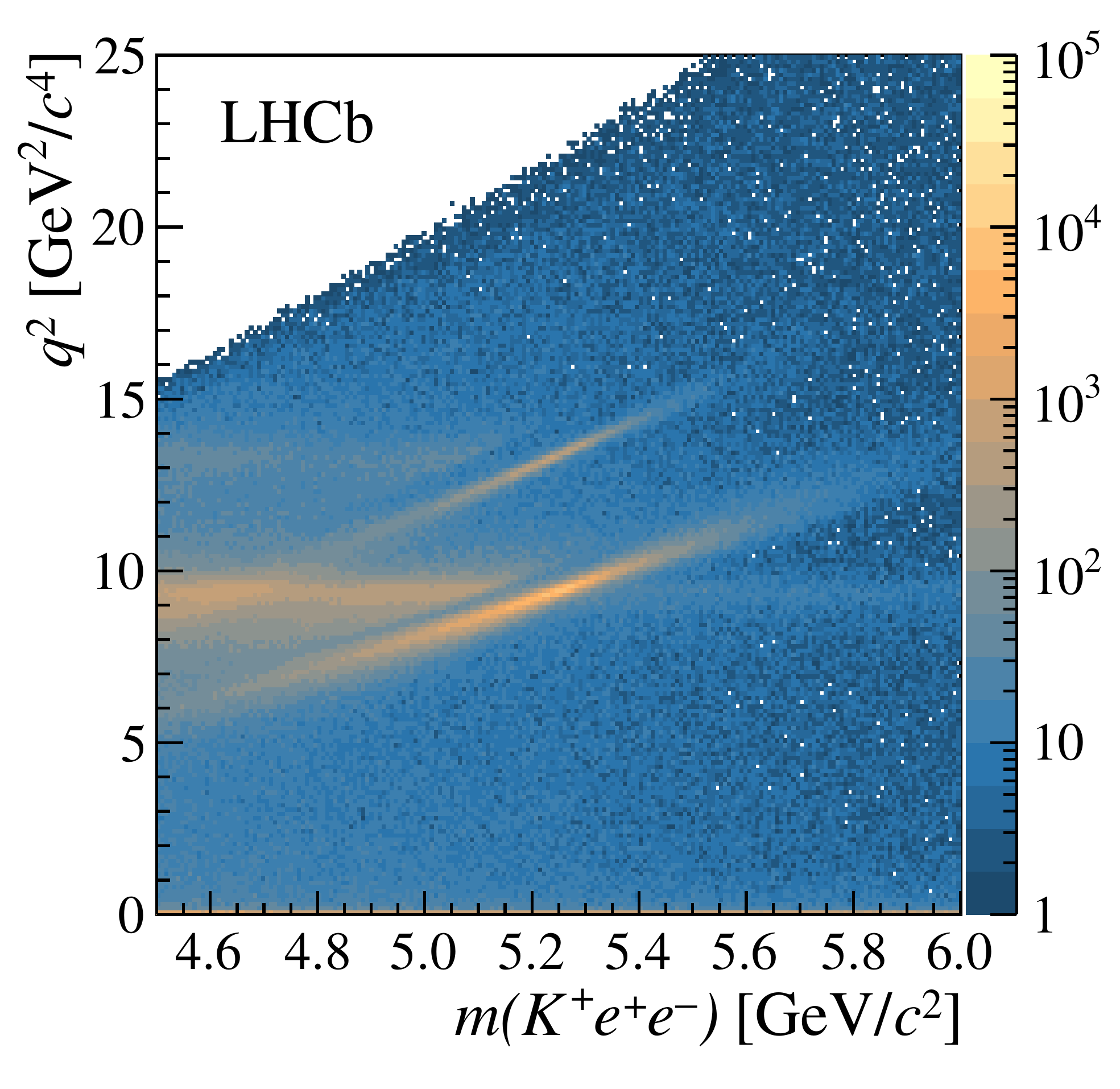}
   \end{center}
   \vspace*{-0.5cm}
   \caption{Two-dimensional distributions of $[\mKll,\qsq]$ for  (left) muon and (right) electron candidates after the application of the pre-selection and trigger requirements but not the multivariate selection.}\label{fig:2Dplot}
\end{figure}

The two-dimensional distributions of $[\mKll,\qsq]$ for  muon  and electron candidates are shown in Fig.~\ref{fig:2Dplot}. For the muon sample, nonresonant candidates can be seen to accumulate in a vertical band around the \Bu meson mass. For the electron candidates, only some of the bremsstrahlung energy is recovered by the procedure described in the Letter and this results in a worse mass resolution and a long tail to lower \Kee masses. The vertical band of signal candidates is then more difficult to discern. The resonant signals from \BuJpsiKll and \BuPsiKll decays are visible as diagonal bands, where the extended tails originate from both radiative and resolution effects, which are especially marked for the electron decay modes.   
As the energy loss affects both \mKll and \qsq measurements, the angle of these bands is fixed and it is not possible for candidates to migrate into the bulk of the signal region in $[\mKll,\qsq]$.  
For the electron mode, the lower radiative tail of \BuJpsiKee decays enters the $1.1 < \qsq < 6.0 \gevgevcccc$ region only at the lower part of the \mKee fit range around 4.9\gevcc (see also the left side of the \BuKee fit projection in Fig.~\ref{fig:fits} of the Letter). 

\begin{figure}[!htbp]
   \begin{center}
      \includegraphics[width=0.32\linewidth]{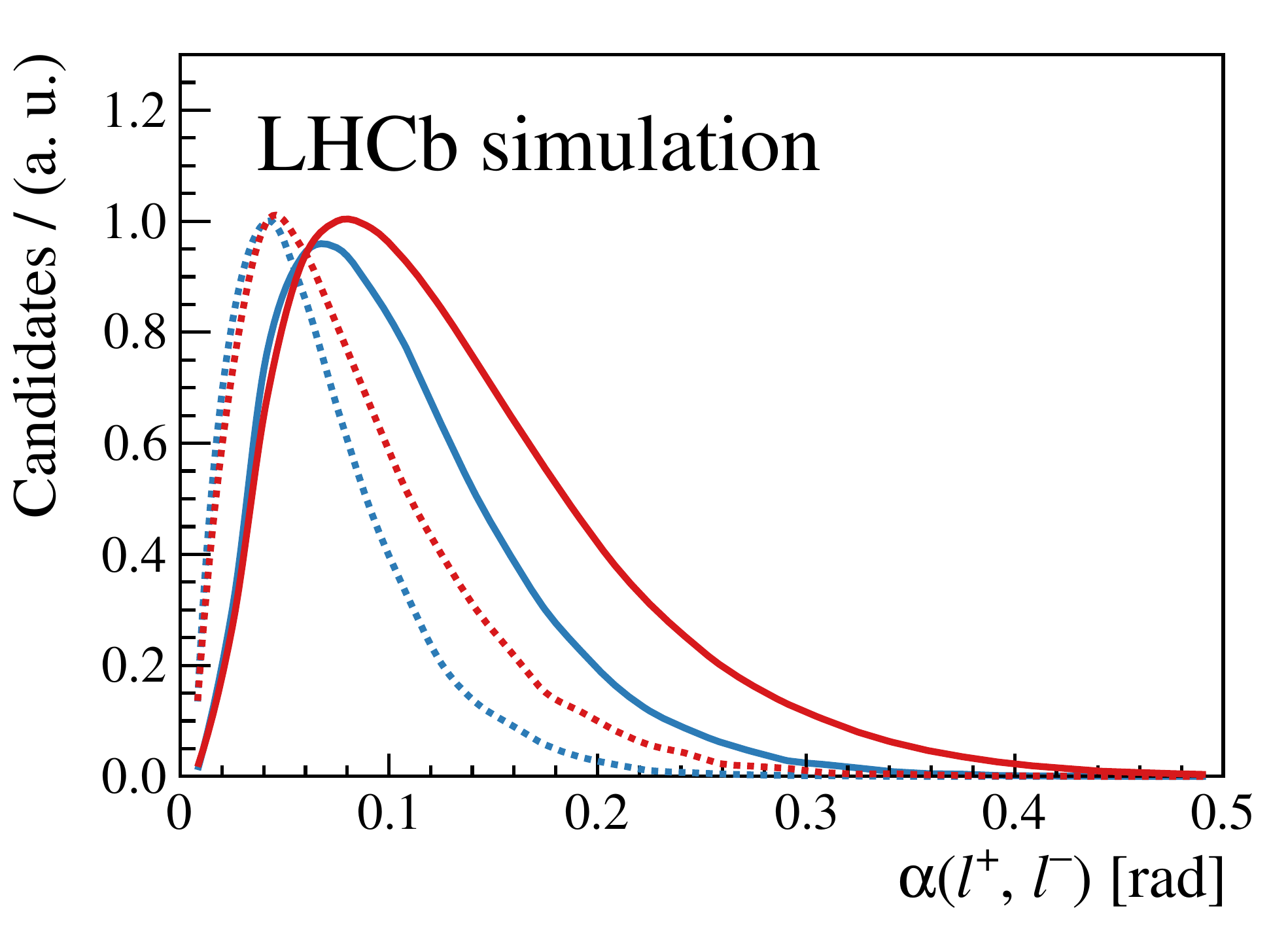}
      \includegraphics[width=0.32\linewidth]{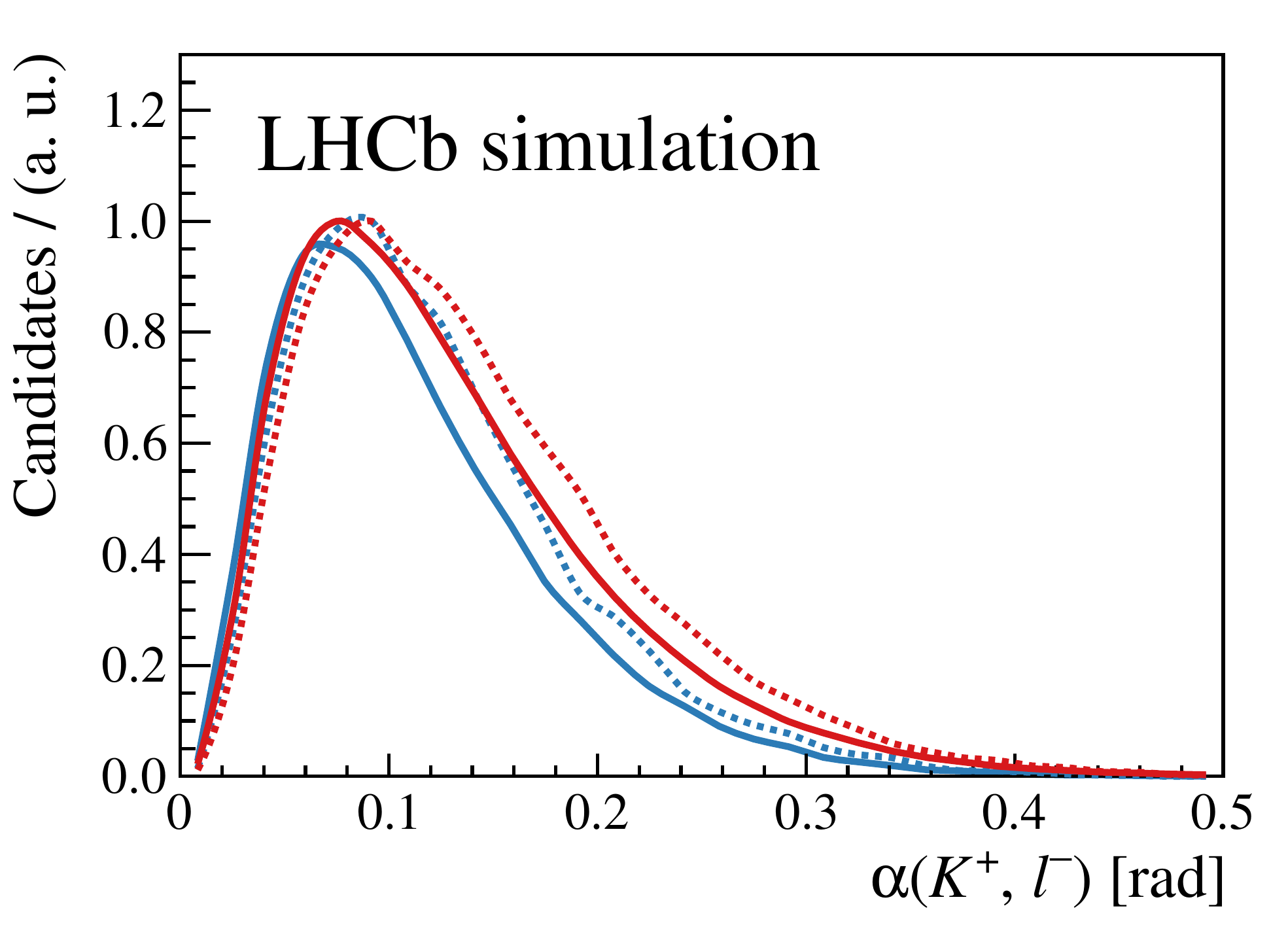}
      \includegraphics[width=0.32\linewidth]{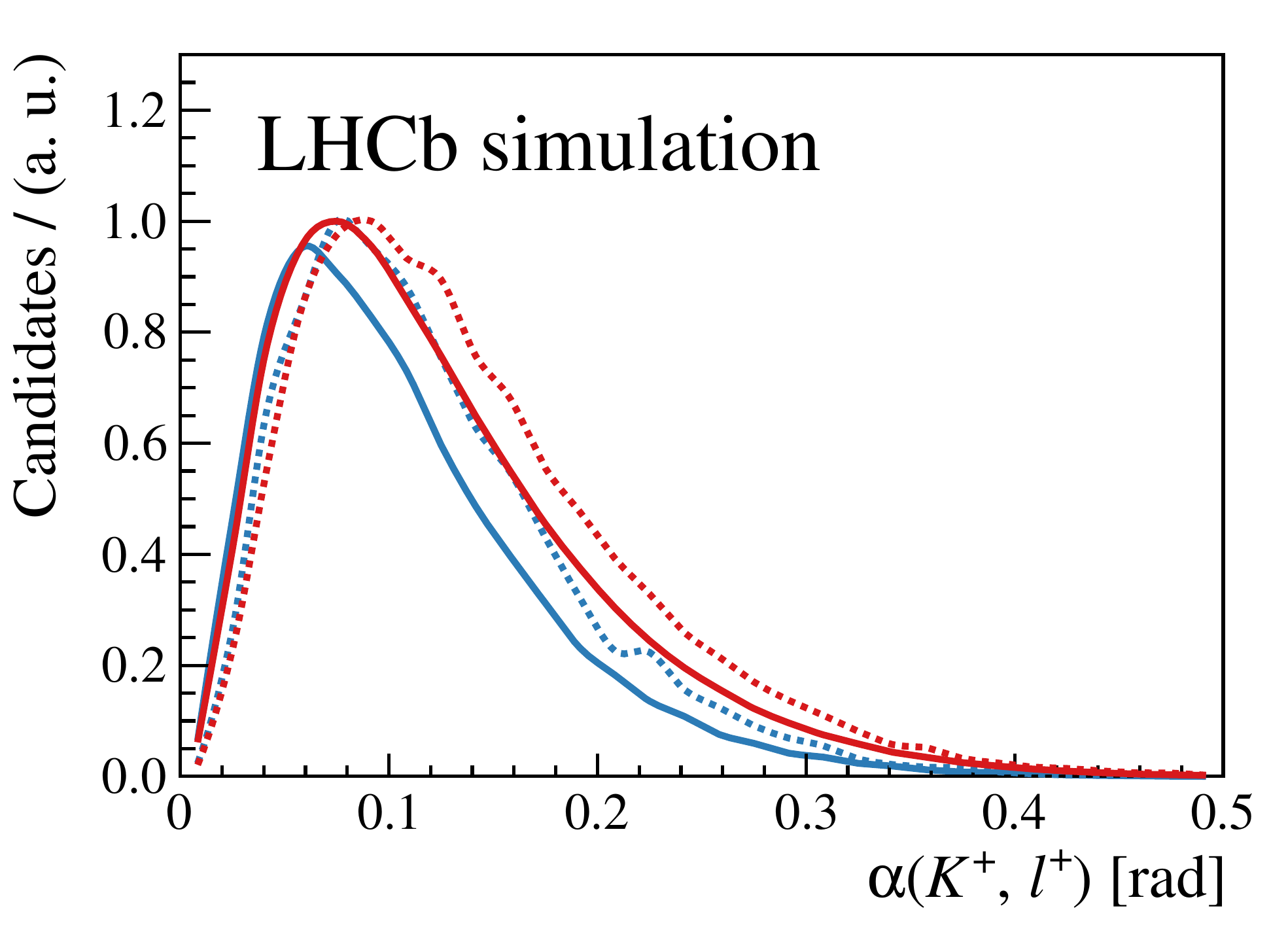}

      \includegraphics[width=0.32\linewidth]{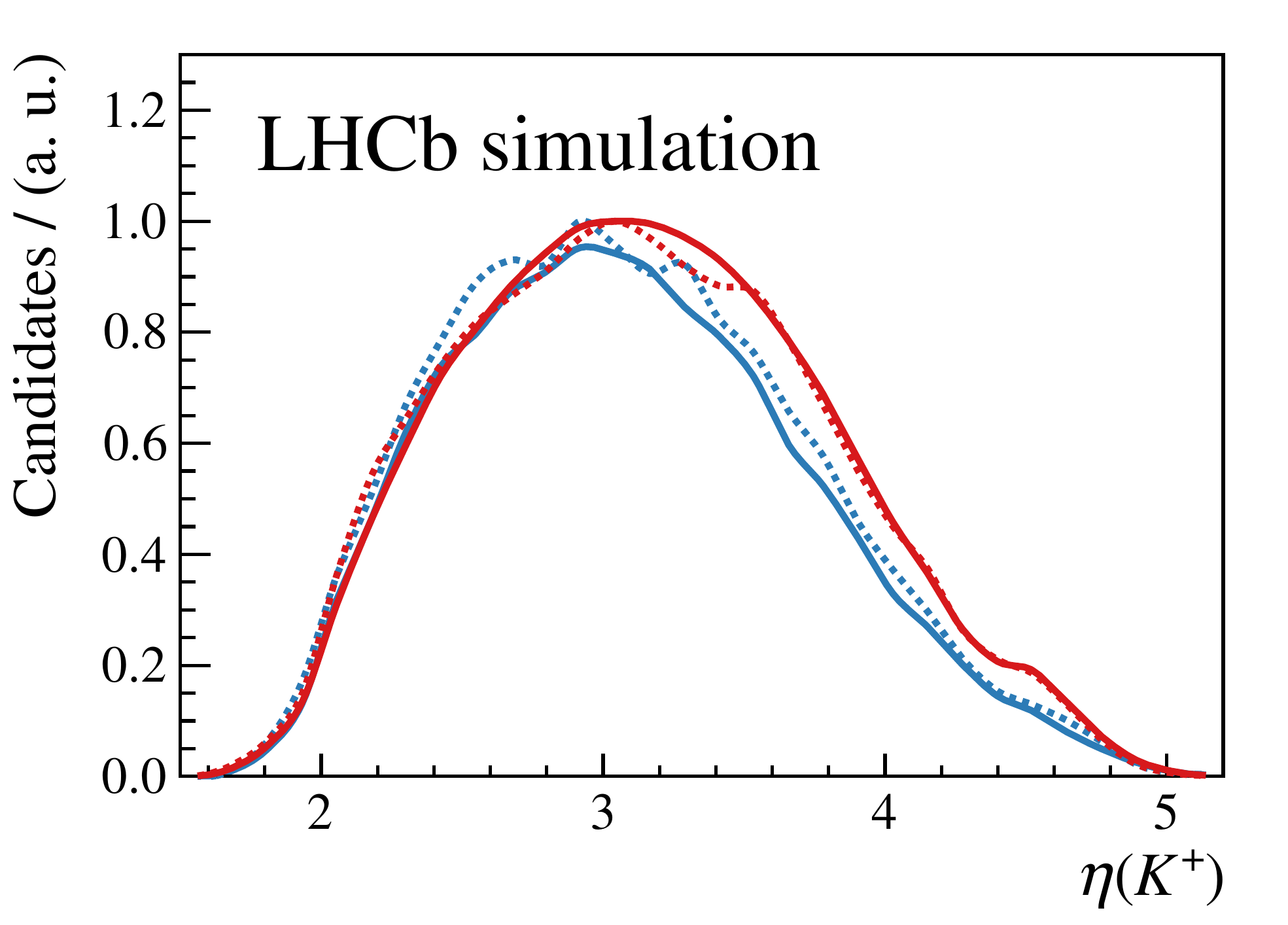}
      \includegraphics[width=0.32\linewidth]{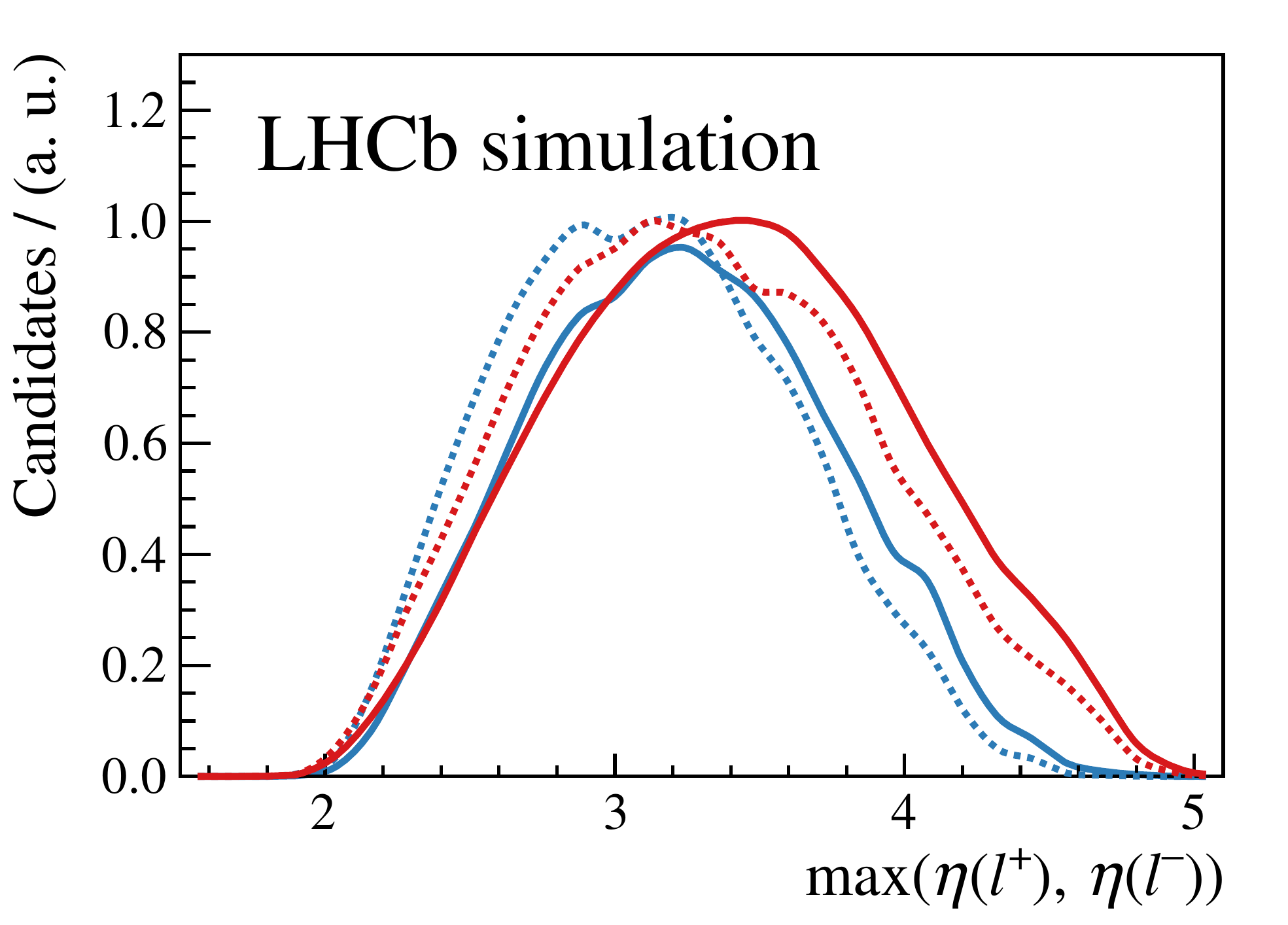}
      \includegraphics[width=0.32\linewidth]{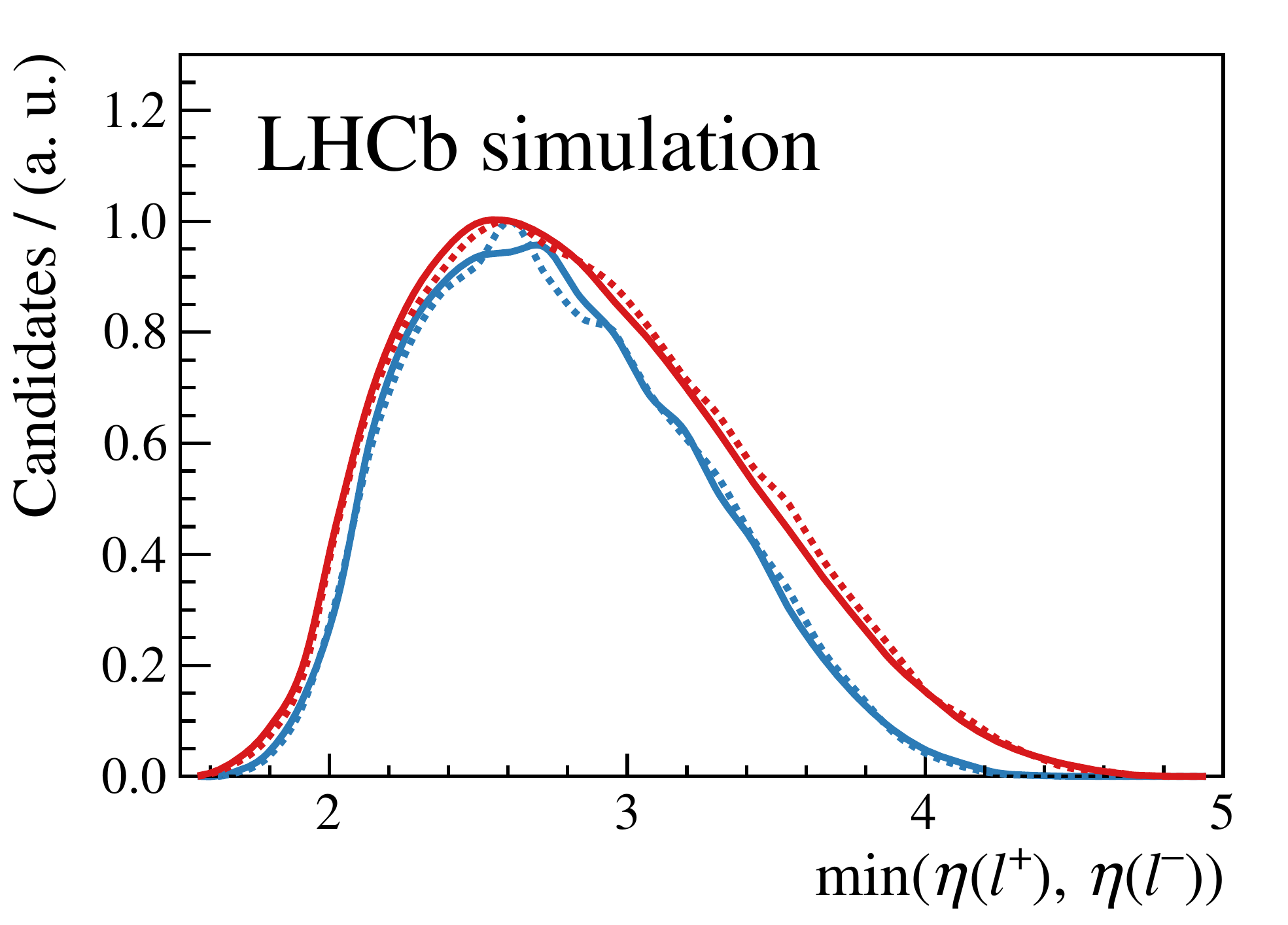}

      \includegraphics[width=0.32\linewidth]{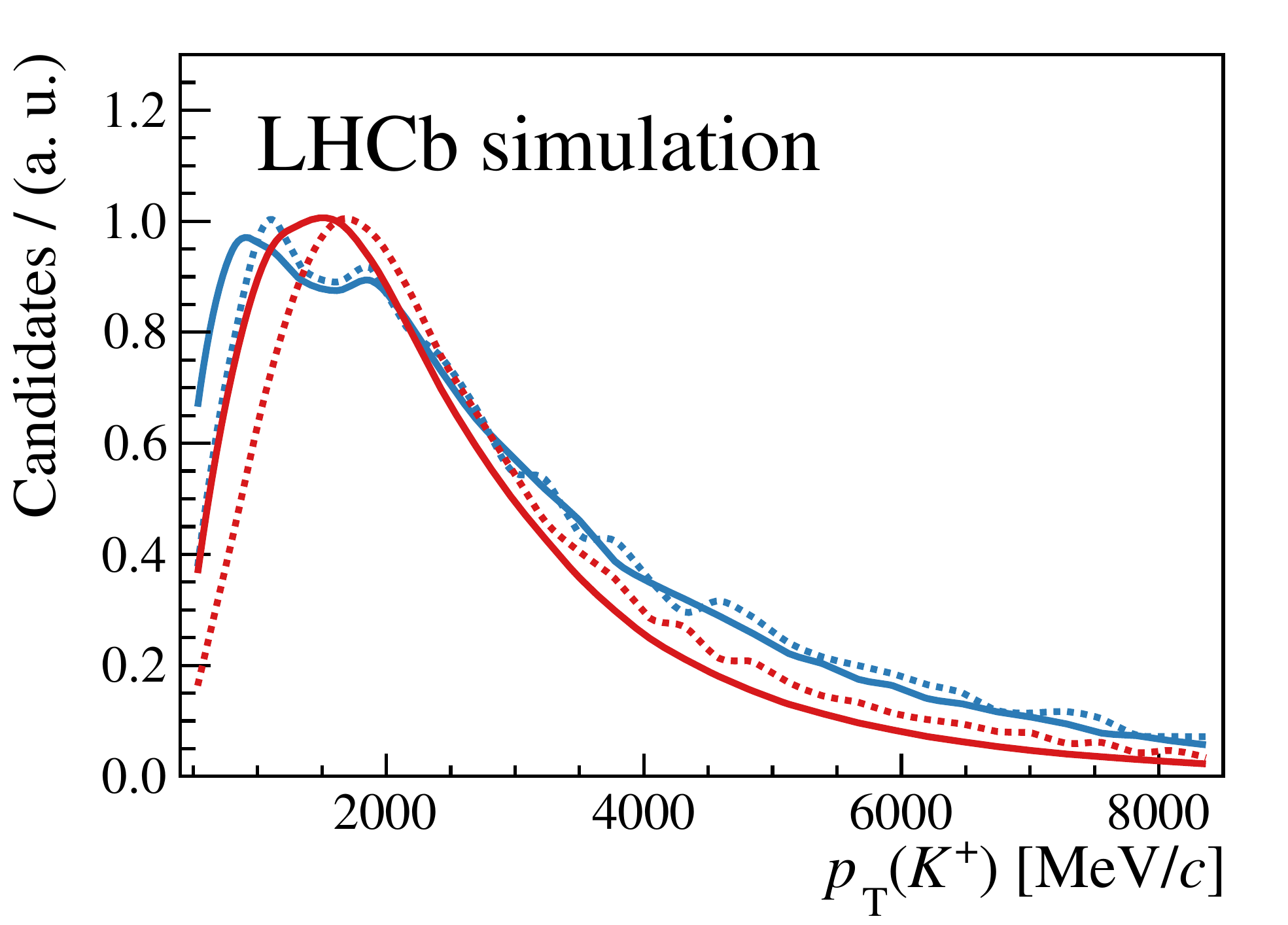}
      \includegraphics[width=0.32\linewidth]{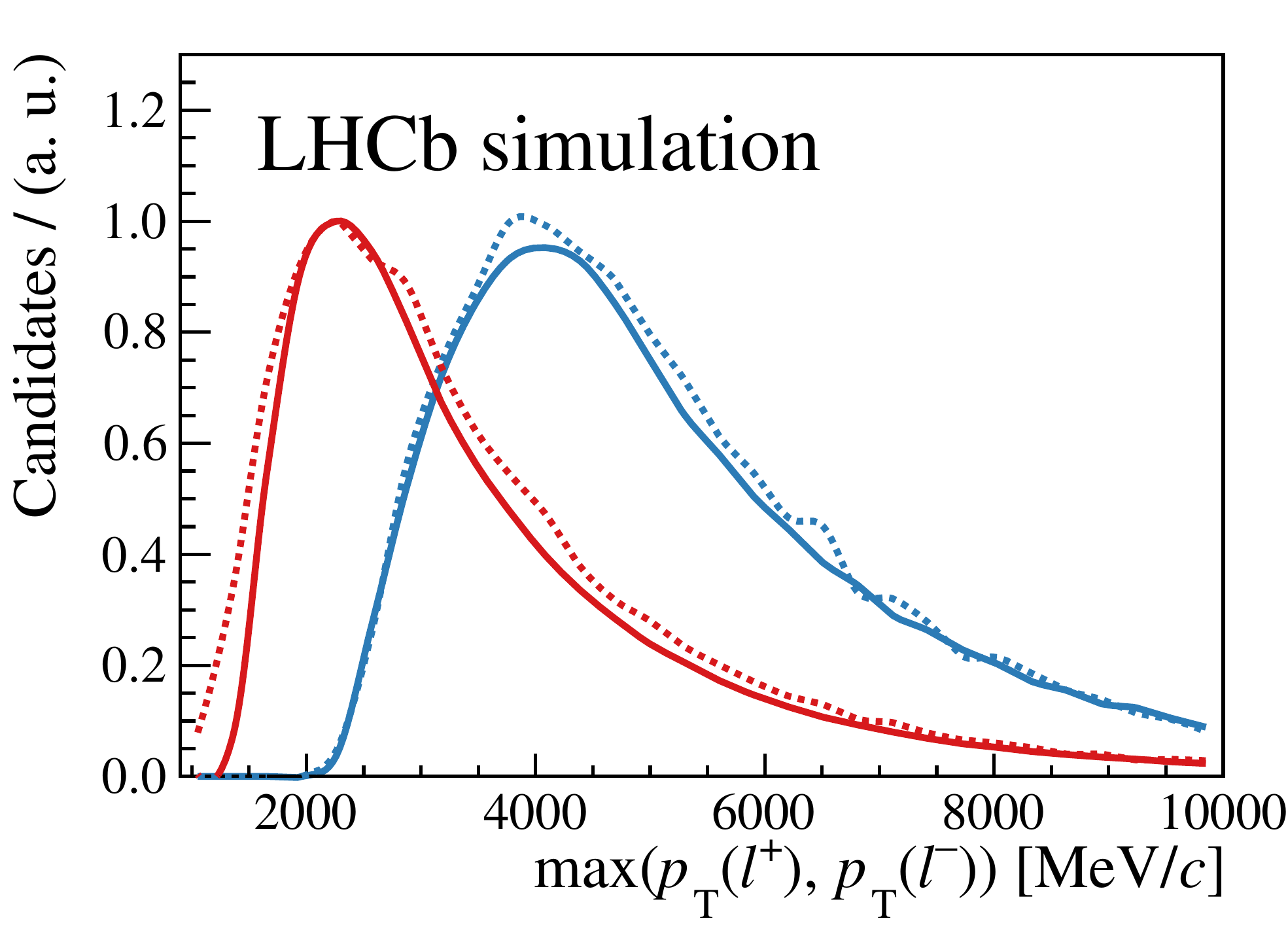}
      \includegraphics[width=0.32\linewidth]{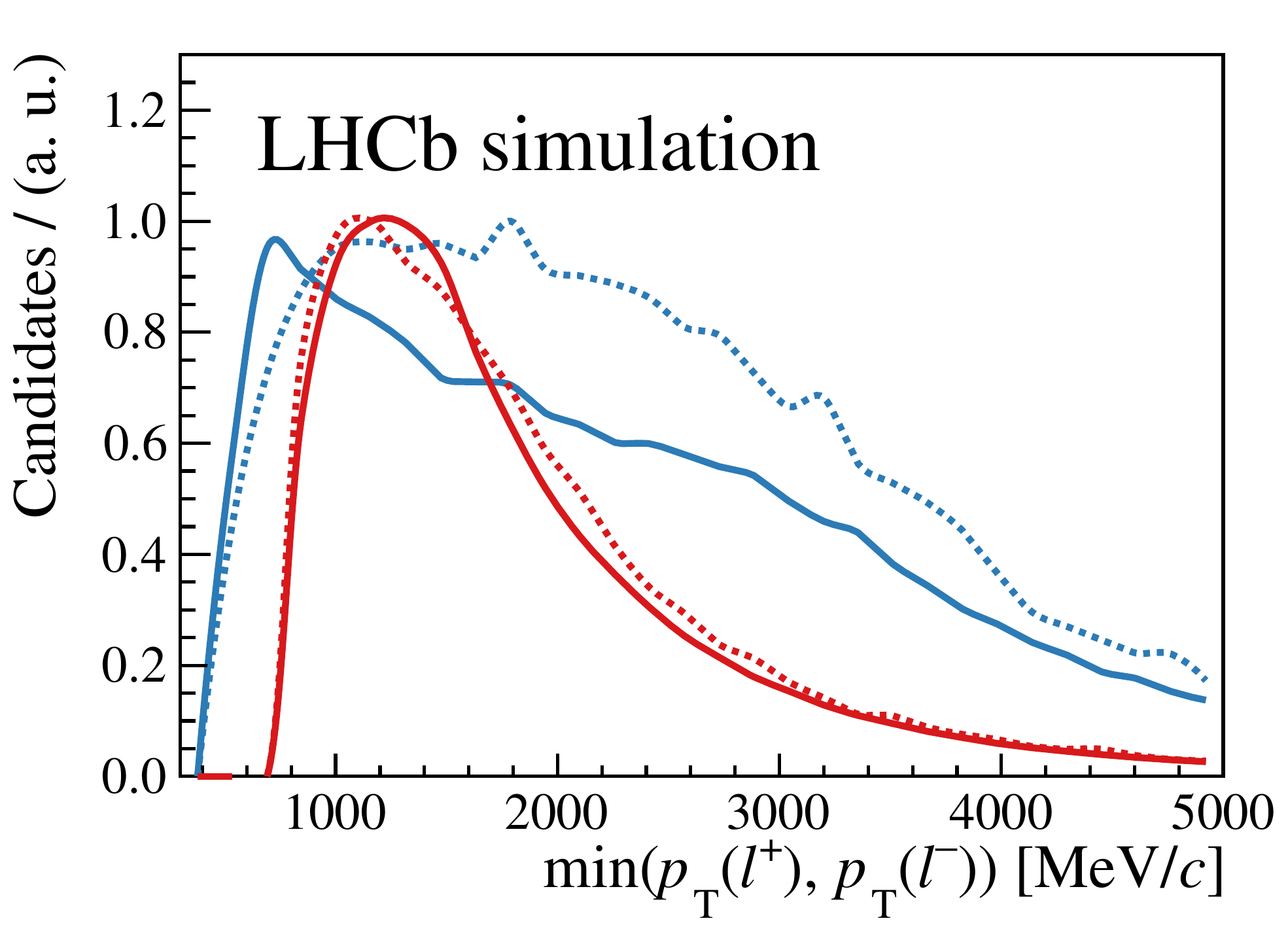}

      \includegraphics[width=0.32\linewidth]{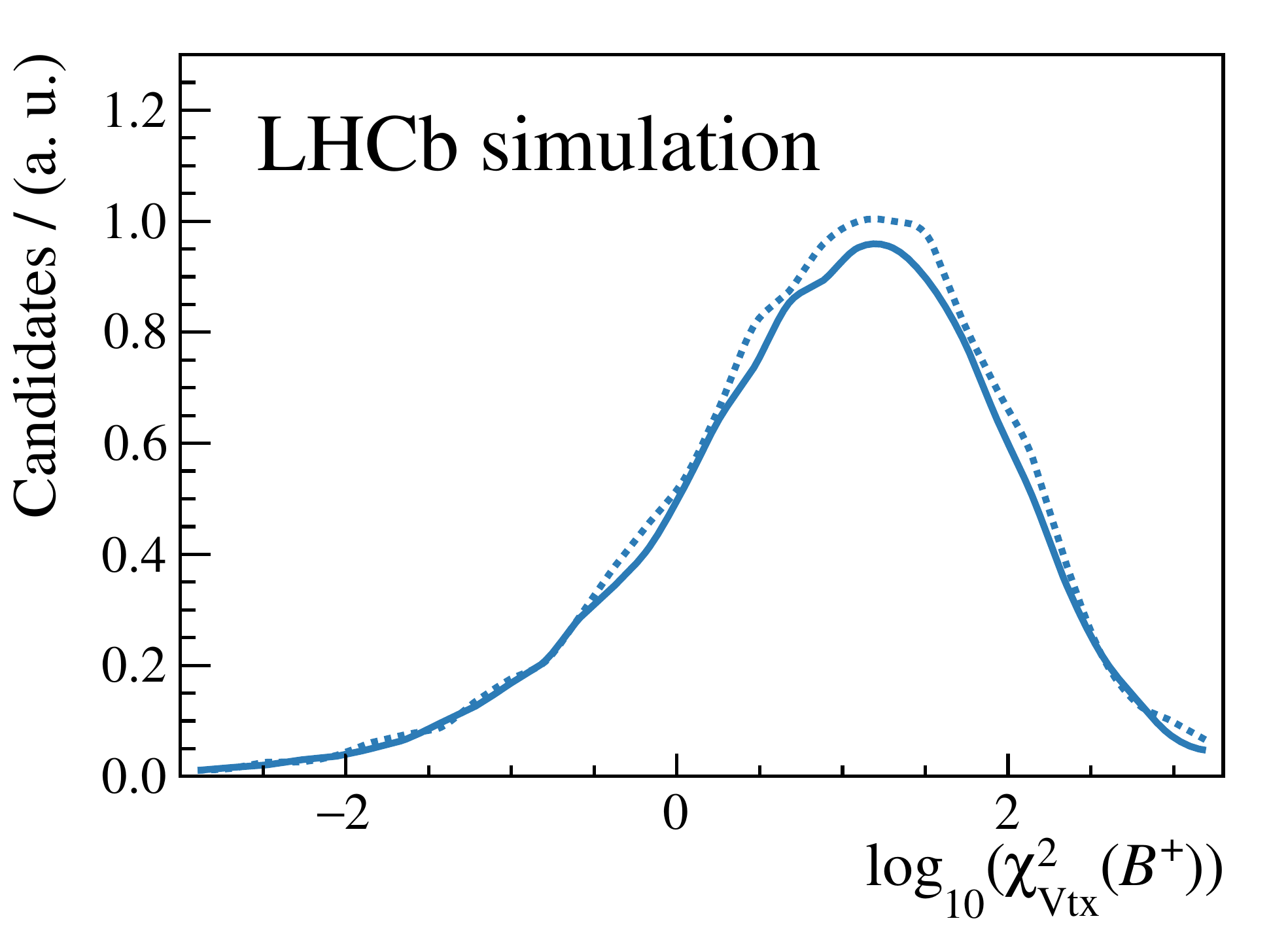}
      \includegraphics[width=0.32\linewidth]{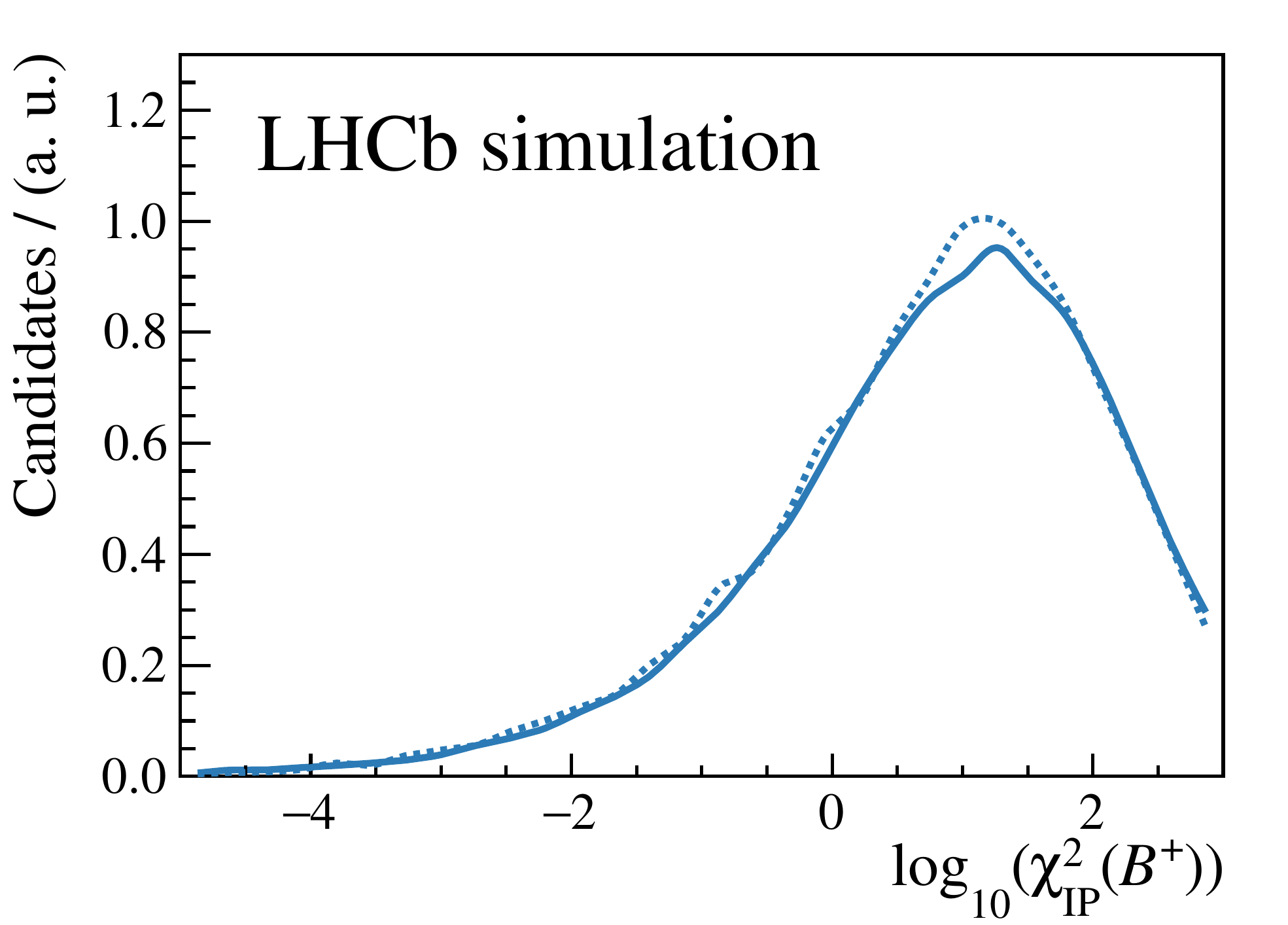}
      \includegraphics[width=0.32\linewidth]{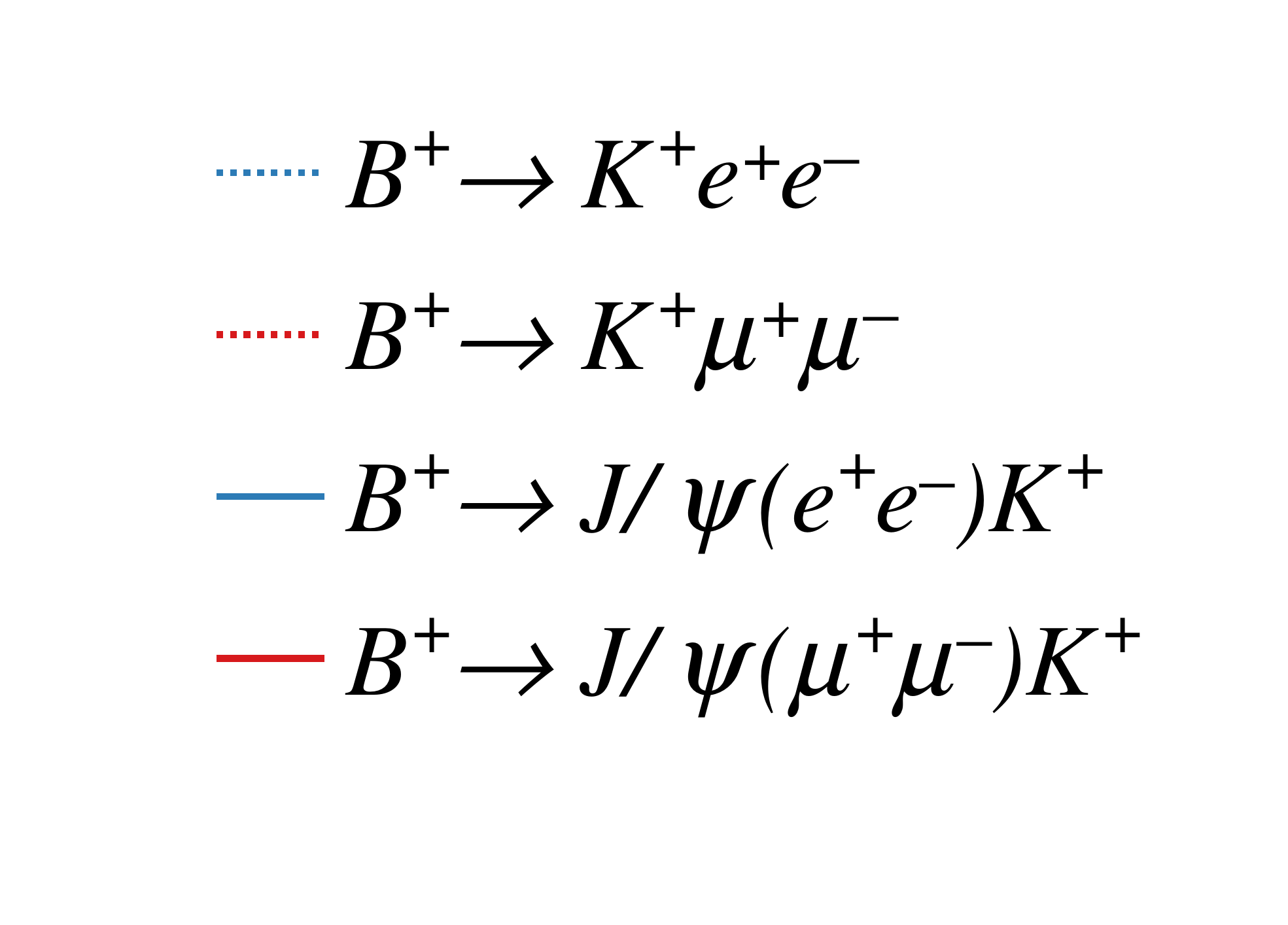}

      \caption{Distributions of various reconstructed properties for simulated decays. The first row  shows the angle between the two leptons, or one lepton and the kaon. 
      The second row shows the rapidity distributions, and the third row the transverse momentum distributions of all the final-state particles.
      The bottom left plot shows the distribution for the quality of the \Bp vertex fit and the bottom right plot shows the $\chisqip(\Bp)$ variable, which quantifies the significance of the \Bp impact parameter. 
      }\label{fig:JPsiVSRare}
   \end{center}
\end{figure}

The reconstructed properties of simulated decays are shown in Fig.~\ref{fig:JPsiVSRare}. The distributions for resonant and nonresonant decays are similar and consequently the determination of the efficiency of each nonresonant decay with respect to its corresponding resonant decay results in the cancellation of systematic effects.

\begin{figure}[!tb]
   \begin{center}
      \includegraphics[width=0.48\linewidth]{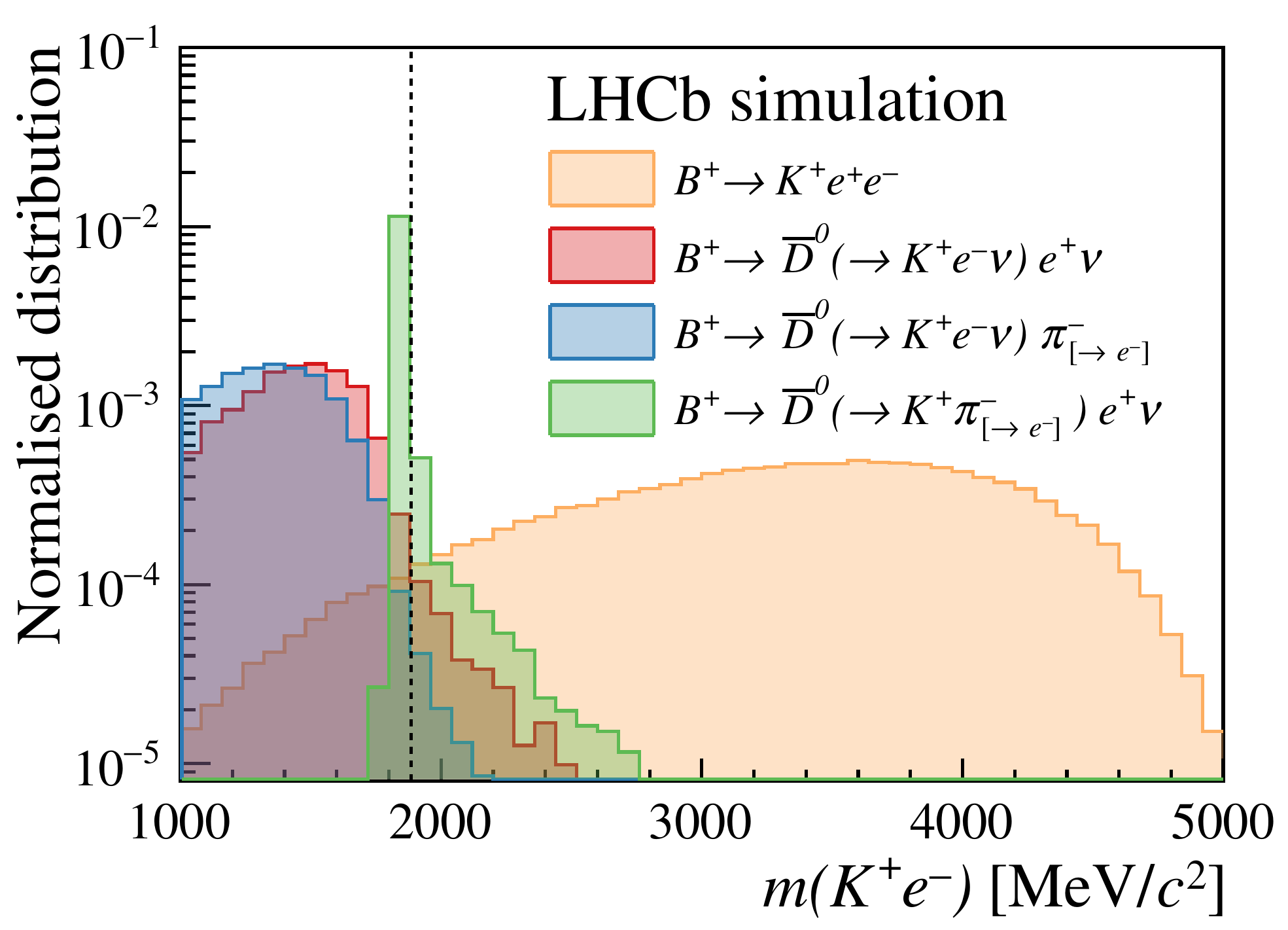}
      \includegraphics[width=0.48\linewidth]{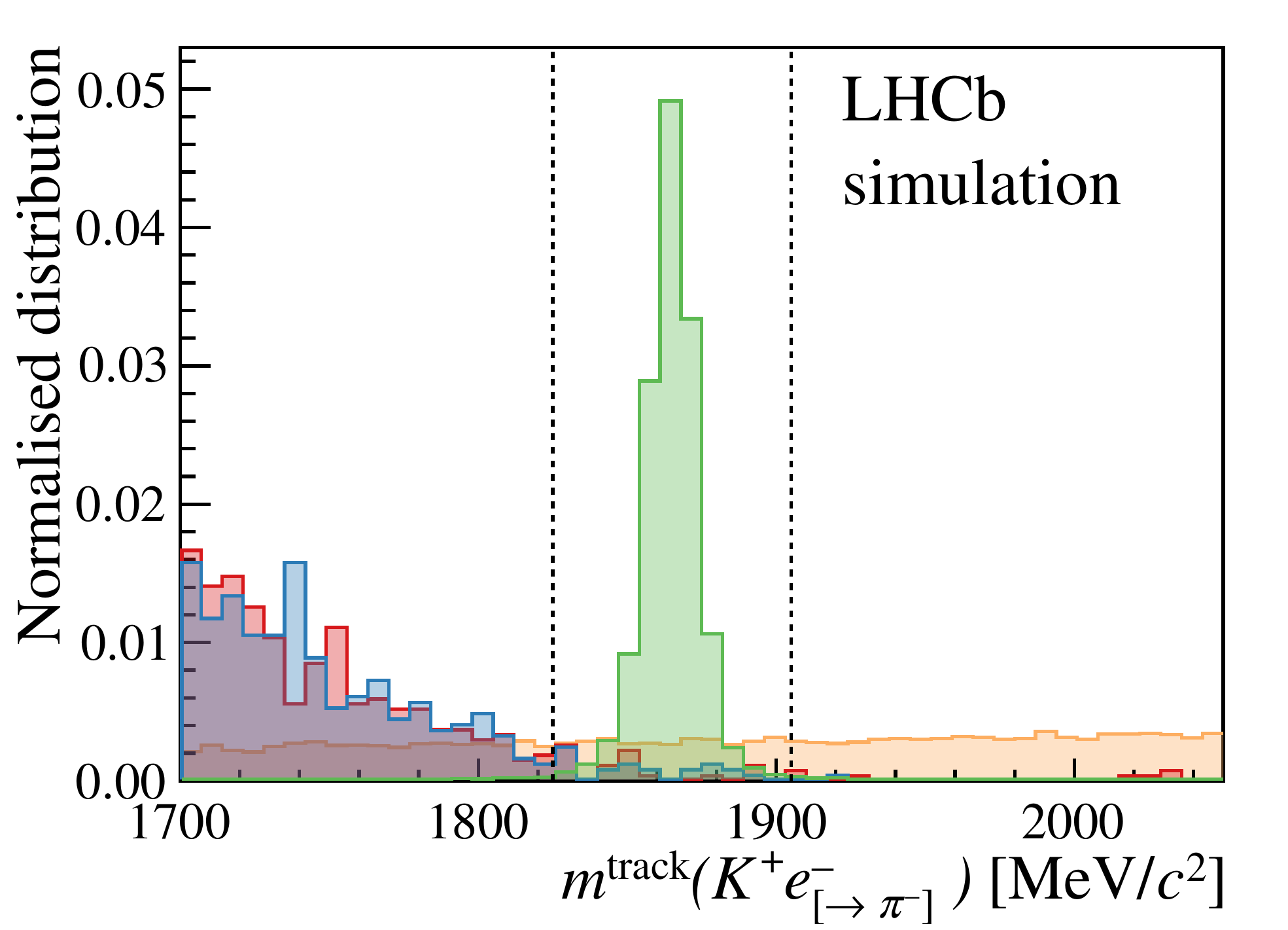}
   \end{center}
    \vspace*{-0.5cm}
   \caption{Simulated $K^+e^-$ mass distributions for  signal and various cascade background samples. The distributions are all normalised to unity.  (Left) 
   the bremsstrahlung correction to the momentum of the electron is taken into account, resulting in a tail to the right. (Right) the mass is computed only from the track information ($m^{\mathrm{track}}$). The notation $\pi_{[\rightarrow e]}$ ($e_{[\rightarrow \pi]}$) is used to denote an electron (pion) that is misidentified as a pion (electron). }\label{fig:cascadeVeto}
\end{figure}

Figure~\ref{fig:cascadeVeto} shows the $m(\Kp e^-)$ mass distribution for \BuKee signal decays and for several cascade background decays. For the mass reconstructed taking into account the bremsstrahlung correction, signal candidates are required to satisfy $m(\Kp e^-)>m_{\Dz}$, suppressing the majority of cascade backgrounds to negligible levels. However, for cascade backgrounds involving $\Dz\to\Kp\pim$ decays, where the $\pip$ is misidentified as an electron, the bremsstrahlung correction gives rise to a long tail of candidates with $m(\Kp e^-)>m_{\Dz}$. Such decays are suppressed by placing an additional veto on the $\Kp e^-$ mass reconstructed without the bremsstrahlung correction, \ie\ based on the measured track momentum alone. This veto removes background around the known \Dz mass, as shown in Fig.~\ref{fig:cascadeVeto}. After the application of both these vetoes, the cascade backgrounds are reduced to a negligible level while retaining 97\% of \BuKmm and 95\% of \BuKee decays passing the remainder of the selection requirements.

\begin{figure}
    \centering
    \includegraphics[width=0.45\textwidth]{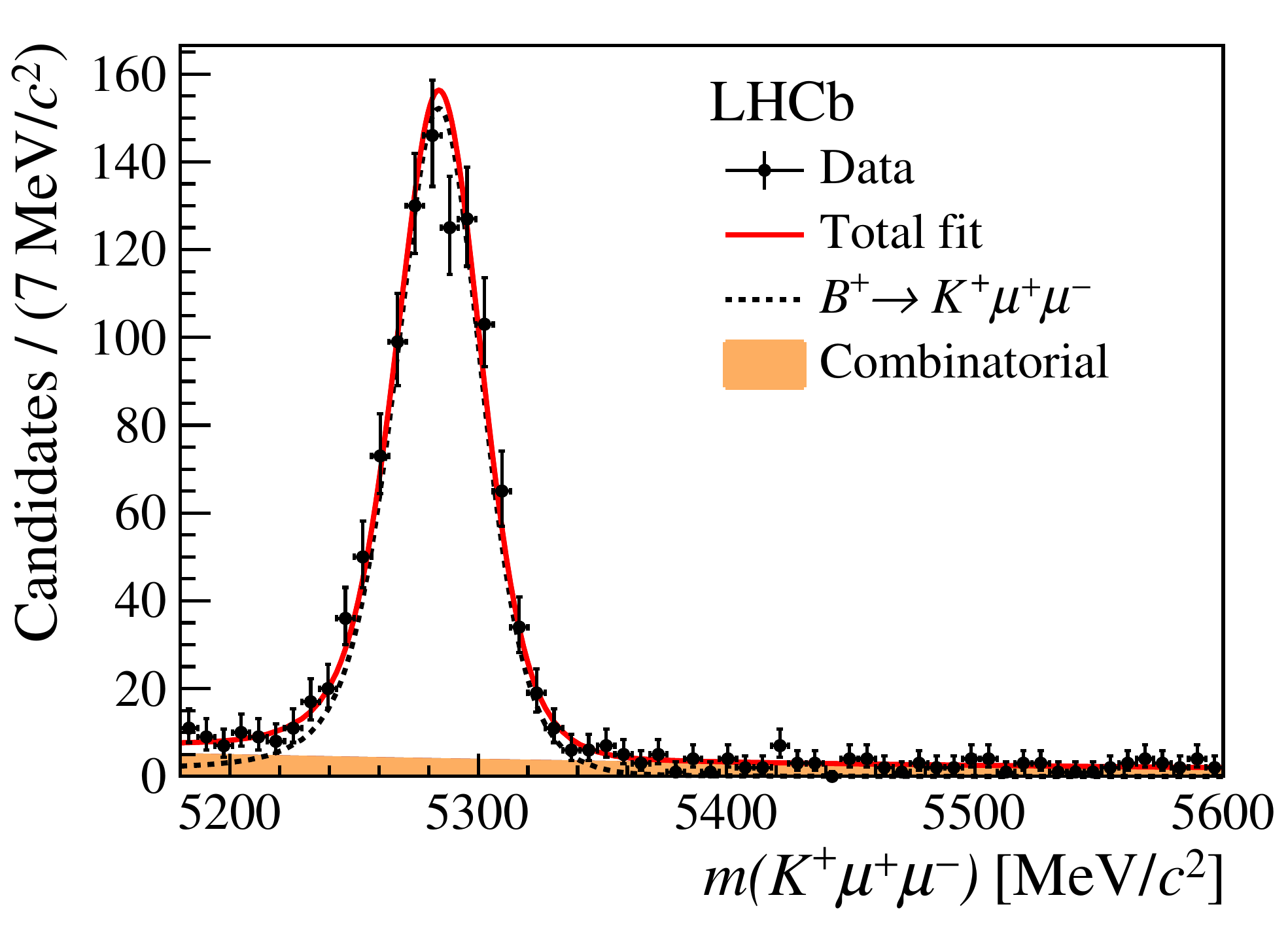}
    \includegraphics[width=0.45\textwidth]{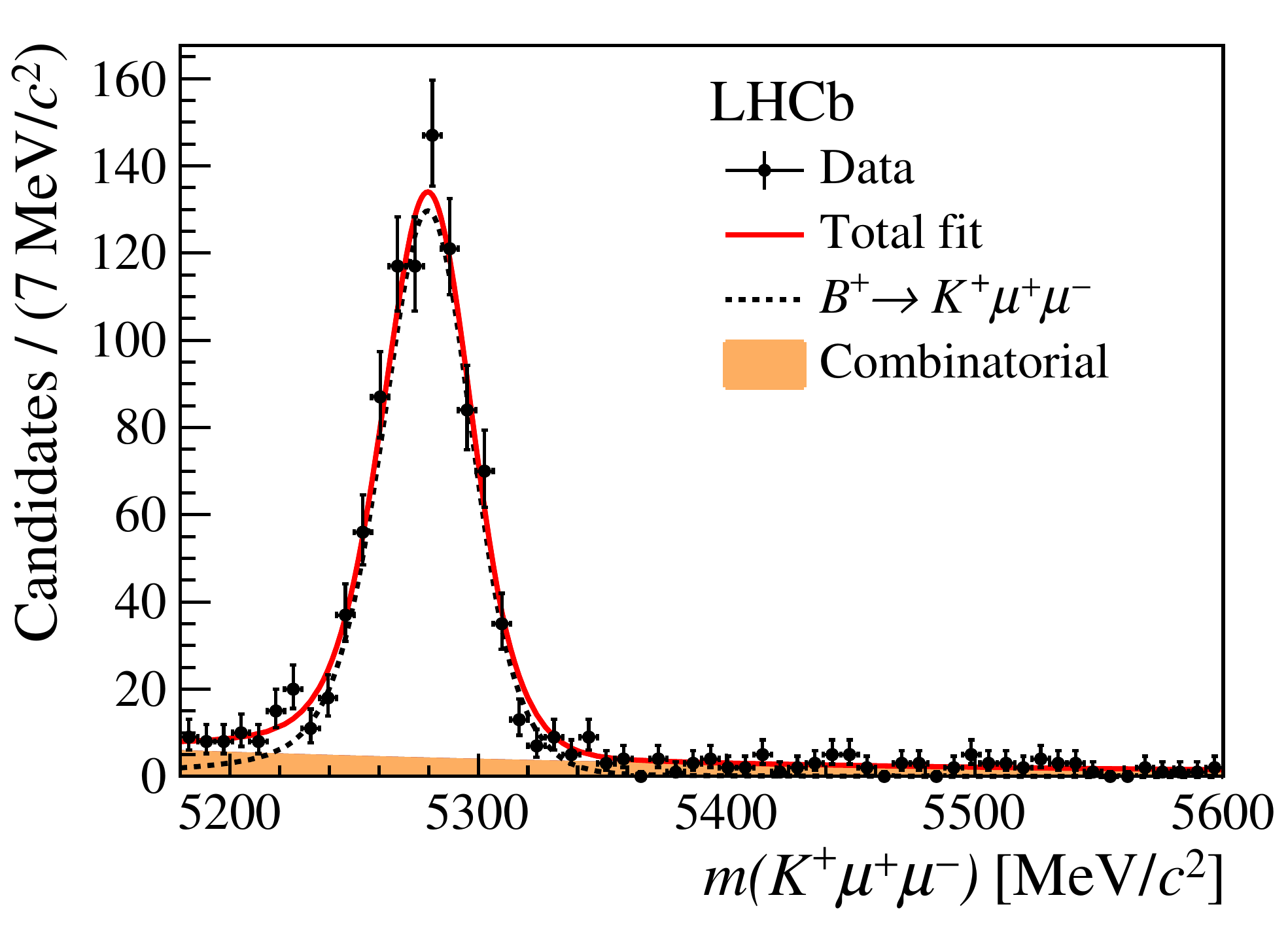}

    \includegraphics[width=0.45\textwidth]{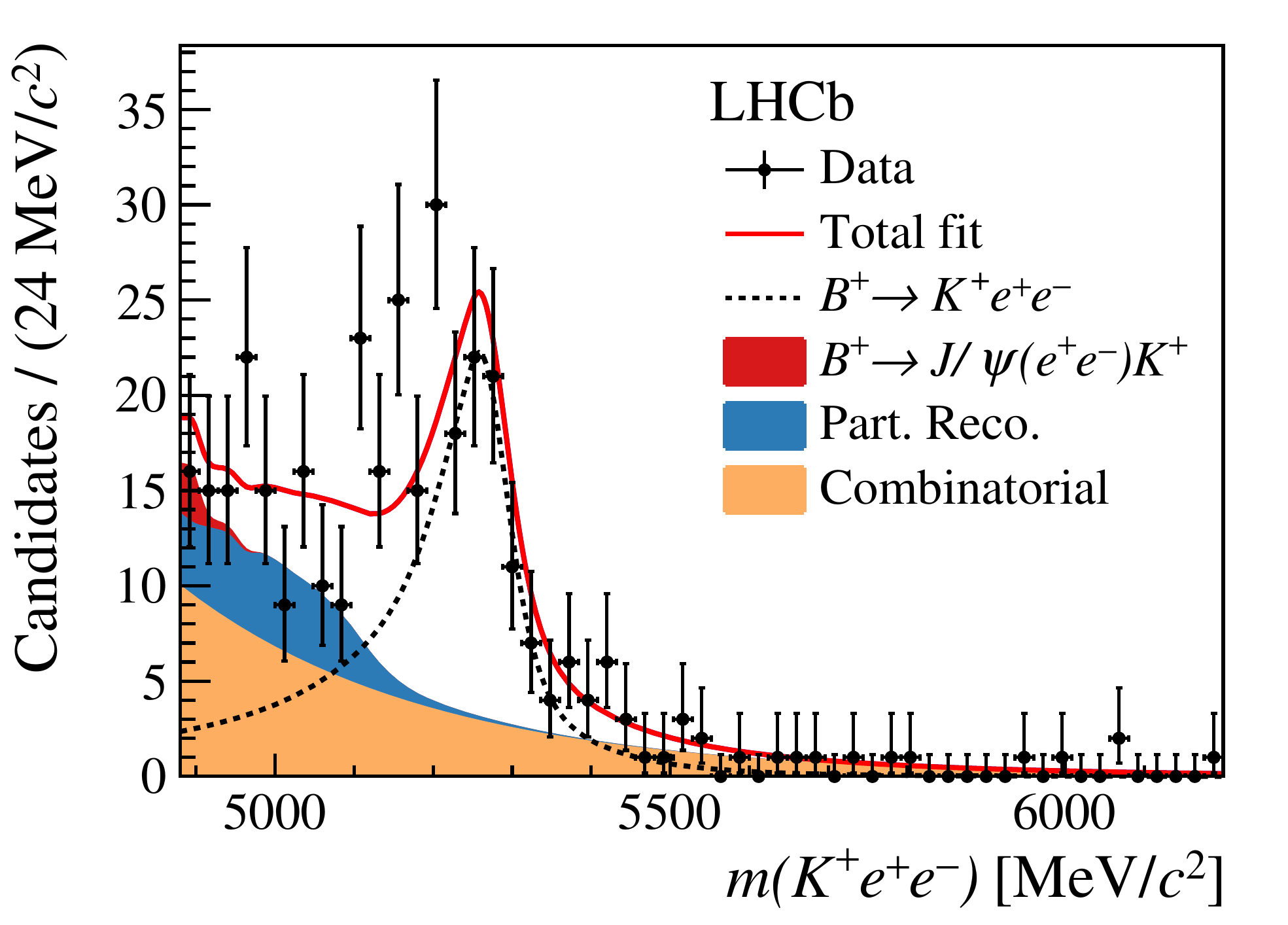}
    \includegraphics[width=0.45\textwidth]{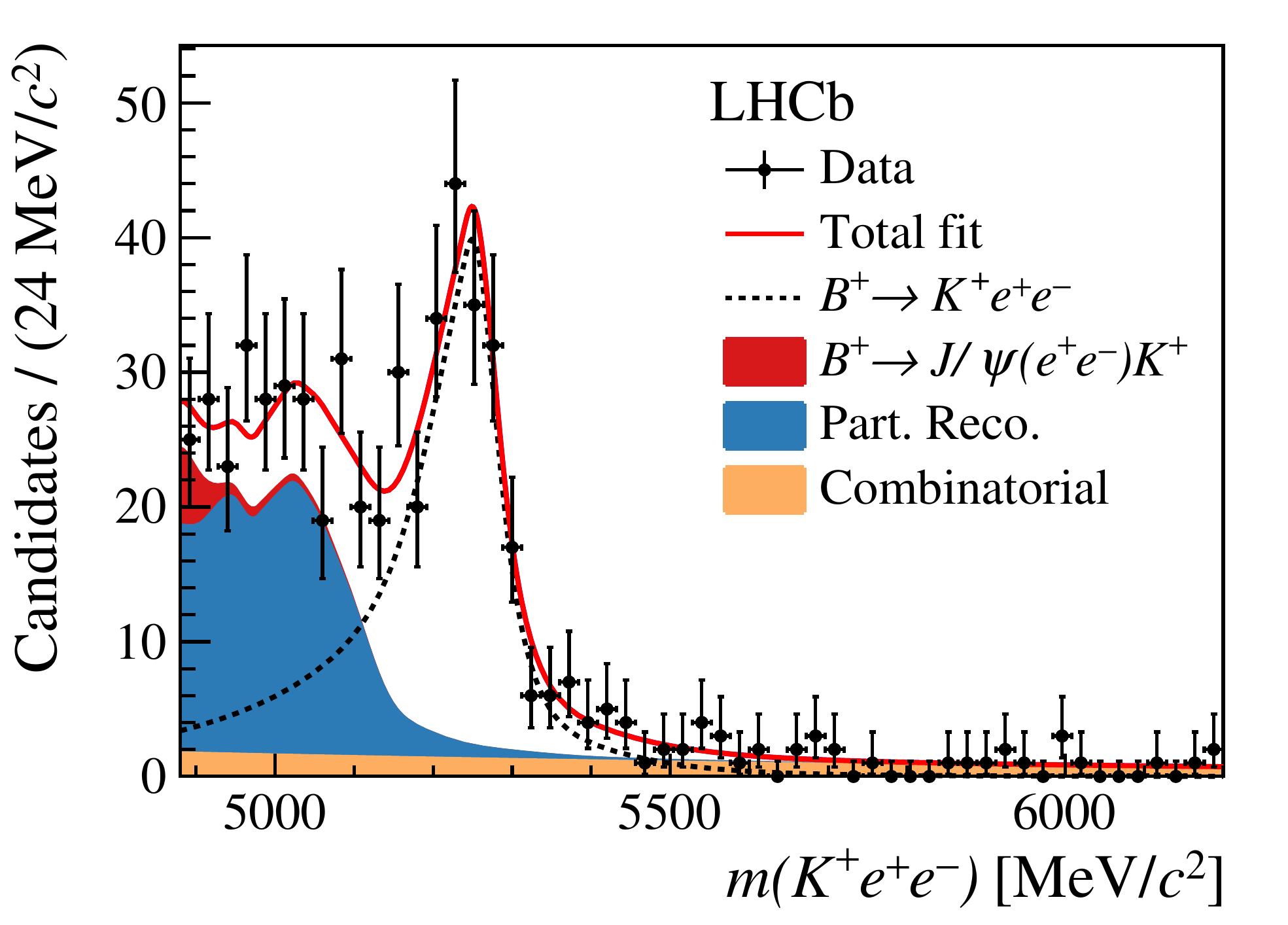}
    
    \includegraphics[width=0.45\textwidth]{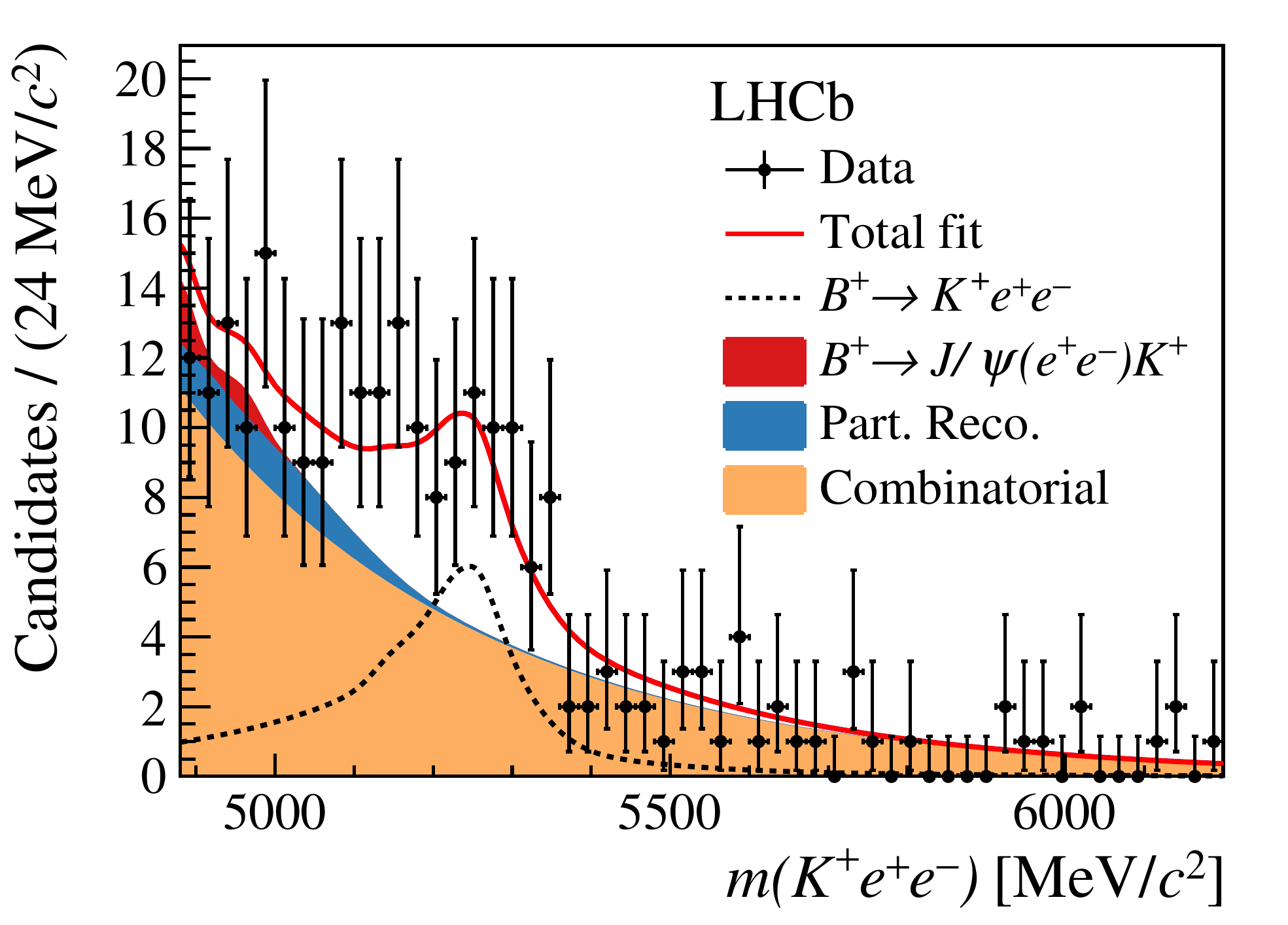}
    \includegraphics[width=0.45\textwidth]{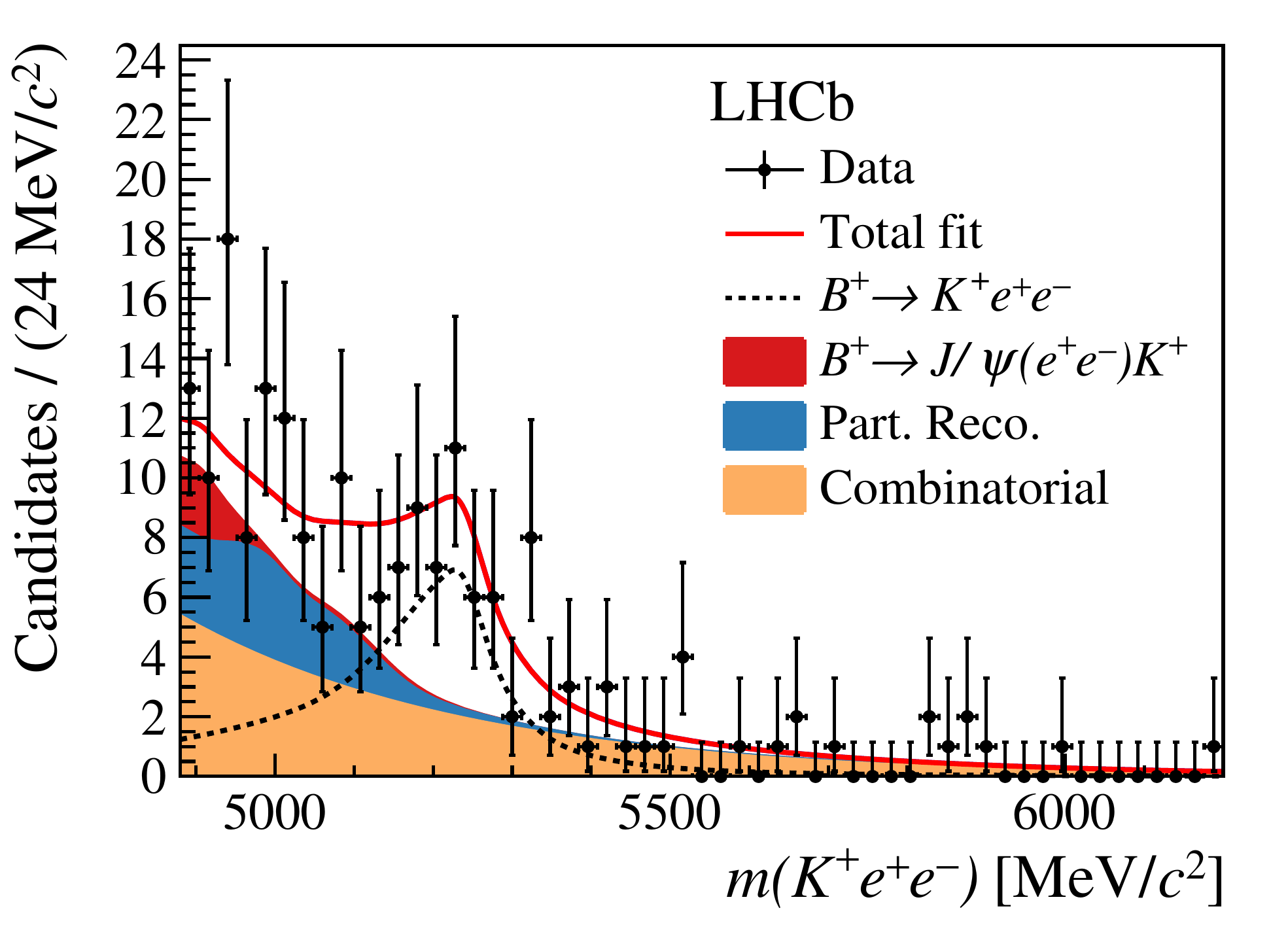}
    
    \includegraphics[width=0.45\textwidth]{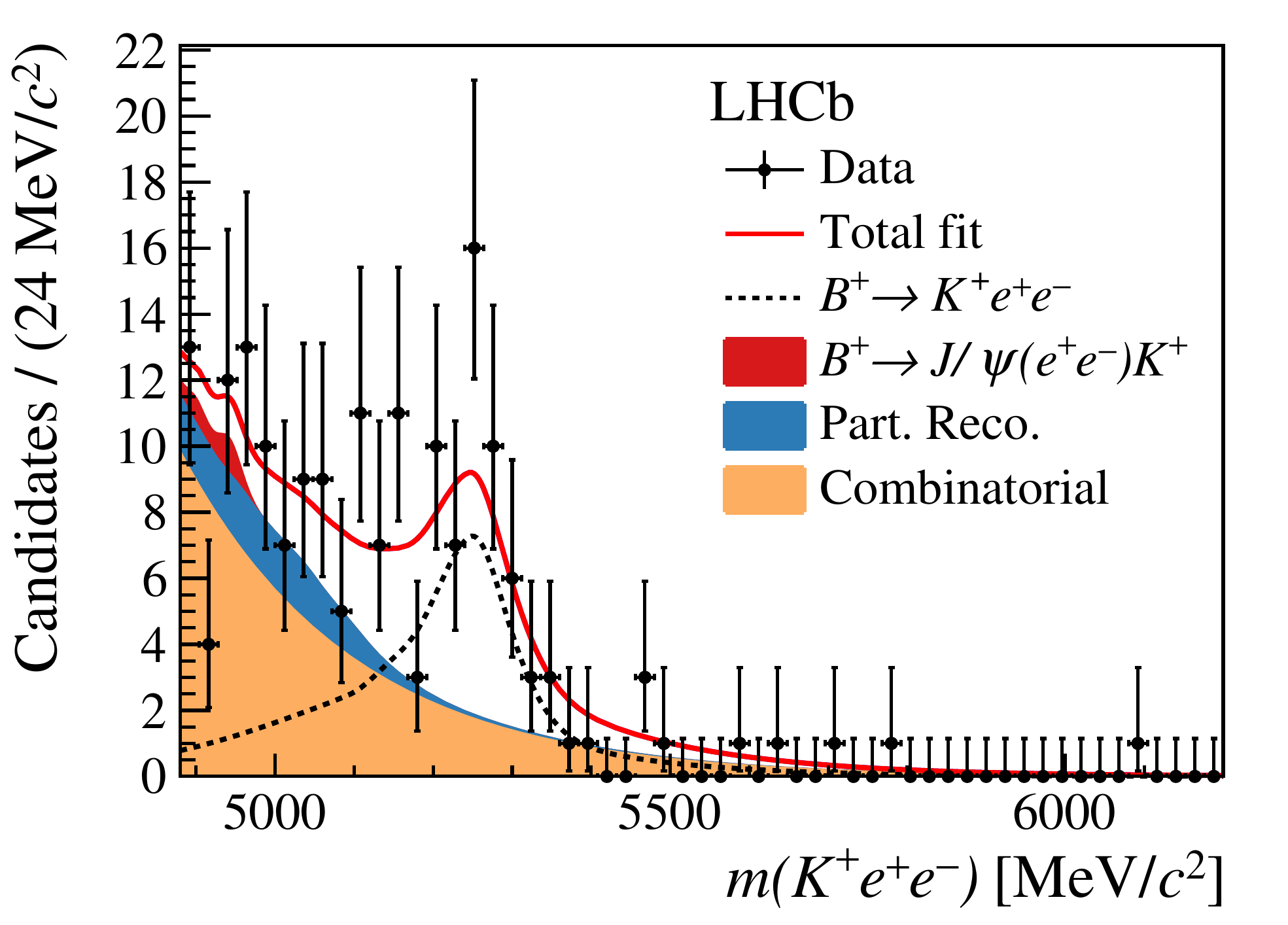}
    \includegraphics[width=0.45\textwidth]{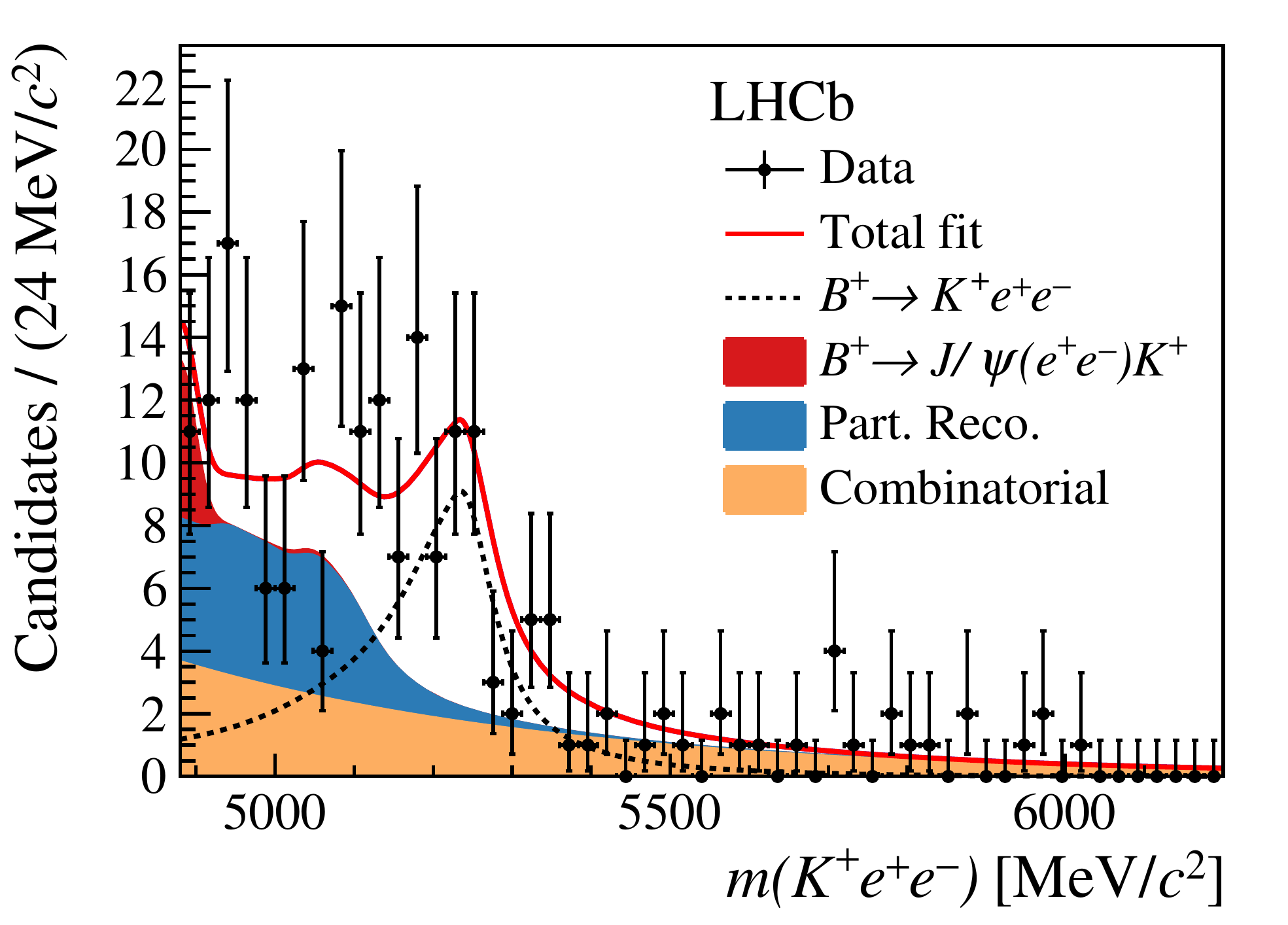}
    \caption{Fit to the \mKll invariant-mass distribution of nonresonant candidates in the (left) 7 and 8\tev and (right) 13\tev data samples. The top row shows the fit to the muon modes and the subsequent rows the fits to the electron modes triggered by (second row) one of the electrons, (third row) the kaon and (last row) by other particles in the event. 
    }
    \label{fig:nonresfits_categories}
\end{figure}

\begin{figure}
    \centering
    \includegraphics[width=0.45\textwidth]{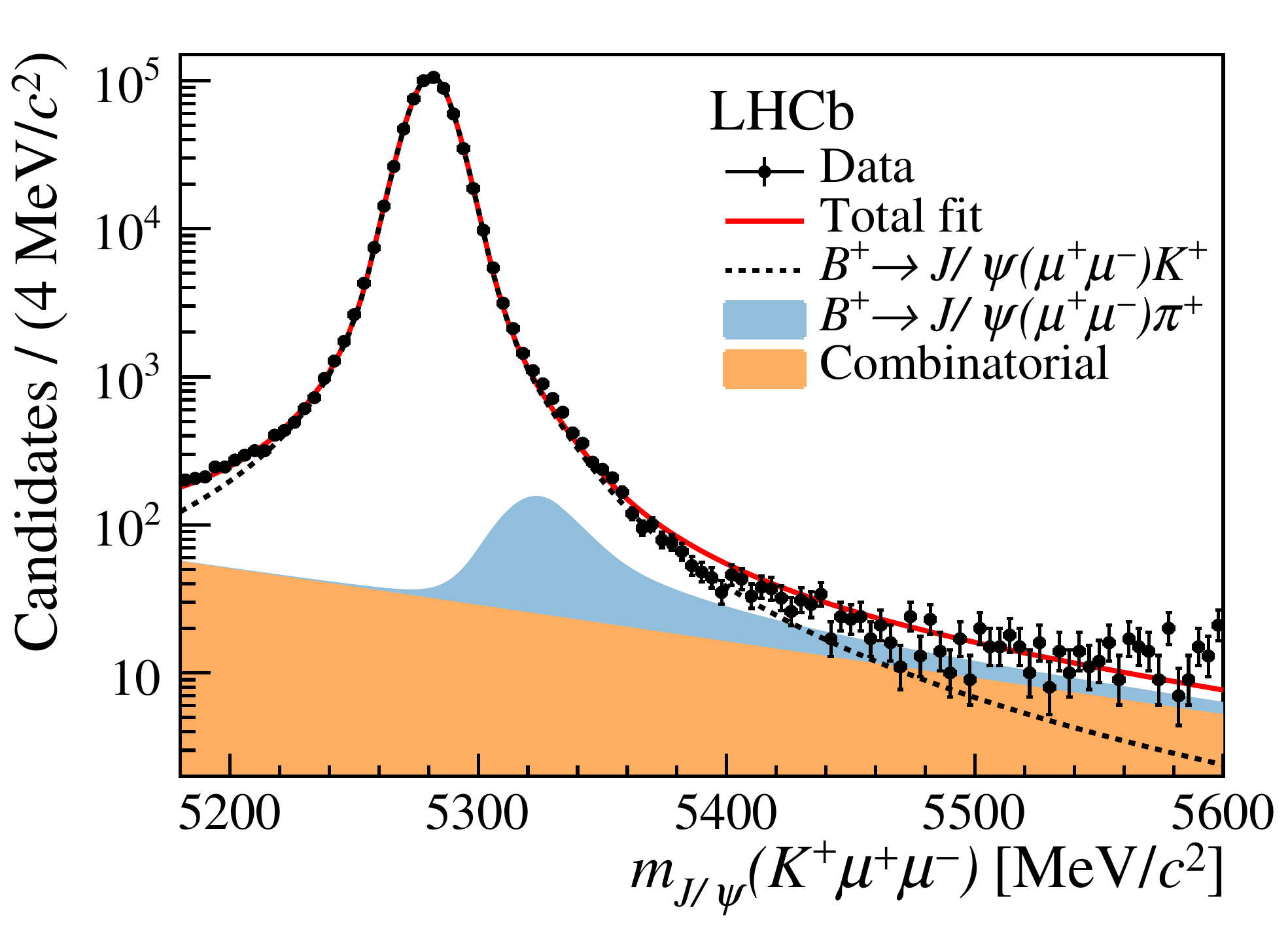}
    \includegraphics[width=0.45\textwidth]{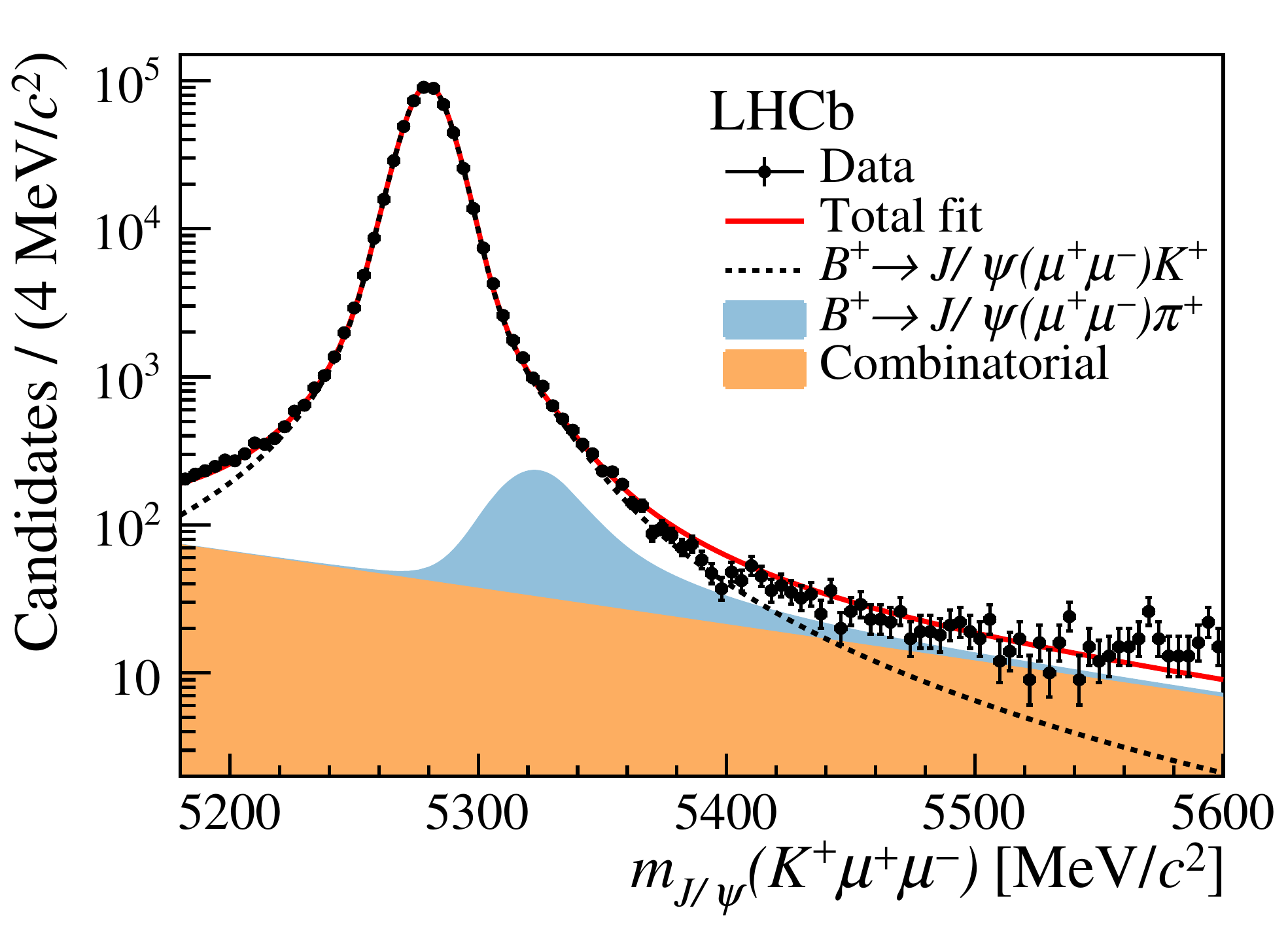}

    \includegraphics[width=0.45\textwidth]{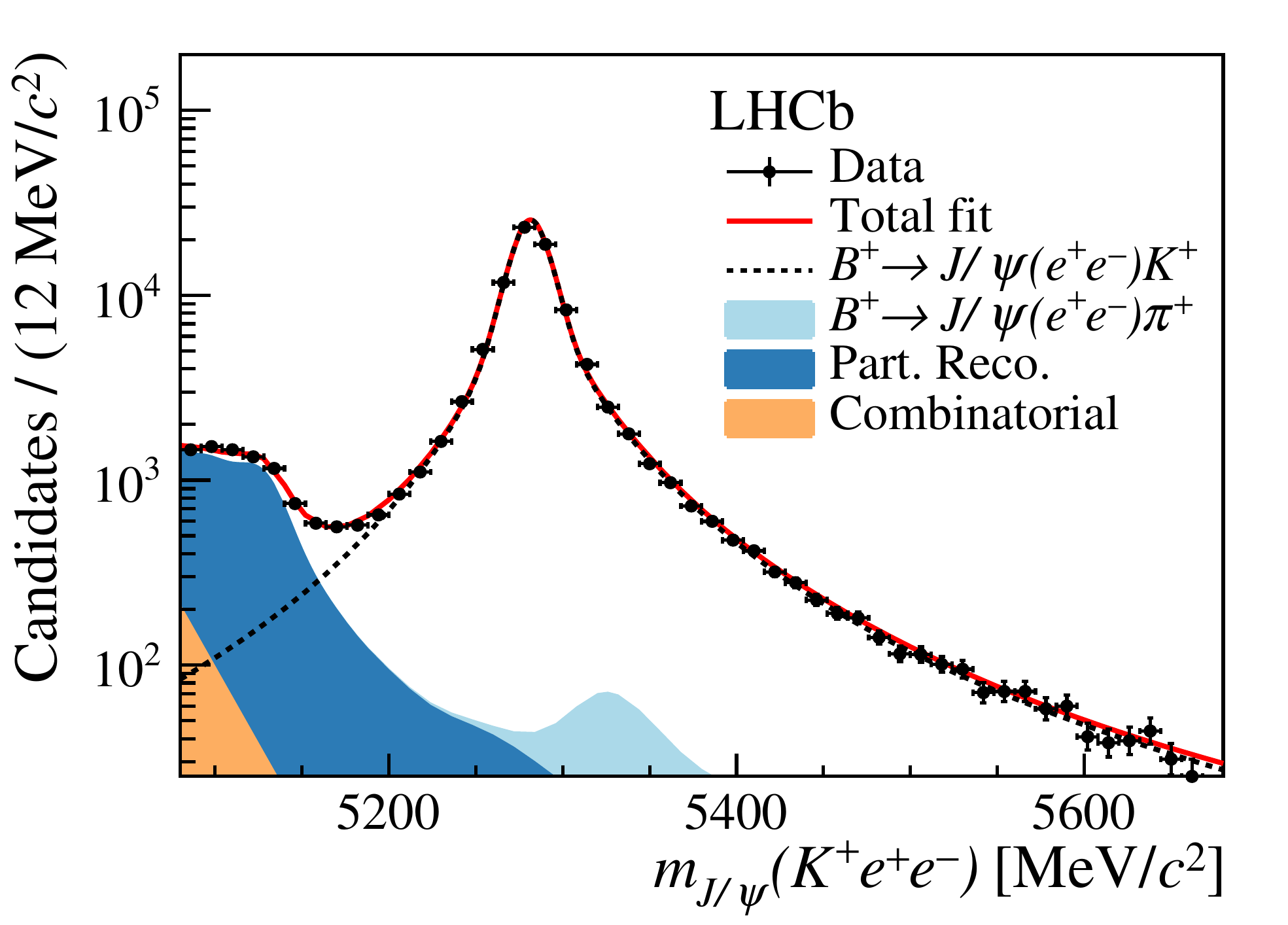}
    \includegraphics[width=0.45\textwidth]{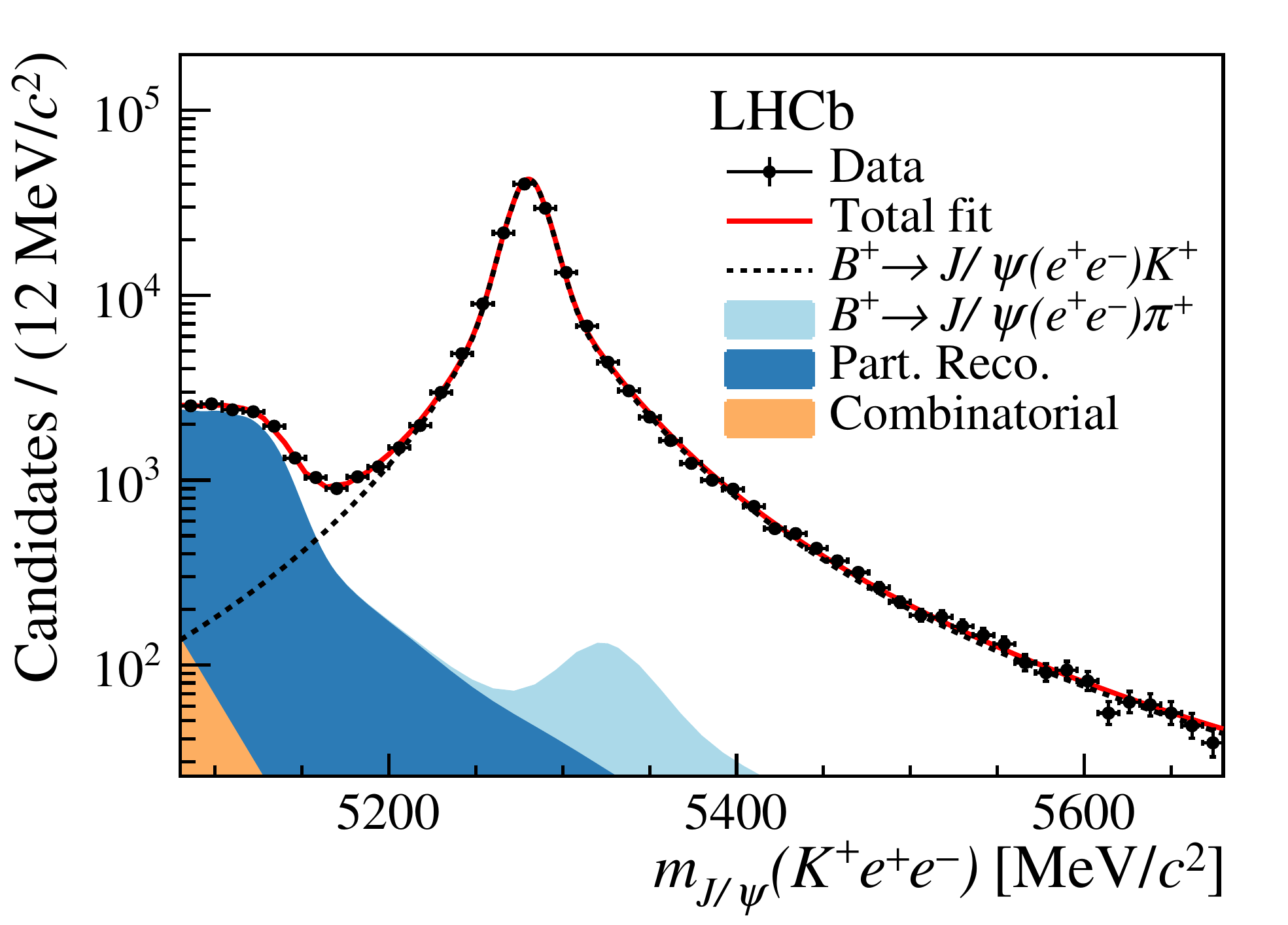}
    
    \includegraphics[width=0.45\textwidth]{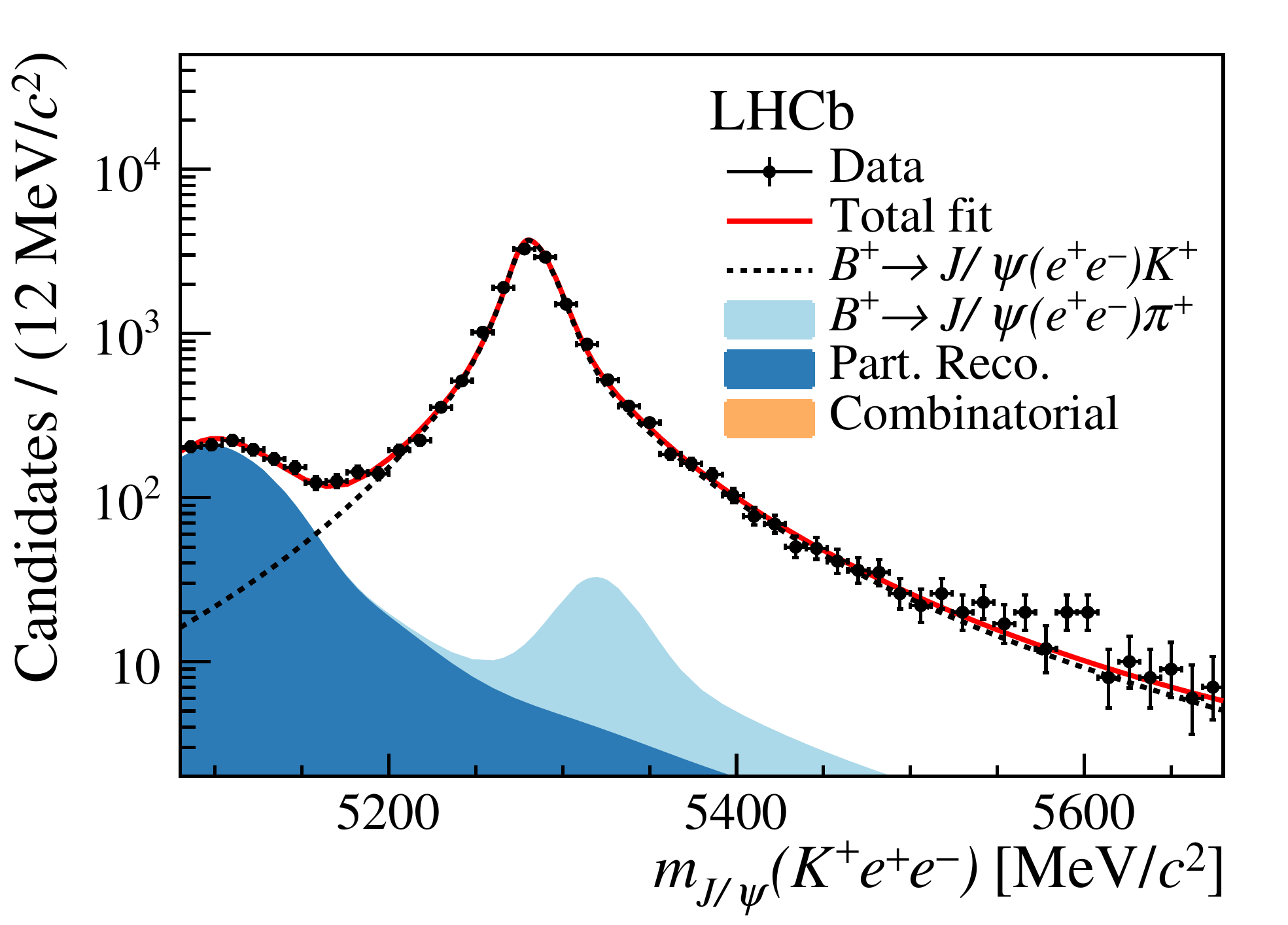}
    \includegraphics[width=0.45\textwidth]{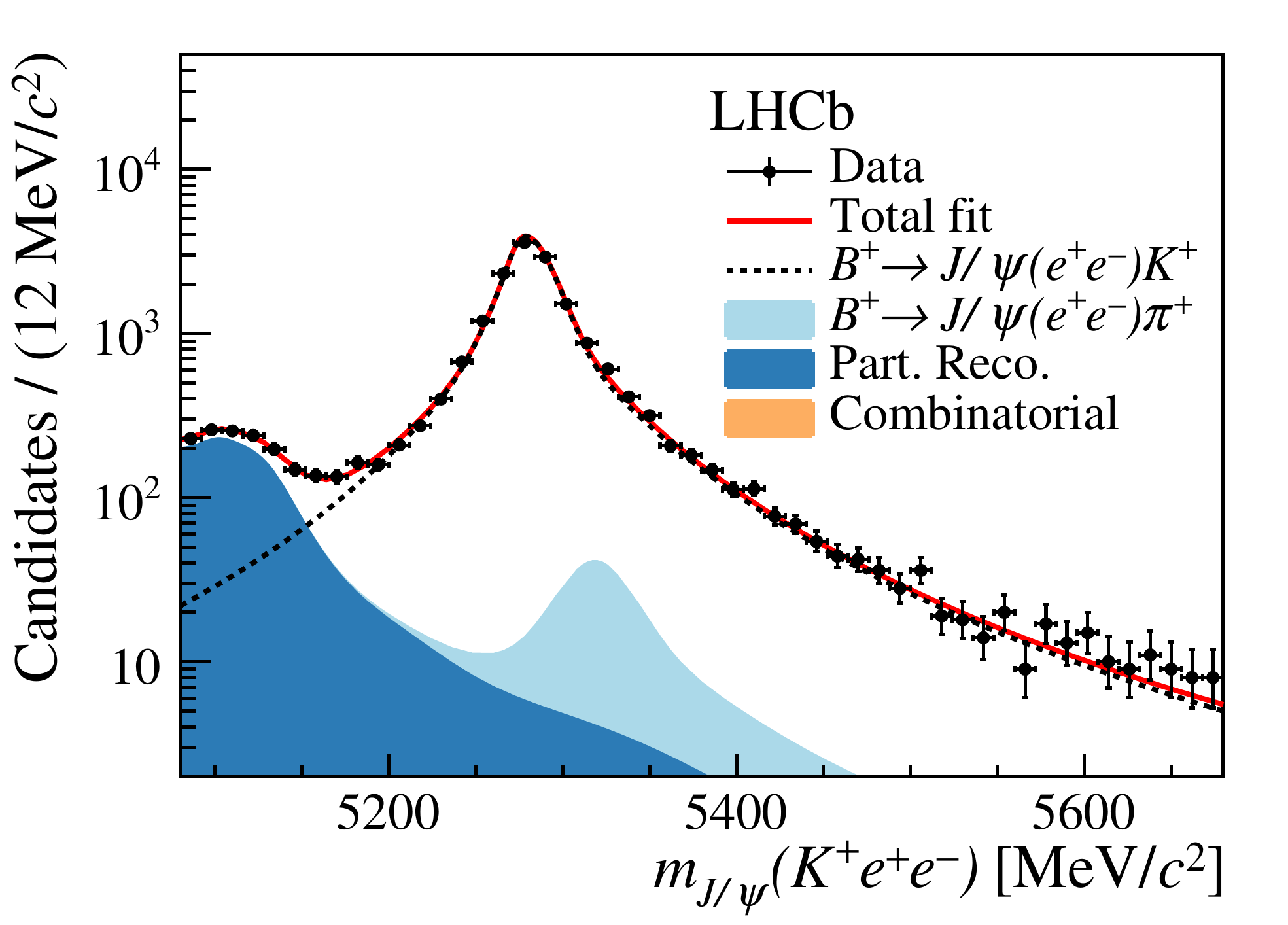}
    
    \includegraphics[width=0.45\textwidth]{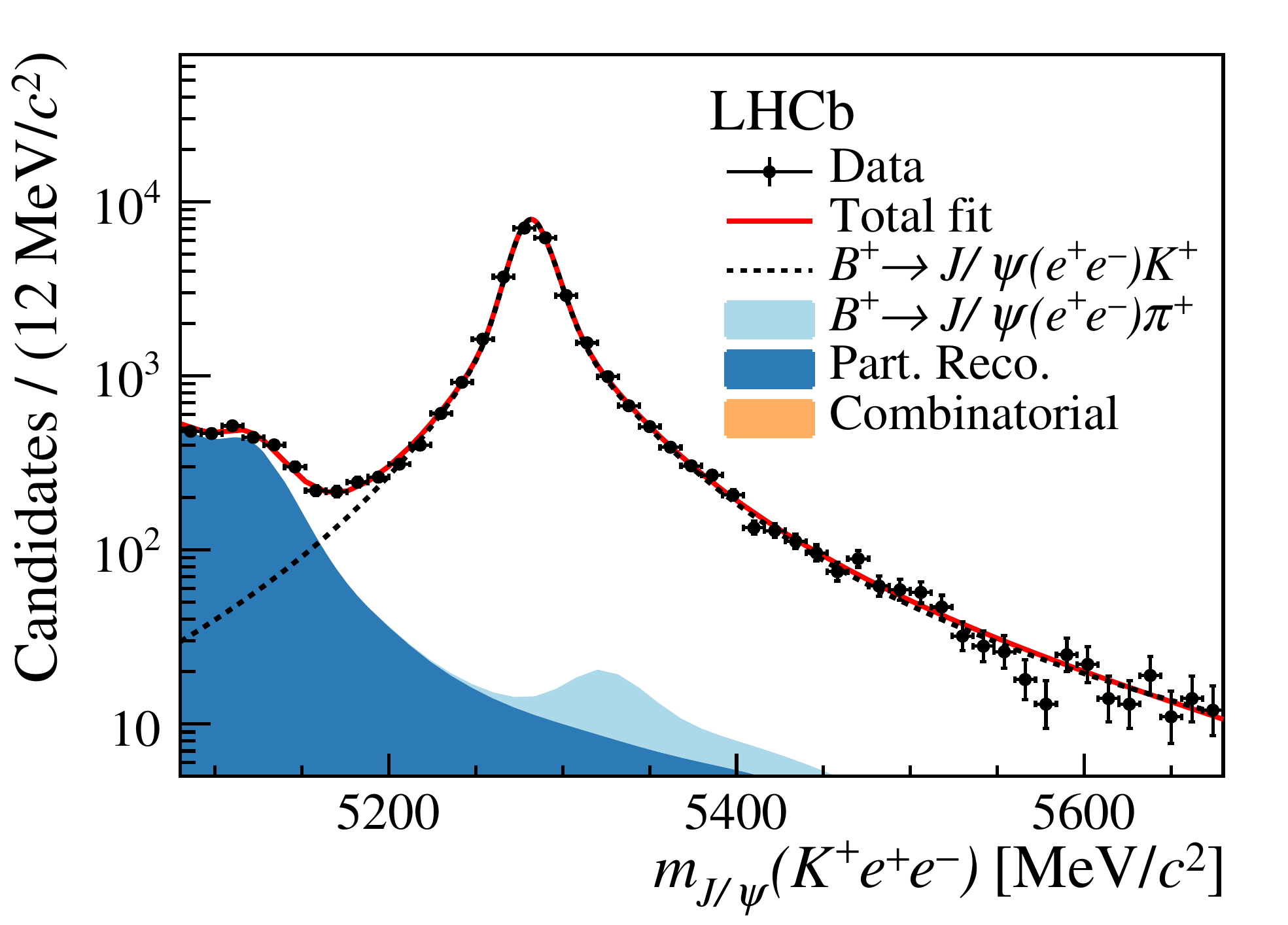}
    \includegraphics[width=0.45\textwidth]{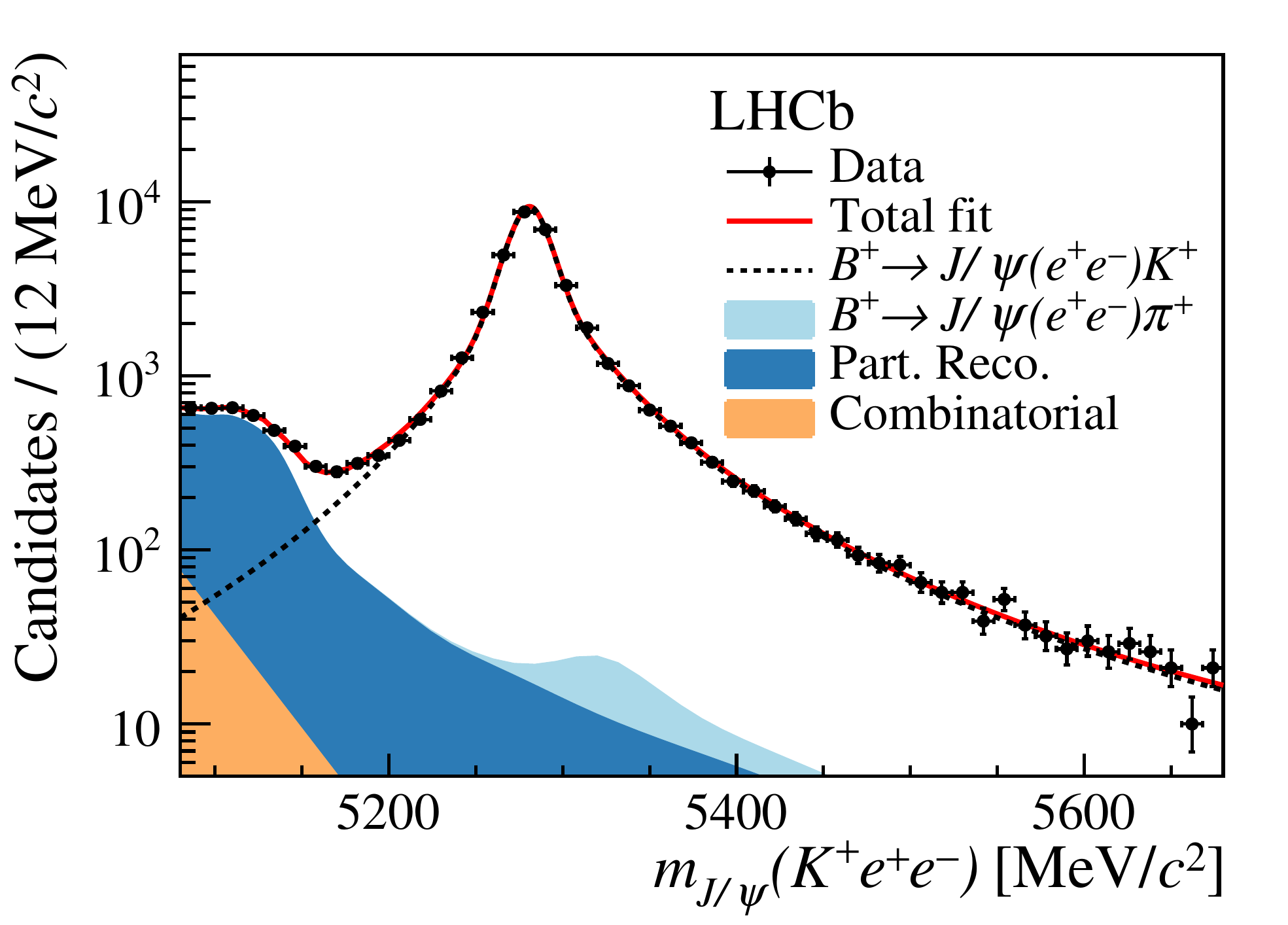}
    
    \caption{Fit to the \mKllconst invariant-mass distribution of resonant candidates in the (left) 7 and 8\tev and (right) 13\tev data samples. The top row shows the fit to the muon modes and the subsequent rows the fits to the electron modes triggered by (second row) one of the electrons, (third row) the kaon and (last row) by other particles in the event. 
   Some large pulls are observed but have a  negligible impact on the yields extracted.}
    \label{fig:resfits_categories}
\end{figure}

The fits to the nonresonant (resonant) decay modes divided into different data-taking periods and trigger categories are shown in Fig.~\ref{fig:nonresfits_categories}  (Fig.~\ref{fig:resfits_categories}).
For the resonant modes these projections come from independent fits to each period/category. The nonresonant figures show the projections from the simultaneous fit that is used to obtain \RK. The total yields for the resonant and nonresonant decays obtained from these fits are given in Table~\ref{tab:yields}. 

\begin{table}[b]
\centering
\caption{Total yields of the decay modes \BuKee, \BuKmm, \BuJpsiKee and \BuJpsiKmm  obtained from the fits to the data.} \label{tab:yields}
\begin{tabular}{lc}
\toprule
Decay Mode  &   Event Yield             \\
\midrule
\BuKee      &   $\phantom{0\,000\,}766 \pm \phantom{0\,0}48$ \\
\BuKmm      &   $\phantom{0\,00}1\,943 \pm \phantom{0\,0}49$ \\
\BuJpsiKee  &   $\phantom{0\,}344\,100 \pm \phantom{0\,}610$ \\
\BuJpsiKmm  &   $1\,161\,800 \pm 1\,100$      \\
\bottomrule
\end{tabular}
\end{table}

\begin{figure}[!htbp]
   \begin{center}
   
   \includegraphics[width=0.45\linewidth,trim={0 0.55cm 0 0}, clip]{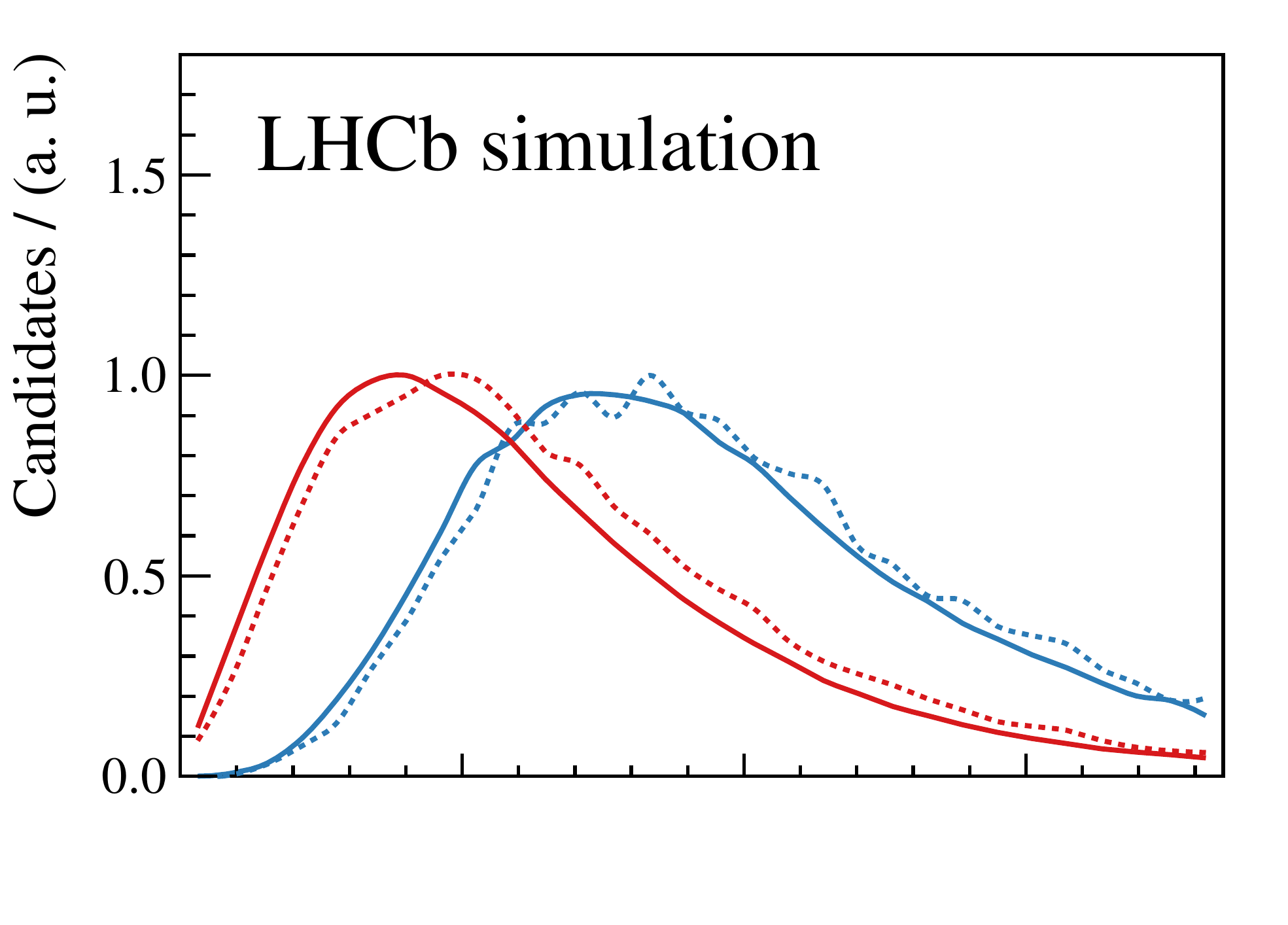}
   \includegraphics[width=0.45\linewidth,trim={0 0.55cm 0 0}, clip]{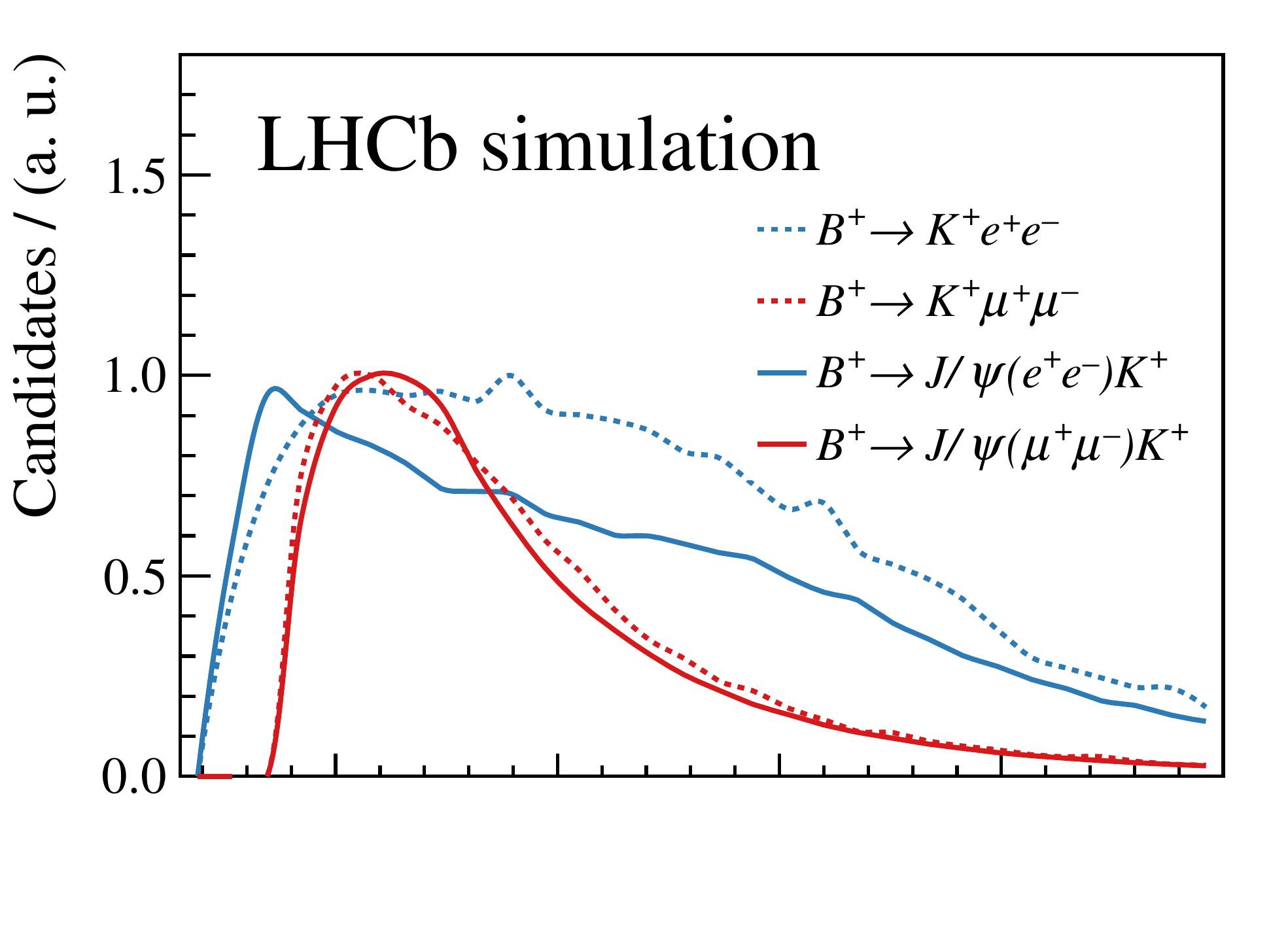}
   
   \includegraphics[width=0.45\linewidth,trim={0 0 0 0.5cm}, clip]{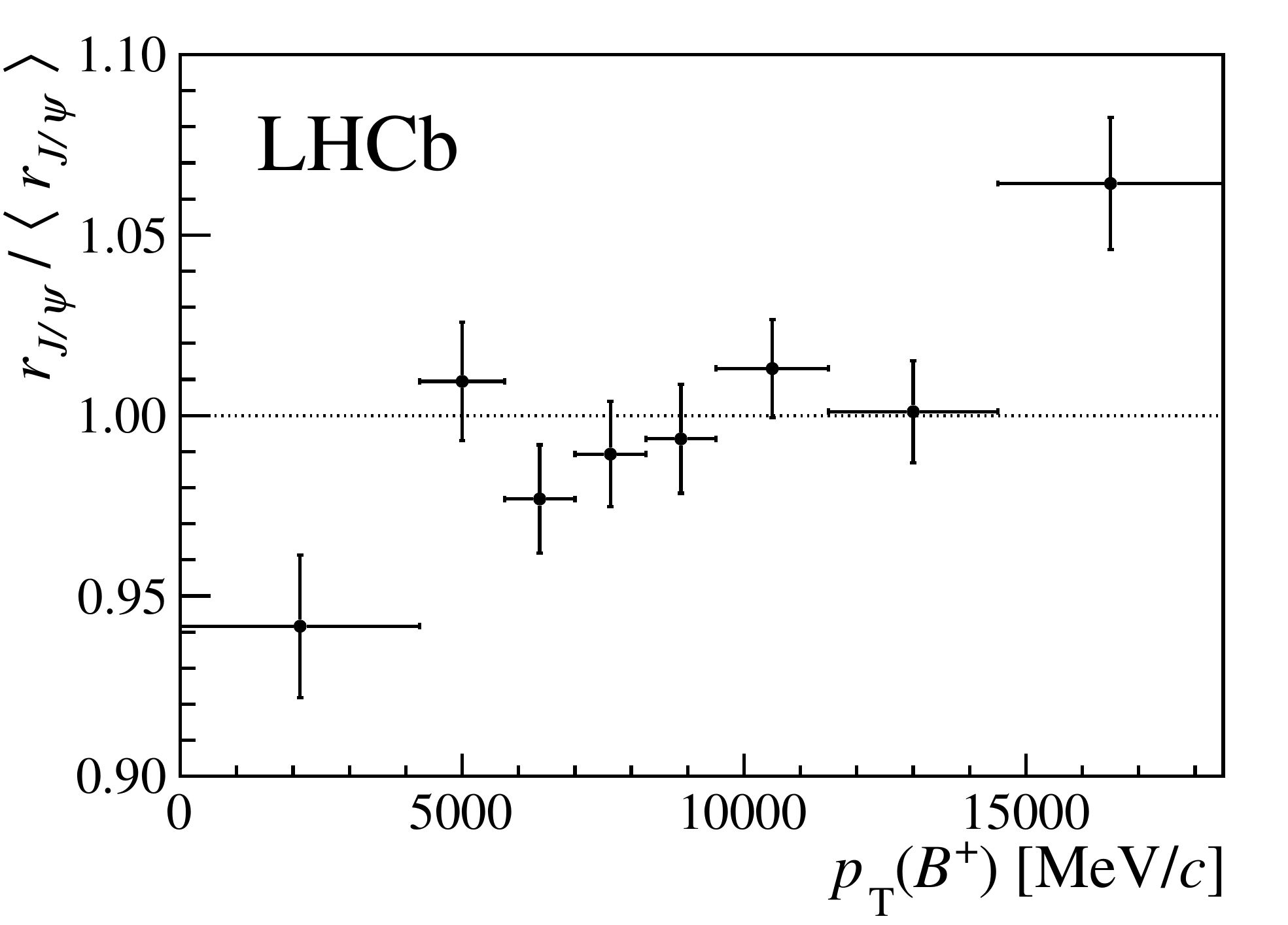}
   \includegraphics[width=0.45\linewidth,trim={0 0 0 0.5cm}, clip]{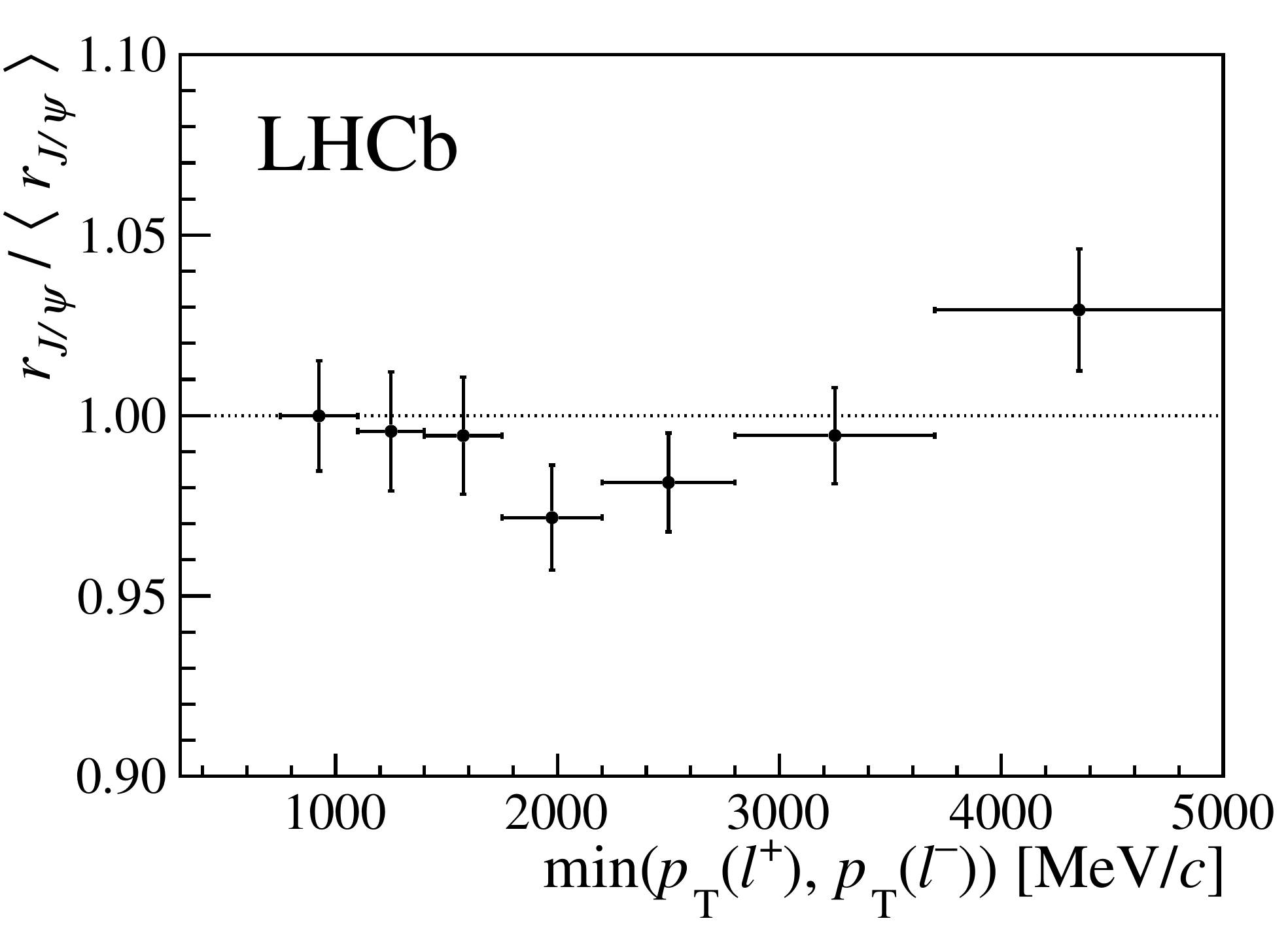}
      \end{center}
     \caption{ 
     (Top) distributions of the spectra of (left) the \Bp transverse momentum and (right) the minimum \pt of the leptons. (Bottom) the single ratio \rjpsi relative to its average value $\left< \rjpsi \right>$ as a function of these variables.  }
    \label{fig:rjpsi_differential}
\end{figure}

\begin{figure}[!htbp]
   \begin{center}
   \includegraphics[width=0.45\linewidth]{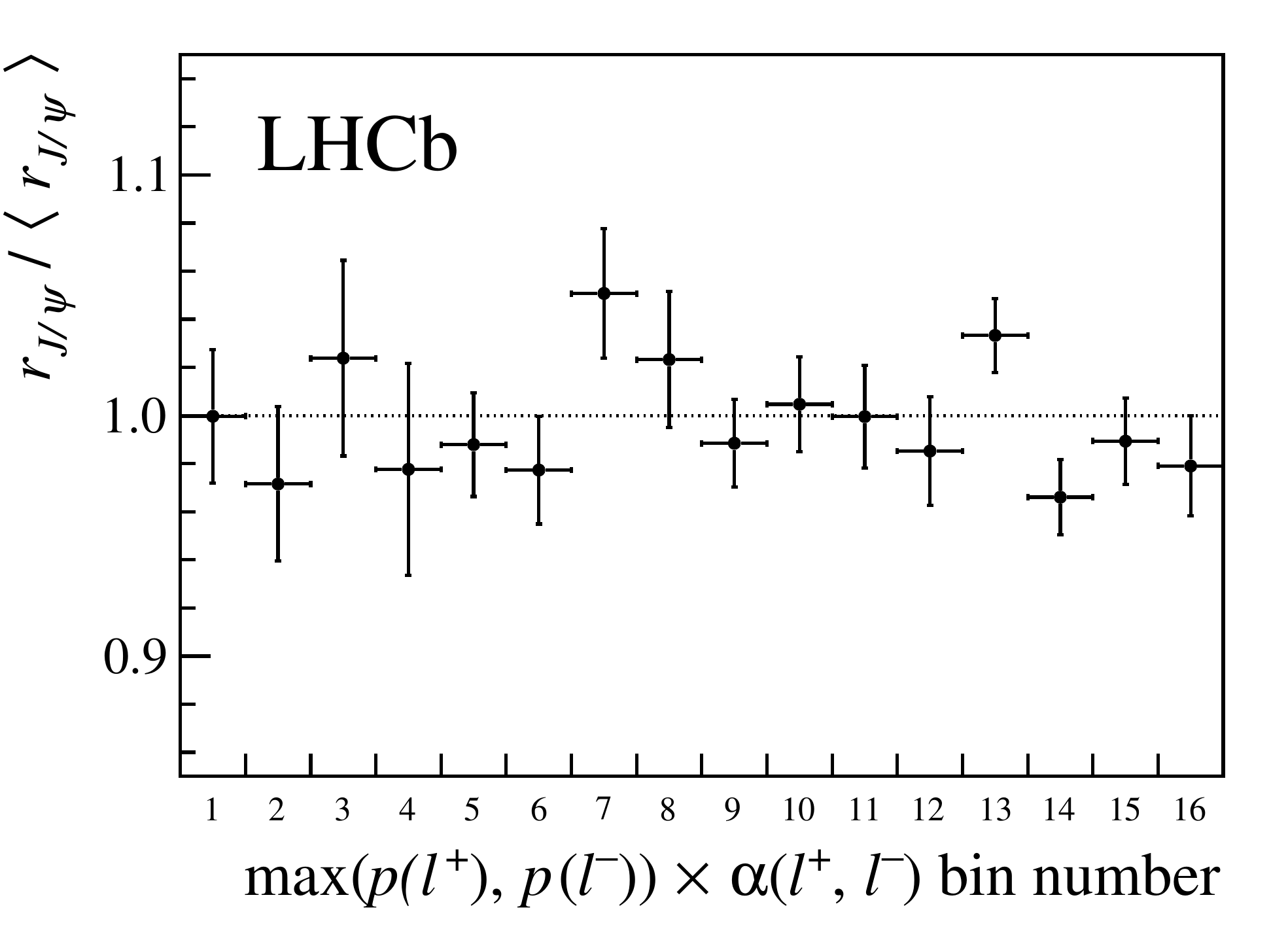}
    \includegraphics[height=0.32\linewidth]{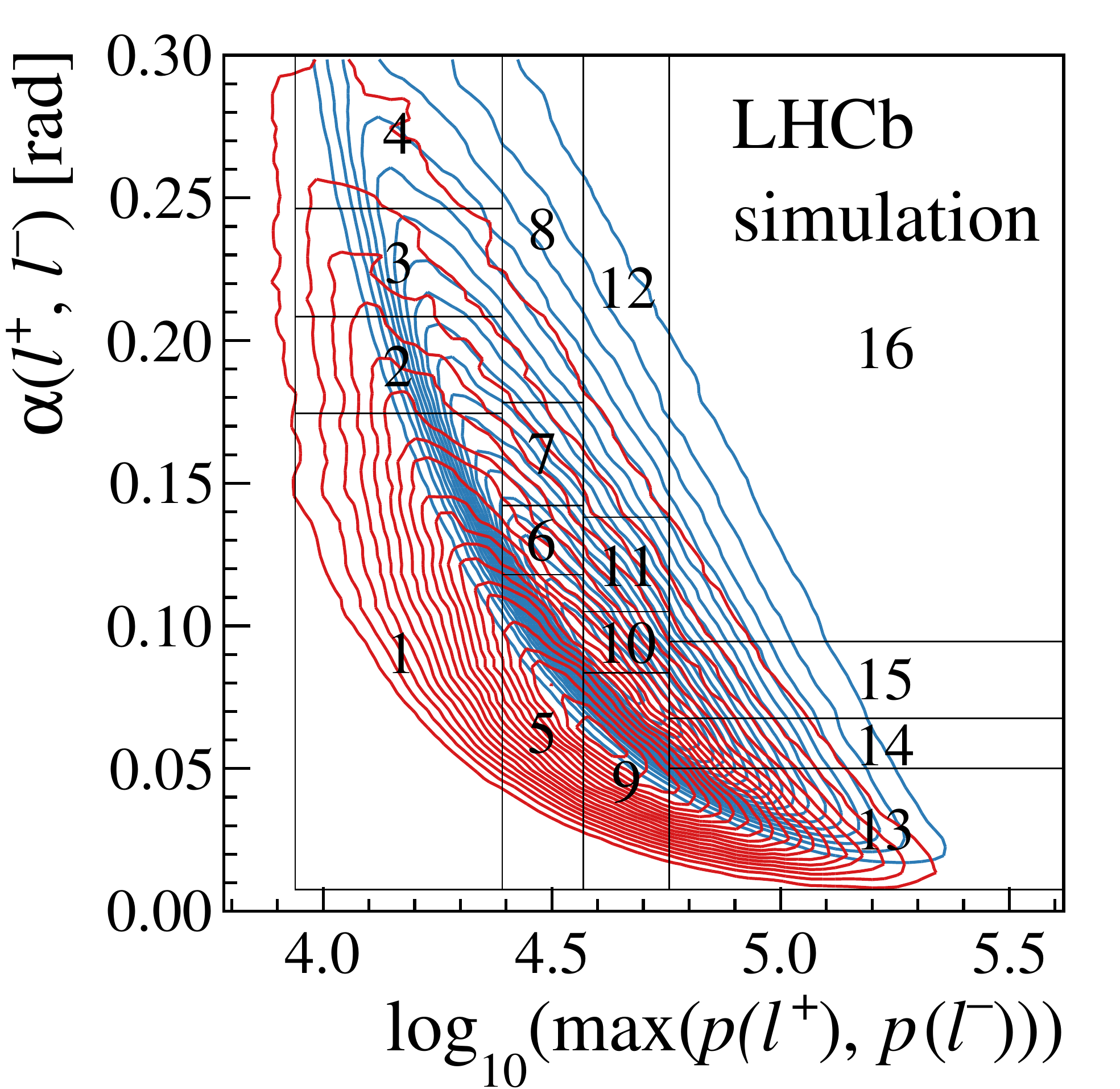}
   \end{center}
     \caption{(Left) the value of \rjpsi, relative to the average value of \rjpsi, measured in two-dimensional bins of the maximum lepton momentum ($p(l)$) and the opening angle between the two leptons ($\alpha(l^+,l^-)$). (Right) the bin definition in this two-dimensional space together with the 
     distribution for \BuKee (\BuJpsiKee) decays depicted as red (blue) contours.}
    \label{fig:rjpsi_bin}
\end{figure}

The distributions of the ratio \rjpsi as a function of the \Bp transverse momentum and the minimum \pt of the leptons are shown in Fig.~\ref{fig:rjpsi_differential}, together with the spectra expected for the resonant and nonresonant decays. 
This single ratio does not benefit from the cancellation of systematic effects that the double ratio exploits in the measurement of \RK, and is therefore a stringent test of the control of the efficiencies.
No significant trend is observed in either \rjpsi distribution and the results are compatible with $\rjpsi=1$. Assuming the deviations observed indicate genuine mismodelling of the efficiencies, rather than fluctuations, and taking into account the spectrum of the relevant variables in the nonresonant decay modes of interest, a total shift of \RK at the level 0.002  would be expected for the \Bp \pt and lepton  minimum \pt. This variation is compatible with the estimated systematic uncertainties on \RK. Similarly, the variations seen in all other reconstructed quantities are compatible with the systematic uncertainties assigned. The ratio \rjpsi is also computed in two- and three-dimensional bins of reconstructed quantities. An example is shown in Fig.~\ref{fig:rjpsi_bin}. Again, no significant trend is seen and the distributions are compatible with $\rjpsi=1$. 

\begin{figure}[!t]
   \begin{center}
      \includegraphics[height=0.25\textheight]{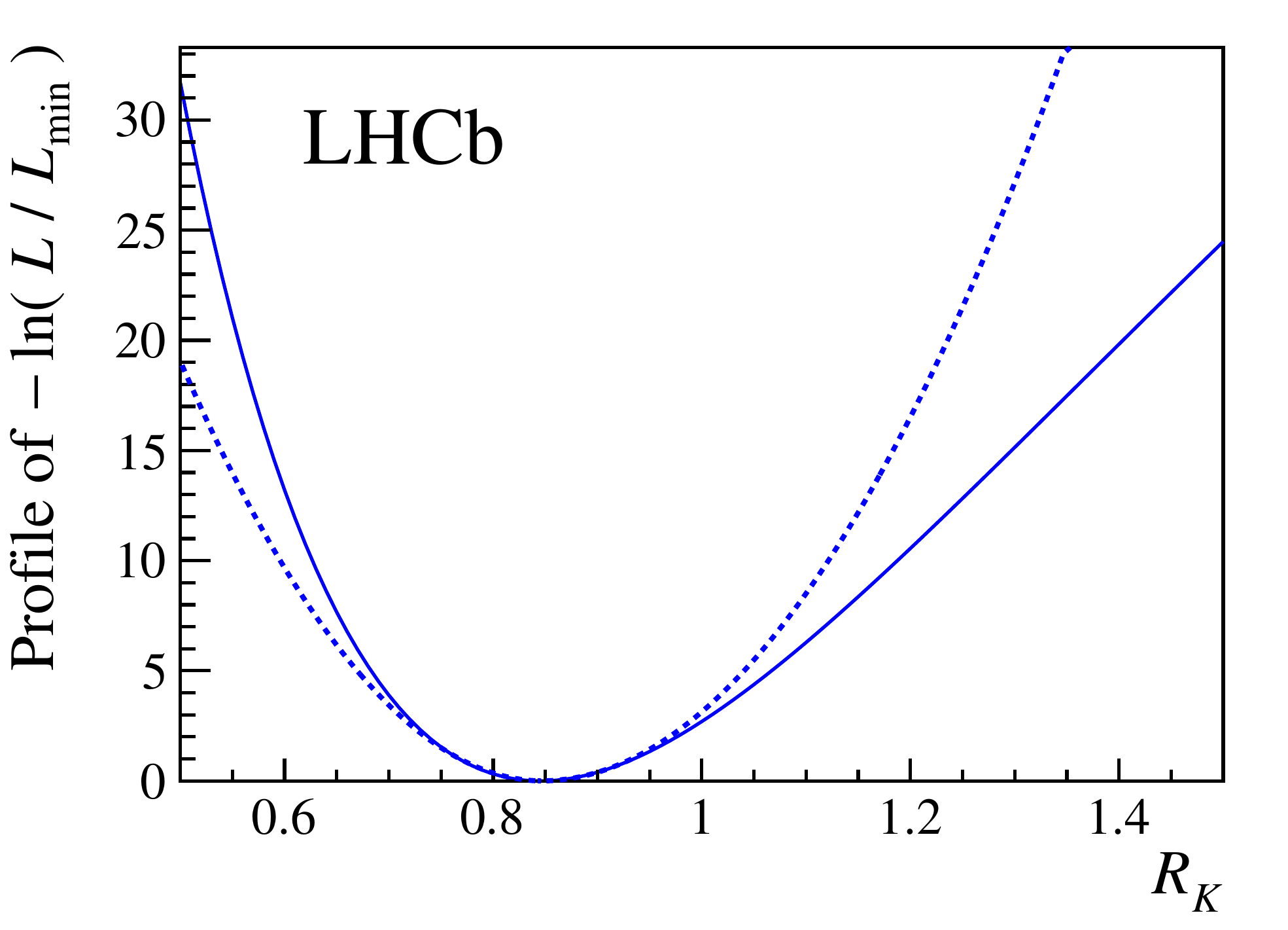}
  \end{center}
     \caption{Likelihood function from the fit to the data profiled as a function of \RK (solid line). The blue dashed line depicts the expected shape of the likelihood profile if the uncertainties were Gaussian.}
    \label{fig:profile_likelihood}
\end{figure}

\clearpage
The profile likelihood for the fit is shown in Fig.~\ref{fig:profile_likelihood}. The likelihood is Gaussian to a reasonable approximation in the range $0.75<\RK<0.95$, but non-Gaussian effects can be seen outside of this range due to the comparatively low yield in the \BuKee decay.  

The \RK values derived from a fit to just the 7 and 8\tev data, and a fit to just the 13\tev data are 
\begin{eqnarray*}
\RKrunone  &=& \RKrunonevalue\,, \\
\RKruntwo  &=& \RKruntwovalue\,,
\end{eqnarray*}
where  the  first  set of uncertainties  are  statistical  and  the  second  systematic. The combination of these values, or a combination of the latter value with the previously published LHCb result~\cite{LHCb-PAPER-2014-024}, requires that correlations are properly taken into account, as is done in the simultaneous fit used to derive the \RK measurement given in the main body of the Letter.

\cleardoublepage
% LHCb Collaboration author list
% Data extracted on March 15th, 2019 at 5:39pm for reference date 12-Feb-2019
\centerline
{\large\bf LHCb collaboration}
\begin
{flushleft}
\small
R.~Aaij$^{29}$,
C.~Abell{\'a}n~Beteta$^{46}$,
B.~Adeva$^{43}$,
M.~Adinolfi$^{50}$,
C.A.~Aidala$^{77}$,
Z.~Ajaltouni$^{7}$,
S.~Akar$^{61}$,
P.~Albicocco$^{20}$,
J.~Albrecht$^{12}$,
F.~Alessio$^{44}$
% ...

{\footnotesize \it

% ...
$ ^{7}$Universit{\'e} Clermont Auvergne, CNRS/IN2P3, LPC, Clermont-Ferrand, France\\
% ...
$ ^{12}$Fakult{\"a}t Physik, Technische Universit{\"a}t Dortmund, Dortmund, Germany\\
% ...
$ ^{20}$INFN Laboratori Nazionali di Frascati, Frascati, Italy\\
% ...
$ ^{29}$Nikhef National Institute for Subatomic Physics, Amsterdam, Netherlands\\
% ...
$ ^{43}$Instituto Galego de F{\'\i}sica de Altas Enerx{\'\i}as (IGFAE), Universidade de Santiago de Compostela, Santiago de Compostela, Spain\\
$ ^{44}$European Organization for Nuclear Research (CERN), Geneva, Switzerland\\
% ...
$ ^{46}$Physik-Institut, Universit{\"a}t Z{\"u}rich, Z{\"u}rich, Switzerland\\
% ...
$ ^{50}$H.H. Wills Physics Laboratory, University of Bristol, Bristol, United Kingdom\\
% ...
$ ^{61}$University of Cincinnati, Cincinnati, OH, United States\\
% ...
$ ^{63}$Syracuse University, Syracuse, NY, United States\\
% ...
$ ^{77}$University of Michigan, Ann Arbor, United States, associated to $^{63}$\\
% ...
}
\end{flushleft}

\end{document}